\documentclass[12pt,a4paper,twoside]{article}

\usepackage[dvips]{graphics}
\usepackage{cite}
\usepackage{bigstrut}
\usepackage{amssymb}
\usepackage{rotating}

\newcommand{\qqbar}  {\ensuremath{\mathrm{q\overline{q}}}}

\newcommand{\lnu}    {\ensuremath{\ell\nu}}

\newcommand{\epem}   {\ensuremath{\mathrm{e^+e^-}}}

\newcommand{\sigtot}     {\ensuremath{\sigma_{\mathrm{tot.}}}}
\newcommand{\as}     {\ensuremath{\alpha_{\mathrm{s}}}}
\newcommand{\assq}     {\ensuremath{\alpha_{\mathrm{s}}^2}}

\newcommand{\asb}    {\ensuremath{\overline{\alpha}_{\mathrm{s}}}}
\newcommand{\asbsq}    {\ensuremath{\overline{\alpha}_{\mathrm{s}}}^2}

\newcommand{\asmz}   {\ensuremath{\alpha_{\mathrm{s}}(\mz)}}
\newcommand{\oa}     {\ensuremath{\mathcal{O}(\alpha_{\mathrm{s}})}}
\newcommand{\oaa}    {\ensuremath{\mathcal{O}(\alpha_{\mathrm{s}}^2)}}
\newcommand{\oaaa}   {\ensuremath{\mathcal{O}(\alpha_{\mathrm{s}}^3)}}

\newcommand{\zzero}  {\ensuremath{\mathrm{Z}}}

\newcommand{\mz}     {\ensuremath{M_{\mathrm{Z}}}}

\newcommand{\nf}     {\ensuremath{n_{\mathrm{F}}}}

\newcommand{\ecm}   {\ensuremath{E_{\mathrm{c.m.}}}}

\newcommand{\bt}     {\ensuremath{B_{\mathrm{T}}}}
\newcommand{\bw}     {\ensuremath{B_{\mathrm{W}}}}
\newcommand{\bn}     {\ensuremath{B_{\mathrm{N}}}}
\newcommand{\cp}     {\ensuremath{C}}
\newcommand{\mh}     {\ensuremath{M_{\mathrm{H}}}}
\newcommand{\ml}     {\ensuremath{M_{\mathrm{L}}}}
\newcommand{\thr}    {\ensuremath{1-T}}
\newcommand{\tma}    {\ensuremath{T_{\mathrm{maj.}}}}
\newcommand{\tmi}    {\ensuremath{T_{\mathrm{min.}}}}
\newcommand{\chisq}  {\ensuremath{\chi^2}}
\newcommand{\chisqd} {\ensuremath{\chi^2/\mathrm{d.o.f.}}}
\newcommand{\momone}[1] {\mbox{\ensuremath{\langle#1\rangle}}}
\newcommand{\momn}[2] {\mbox{\ensuremath{\langle#1^{#2}\rangle}}}
\newcommand{\xmu}    {\ensuremath{x_{\mu}}}
\newcommand{\xl}     {\ensuremath{x_{\scriptscriptstyle L}}}

\newcommand{\ycut}   {\ensuremath{y_{\mathrm{cut}}}}

\newcommand{\ytwothree}   {\ensuremath{y^{\mathrm{D}}_{23}}}

\newcommand{\twothirds}   {\ensuremath{\frac{2}{3}}}
\newcommand{\rs}     {\ensuremath{\sqrt{s}}}

\newcommand{\ww}     {\ensuremath{\mathrm{W^+W^-}}}

\newcommand{\bm}[1]  {\mbox{\boldmath\ensuremath{#1}}}

\newcommand{\dd}    {\ensuremath{\mathrm{d}}}
\newcommand{\Opal}{\mbox{\rm OPAL}}
\newcommand{\Aleph}{\mbox{\rm ALEPH}}
\newcommand{\Delphi}{\mbox{\rm DELPHI}}
\newcommand{\Lep}{\mbox{LEP}}
\newcommand{\Cern}{\mbox{CERN}}
\newcommand{\isr}{\mbox{ISR}}
\newcommand{\qcd}{\mbox{QCD}}
\newcommand{\nlla}{\mbox{NLLA}}
\newcommand{\lo}{\mbox{LO}}
\newcommand{\nlo}{\mbox{NLO}}

\newcommand{\dof}{\mbox{d.o.f.}}
\newcommand{\Leptwo}{\mbox{\rm LEP II}}
\newcommand{\Lepone}{\mbox{LEP I }}
\newcommand{\Herwig}{\mbox{HERWIG}}
\newcommand{\Pythia}{\mbox{PYTHIA}}
\newcommand{\Jetset}{\mbox{JETSET}}
\newcommand{\Event}{\mbox{EVENT}}
\newcommand{\Koralw}{\mbox{KORALW}}
\newcommand{\kkff}{\mbox{$\mathcal{KK}$2f}}
\newcommand{\Ariadne}{\mbox{ARIADNE}}
\newcommand{\grcff}{\mbox{grc4f}}
\newcounter{hours}
\newcounter{minutes}

\newcommand{\Printtime}{%
  \setcounter{hours}{\time/60}%
  \setcounter{minutes}{\time-\value{hours}*60}%
  \ifthenelse{\value{hours}<10}{0}{}\thehours:%
  \ifthenelse{\value{minutes}<10}{0}{}\theminutes}

\begin{document}

%
\begin{titlepage}
\begin{center}{\large   EUROPEAN ORGANIZATION FOR NUCLEAR RESEARCH
}\end{center}\bigskip
\begin{flushright}
       CERN-EP/04-044   \\ 10$^{\mathrm{th}}$~August~2004
\\ Journal version 25$^{\mathrm{th}}$~November~2004
\end{flushright}
\bigskip\bigskip\bigskip\bigskip\bigskip
\begin{center}
{\huge\bf
Measurement of event shape distributions 
and moments \\ in \bm{\epem\rightarrow\;}hadrons  
at 91--209 GeV \\\vspace{3mm} and a determination of \bm{\as}
}
\end{center}
\bigskip\bigskip
\begin{center}
{\LARGE The OPAL Collaboration}
\end{center}
\bigskip\bigskip\bigskip
\begin{center}{\large  Abstract}\end{center}
We have studied hadronic events from \epem\ annihilation data at
centre-of-mass energies from 91 to 209~GeV. 
We present distributions of event shape observables 
and their moments
at each energy and compare with \qcd\ Monte Carlo models.
From the event shape distributions we extract the
strong coupling \as\ and test its evolution with energy scale.
The results are consistent with the running of \as\
expected from \qcd.
Combining all data, the value of \asmz\ is determined to be
\[ \asmz =  0.1191
\pm0.0005~\mathrm{(stat.)}
\pm0.0010~\mathrm{(expt.)}
\pm0.0011~\mathrm{(hadr.)}
\pm0.0044~\mathrm{(theo.)}\;. \]
The energy evolution of the moments is also used to determine 
a value of \as\ with slightly larger errors:
$\asmz=0.1223\pm 0.0005(\mathrm{stat.})\pm 0.0014(\mathrm{expt.})\pm 0.0016(\mathrm{hadr.}) ^{+0.0054}_{-0.0036}(\mathrm{theo.})$.

\bigskip\bigskip
%
\bigskip
\begin{center}
{\large 
(Submitted to European Physical Journal C) }
\end{center}
 
%
\begin{center}
%





\end{center}

\end{titlepage}

\begin{center}{\Large        The OPAL Collaboration
}\end{center}\bigskip
\begin{center}{
G.\thinspace Abbiendi$^{  2}$,
C.\thinspace Ainsley$^{  5}$,
P.F.\thinspace {\AA}kesson$^{  3,  y}$,
G.\thinspace Alexander$^{ 22}$,
J.\thinspace Allison$^{ 16}$,
P.\thinspace Amaral$^{  9}$,
G.\thinspace Anagnostou$^{  1}$,
K.J.\thinspace Anderson$^{  9}$,
S.\thinspace Asai$^{ 23}$,
D.\thinspace Axen$^{ 27}$
% ...
}\end{center}\bigskip
\bigskip
$^{  1}$School of Physics and Astronomy, University of Birmingham,
Birmingham B15 2TT, UK
\newline
$^{  2}$Dipartimento di Fisica dell' Universit\`a di Bologna and INFN,
I-40126 Bologna, Italy
\newline
$^{  3}$Physikalisches Institut, Universit\"at Bonn,
D-53115 Bonn, Germany
\newline
% ...
$^{  5}$Cavendish Laboratory, Cambridge CB3 0HE, UK
\newline
% ...
$^{  9}$Enrico Fermi Institute and Department of Physics,
University of Chicago, Chicago IL 60637, USA
\newline
% ...
$^{ 16}$Department of Physics, Schuster Laboratory, The University,
Manchester M13 9PL, UK
\newline
% ...
$^{ 22}$Department of Physics and Astronomy, Tel Aviv University,
Tel Aviv 69978, Israel
\newline
$^{ 23}$International Centre for Elementary Particle Physics and
Department of Physics, University of Tokyo, Tokyo 113-0033, and
Kobe University, Kobe 657-8501, Japan
\newline
% ...
$^{ 27}$University of British Columbia, Department of Physics,
Vancouver BC V6T 1Z1, Canada
\newline
% ...
$^{  y}$ now at CERN
\newline
% ...
\newpage

%
%
%

\newpage
\section{Introduction}
Hadronic final states produced in the process
$\epem\rightarrow\qqbar$ are a valuable testing ground for the
theory of the strong interaction in the Standard Model,
Quantum Chromodynamics (\qcd).  The hadronic system 
in the energy range considered here
is complex, consisting of typically 20--50 hadrons.
Many ``event shape'' observables have been devised which provide 
a convenient way of characterizing the main features of
such events. Analytic \qcd\ predictions of the distributions
of several of these event shape observables have been presented
in the literature (see e.g.\ ref.~\cite{ESW}), 
and can be used to determine the crucial 
free parameter of \qcd\ --- the coupling strength \as.
These predictions describe the distributions of quarks and gluons, 
while the distributions of hadrons are measured in the data.
In confronting the data with theory, Monte Carlo models of the
hadronization process are commonly used to relate the partons and hadrons.
Analytic \qcd\ predictions have also been made
for the moments of event shape distributions, whose
evolution with centre-of-mass (c.m.) energy permit
complementary determinations of \as. 
The determination of \as\ from many different observables
provides an important test of the consistency of \qcd.
In addition, measurements of event shape distributions
have proved invaluable for testing and tuning
Monte Carlo models of hadron production in $\epem\rightarrow\;$hadrons. 

In this paper we present a coherent analysis of event shape
distributions and moments using data collected by 
the \Opal\ detector at 12 c.m.\ energy points covering 
the \Lep\ c.m.\ energy range of $\rs\equiv\ecm=91$--209~GeV.
Results at 192--209~GeV 
are published for the first time here.
Partial results at 91--189 GeV have been published by \Opal\
previously~\cite{OPALPR075,OPALPR158,OPALPR197,OPALPR303};
these are superseded by the present measurements
in order that the data at all energies can be analysed and 
interpreted in a consistent manner.
The results  presented here  
use identical analysis procedures throughout, 
and some event shape observables are included for the first time.  
In several cases, improved theoretical calculations are now available,
as described in Sect.~\ref{sec_theo}. 
The results at 91~GeV are based on
calibration data taken during the \Leptwo\ running period
(the period from 1996 onwards when \Lep\ operated well above the
\zzero\ mass);
these share the same detector configuration (slightly different from 
that used in the earlier \Lepone phase 
when \Lep\ operated close to the \zzero\ peak)
and reconstruction code as the higher energy data, 
which means that we can compare results over a wide energy 
range with minimal systematic differences between energies. 
Similar results from other \Lep\ collaborations can be
found in refs.~\cite{alephas133,alephas209,delphias133,delphias183,delphias209,l3as133,l3as161172,l3as183,l3as189,l3as209}.
Another recent \Opal\ paper~\cite{OPALjets} uses the same data sample
as the present study to measure jet rates.

The structure of the paper is as follows.
In Sect.~\ref{sec_detector} we give a brief description of the \Opal\
detector, and in Sect.~\ref{sec_datamc} we summarize the data and
Monte Carlo samples used.  The theoretical background to the work is 
outlined in Sect.~\ref{sec_theo}.  The experimental analysis 
techniques are explained in Sect.~\ref{sec_anal} before the measurements
are presented and compared with theory in Sect.~\ref{sec_results}.

\section{ The OPAL detector}
\label{sec_detector}

The \Opal\ 
detector was operated at the \Lep\ \epem\ collider at \Cern\ 
from 1989 to 2000.  A detailed
description can be found in ref.~\cite{opaltechnicalpaper}. The
analysis presented here relies mainly on the measurements of momenta
and directions of charged particles in the tracking chambers and of
energy deposited in the electromagnetic calorimeters of
the detector.
 
All tracking systems were located inside a solenoidal magnet which
provided a uniform axial magnetic field of 0.435~T along the beam
axis\footnote{In the \Opal\ coordinate system the $x$-axis points
  towards the centre of the \Lep\ ring, the $y$-axis points 
  approximately upwards and
  the $z$-axis points in the direction of the electron beam.  The
  polar angle $\theta$ and the azimuthal angle $\phi$ are defined
  w.r.t.\ $z$ and $x$, respectively, while $r$ is the distance from the
  $z$-axis.}.
The magnet was surrounded by a lead glass electromagnetic
calorimeter and a hadron calorimeter of the sampling type.  Outside
the hadron calorimeter, the detector was surrounded by a system of muon
chambers.  There were similar layers of detectors in the forward and
backward endcaps.
 
The main tracking detector was the central jet chamber. This device was
approximately 4~m long and had an outer radius of about 1.85~m. It had 24
sectors with radial planes of 159 sense wires spaced by 1~cm. 
The electromagnetic calorimeters in the barrel and the endcap sections
of the detector consisted of 11704 lead glass blocks with a depth of
$24.6$ radiation lengths in the barrel and more than $22$ 
radiation lengths in the endcaps.
 
\section{ Data and Monte Carlo samples }
\label{sec_datamc}
The data used here were recorded from 1995--2000 using the 
\Opal\ detector at \Lep.  In 1995 the \Lep\ c.m.\ energy was 
increased above the vicinity of the \zzero\ peak in runs at
$\ecm=130$ and 136~GeV.  By 2000, the maximum c.m.\ 
energy had reached 209~GeV.  All of the data recorded above the 
\zzero\ peak are analysed in the present study.  In addition, 
interspersed at various points during the high energy \Lep\ running, 
calibration runs were taken on the \zzero\ peak, at $\sqrt{s}=91.3$~GeV.
These data were recorded with identical detector configuration
and performance, and reconstructed with the same code, 
as the high energy data.
For the purpose of analysis, the
data have been grouped into small energy ranges, which are often
merged into larger ranges for clarity of presentation.
Table~\ref{datasummary} summarizes the c.m.\ energy points used,
the integrated luminosities at each point and the 
numbers of events employed for analysis after the selection
described in Sect.~\ref{sec_selec}.

Samples of Monte Carlo simulated events were used to
correct the data for experimental acceptance, efficiency and backgrounds.
The process $\epem\rightarrow\qqbar$ was simulated using
\Jetset~7.4~\cite{jetset3} at $\sqrt{s}=91.2$~GeV, and at higher energies
using \kkff~4.01 or \kkff~4.13~\cite{kk2f} with 
fragmentation performed using \Pythia~6.150 or \Pythia~6.158~\cite{jetset3}.
Corresponding samples using \Herwig~6.2\cite{herwig} or 
\kkff\ with  \Herwig~6.2 fragmentation were used for systematic checks.  
Four-fermion background processes were
simulated using \grcff~2.1~\cite{grc4f} or \Koralw~1.42~\cite{koralw} 
with \grcff~\cite{grc4f} matrix elements and with 
 fragmentation performed using 
\Pythia.  The above samples, generated at each energy point studied, 
were processed through a full simulation of the 
\Opal\ detector~\cite{gopal}, and 
reconstructed in the same way as real data.
In addition, for comparisons with the corrected data, and when 
correcting for the effects of fragmentation, large samples
of generator-level Monte Carlo events were employed, using the
parton shower 
models \Pythia~6.158, \Herwig~6.2 and 
\Ariadne~4.11\cite{ariadne3}.
Each of these fragmentation 
models contains a number of tunable parameters; 
these were adjusted by tuning to previously published \Opal\ data at 
$\sqrt{s}\sim91$~GeV as described in ref.~\cite{OPALPR141} for 
\Pythia/\Jetset\ and in ref.~\cite{OPALPR379} for \Herwig\ and \Ariadne.

\section{ Theoretical background}
\label{sec_theo}
\subsection{Event shape observables}
\label{sec_evshapedef}

The properties of hadronic events may be described by a set of
event shape observables.  These may be used to characterize the
distribution of particles in an event as ``pencil-like'',
planar, spherical, etc.
They can be computed either using the measured charged particles
and calorimeter clusters, or using the true hadrons or partons in
simulated events.
The following event shapes are considered here:
\begin{description}
\item[Thrust \bm{T}:]
  defined by the expression\cite{thrust1,thrust2}
  \begin{equation}
  T= \max_{\vec{n}}\left(\frac{\sum_i|p_i\cdot\vec{n}|}
                    {\sum_i|p_i|}\right)\;,
  \label{equ_thrust}
  \end{equation}
  where $p_i$ is the three-momentum of particle $i$ and
  the summation runs over all particles, which may be 
  the measured particles, or the true hadrons 
  or partons in Monte Carlo events.  
  The thrust axis $\vec{n}_T$ is the direction $\vec{n}$ which
  maximises the expression in parentheses.  A plane through the origin
  and perpendicular to $\vec{n}_T$ divides the event into two
  hemispheres $H_1$ and $H_2$.
\item[Thrust major \bm{\tma}:]
The maximization in equation~(\ref{equ_thrust}) is performed subject to 
 the constraint that
  $\vec{n}$ must lie in the plane perpendicular to $\vec{n}_T$. The
  resulting vector is called $\vec{n}_{\tma}$.
\item[Thrust minor \bm{\tmi}:]
  The expression in parentheses in equation~(\ref{equ_thrust})
 is evaluated for the vector
  $\vec{n}_{\tmi}$ which is perpendicular to both $\vec{n}_T$ and 
  $\vec{n}_{\tma}$.
\item[Oblateness \bm{O}:] 
  This observable is defined by $O=\tma-\tmi$ \cite{def_o}.
\item[Sphericity \bm{S} and Aplanarity \bm{A}:] 
  These observables are based on the momentum tensor
  \begin{equation}
    S^{\alpha\beta}= \frac{\sum_ip_i^{\alpha}p_i^{\beta}}{\sum_ip_i^2}\;,
    \;\;\;\alpha,\beta= 1,2,3\;,
  \end{equation}
  where the sum runs over particles, $i$, and $\alpha$ and $\beta$ denote 
  the cartesian coordinates of the momentum vector. 
  The three eigenvalues $Q_j$ of $S^{\alpha\beta}$ are ordered such
  that $Q_1<Q_2<Q_3$. These then define $S$ \cite{def_s1,def_s2} and $A$
  \cite{def_a} by
 \begin{equation}
    S= \frac{3}{2}(Q_1+Q_2)\;\;\;\mathrm{and}\;\;\; A= \frac{3}{2}Q_1\;.
  \end{equation}
\item[\bm{C}- and \bm{D}-parameters:]
  The momentum tensor $S^{\alpha\beta}$ is modified to become
  \begin{equation}
    \Theta^{\alpha\beta}= \frac{\sum_i(p_i^{\alpha}p_i^{\beta})/|p_i|}
                               {\sum_i|p_i|}\;\;\;,
                           \;\;\;\alpha,\beta= 1,2,3\;.
  \end{equation}
  The three eigenvalues $\lambda_j$ of this tensor define \cp
  \cite{def_c} through
 \begin{equation}
    \cp= 3(\lambda_1\lambda_2+\lambda_2\lambda_3+\lambda_3\lambda_1)\;
  \end{equation}
and $D$ through
\begin{equation}D=27\lambda_1\lambda_2\lambda_3 \;.\end{equation}
\item[Jet Masses \bm{\mh} and \bm{\ml}:] The hemisphere 
invariant masses are calculated using  the particles
  in the two hemispheres $H_1$ and $H_2$.   We define
  \mh\ \cite{def_mh1,def_mh2} as the heavier mass, divided by $\rs$,
and  \ml\ as the lighter mass, likewise divided by $\rs$.
\item[Jet Broadening observables \bm{\bt}, \bm{\bn} and \bm{\bw}:] 
  These are defined by computing the quantity
  \begin{equation}
    B_k= \left(\frac{\sum_{i\in H_k}|p_i\times\vec{n}_T|}
                    {2\sum_i|p_i|}\right)
  \label{equ_btbw}
  \end{equation}
  for each of the two event hemispheres, $H_k$,  defined above.
  The three observables \cite{nllabtbw} are defined by
  \begin{equation}
    \bt= B_1+B_2\;,\;\;\bn= \min(B_1,B_2)\;\;\;\mathrm{and}
\;\;\;\bw= \max(B_1,B_2)\;
  \end{equation}
  where \bt\ is the total, \bn\ is the narrow and 
\bw\ is the wide jet broadening.
\item[Transition value between 2 and 3 jets {\boldmath \ytwothree}:]
  The value of the jet resolution parameter, \ycut, 
  at which the
  event makes a transition between a  2-jet and a 3-jet assignment,
  for the Durham jet finding scheme~\cite{durham}.
\end{description}
 
In the following discussion, whenever we wish to
 refer to  a generic event shape observable we use the symbol $y$. 
In almost all cases, 
larger values of $y$ indicate regions
dominated by the radiation of hard gluons and small values of $y$
indicate the region influenced by multiple soft gluon radiation. 
Note that thrust $T$ forms an exception to this rule, as the value
of $T$ reaches unity for events consisting of two collimated
back-to-back jets. We therefore use $y=\thr$ instead. 

For all of these event shapes, a perfectly collimated 
(``pencil-like'') two-jet
final state will have $y=0$.  
\oa\ \qcd\ processes generate planar $\qqbar\mathrm{g}$
configurations; for most of the observables, 
these processes will generate contributions
at $y\neq0$ --- these are sometimes referred to as ``three-jet'' 
observables.  However,
five of the observables (\tmi, \ml, \bn, $D$ and $A$), 
are still zero at \oa, i.e.\ for planar events, 
and receive their leading contributions 
at \oaa\ --- these are referred to as ``four-jet'' observables.  

\subsection{QCD predictions for event shape distributions}
\label{sec-theocalc}
\qcd\ perturbation theory may be used to make predictions
for event shape observables~\cite{ESW}.  In order that these predictions
be reliable, it is necessary that the value of the observable be 
infra-red stable (i.e.\ unaltered under the emission of soft gluons)
and collinear stable (i.e.\ unaltered under collinear parton branchings).

The \qcd\ matrix elements in \epem\ annihilations are fully known to
\oaa~\cite{ERT}, i.e.\ to next-to-leading order (\nlo) for 
those observables dominated by three parton final states.
In the two-jet (low $y$) region, however, the effect of soft and 
collinear emissions introduces large logarithmic 
effects depending on $L=\log(1/y)$, such that the leading dependence
on \as\ and $L$ at each order, $n$, is proportional to $\as^n L^{n+1}$. 
For six of the 
three-jet observables studied here ($(1-T)$, \mh, \bt, \bw, $C$ and
\ytwothree), these large logarithms can
be resummed to next-to-leading order, referred to as the 
next-to-leading-logarithmic approximation (\nlla).  
The most complete \qcd\ predictions come from combining the
\oaa\ and \nlla\ predictions, taking care not to double count those 
terms which are in common between them.
Further details may be found in ref.~\cite{LEPQCDWG1,OPALPR075}, and only a brief 
outline of the procedure is given below.

The \qcd\ calculations make predictions for the cumulative cross-section
\begin{equation}
R(y) \equiv \int_0^y \frac{1}{\sigma}\frac{\dd\sigma}{\dd y} \dd y
\label{eq-match1}
\end{equation}
which take the following form for the \nlla\ 
calculations~\cite{nllathmh,nllay23,nllabtbw2,nllacp}:
\begin{equation}
R_{\mathrm{NLLA}}(y)=(1+C_1\asb+C_2\asbsq) \exp [
L g_1(\as L) + g_2(\as L) ] \;\;,
\label{eq-match4}
\end{equation}
where for brevity we write $\asb$ for $(\as/2\pi)$.
The corresponding formula for the \oaa\ prediction is:
\begin{equation}
R_{\oaa}(y)= 1+ \mathcal{A}(y) \asb + \mathcal{B}(y) \asbsq \;\; .
\label{eq-match5}
\end{equation}

For the analysis presented here, the ``$\log(R)$'' matching 
scheme~\cite{nllathmh} is adopted for combining 
the \oaa\ and \nlla\ predictions.
This matching scheme involves taking the logarithm of 
equation~(\ref{eq-match5}) and expanding as a power series, yielding:
\begin{equation}
\ln R_{\oaa}(y)= \mathcal{ A}(y) \asb + 
\left[\mathcal{ B}(y) - \frac{1}{2} \mathcal{ A}(y)^2\right] \asbsq + \oaaa \;\;,
\label{eq-match6}
\end{equation}
and similarly rewriting equation~(\ref{eq-match4}) as:
\begin{equation}
\ln R_{\mathrm{NLLA}}(y)= L g_1(\as L) + g_2(\as L)
+ C_1 \asb + [ C_2 - \frac{1}{2}  C_1^2] \asbsq + \oaaa \;\;.
\label{eq-match6a}
\end{equation}
In the $\log(R)$ matching scheme the terms up to \oaa\ 
in the \nlla\ expression, equation~(\ref{eq-match6a}), are replaced
by the \oaa\ terms from equation~(\ref{eq-match6}).

Both the \oaa\ and \nlla\ \qcd\ predictions depend on the 
choice of renormalization scale, $\mu$ (see ref.~\cite{ESW} for example). 
The {\it renormalization scale factor}
is defined as  \mbox{$\xmu=\mu/\ecm$} where $\as(\mu)$ is the 
expansion parameter which appears 
in the \nlo\ perturbative predictions above.
Na\"{\i}vely $\mu$ would be expected to be of order of, 
but not necessarily equal to, \ecm. 
A \qcd\ calculation to all orders
should be independent of $\xmu$, but a truncated fixed order calculation 
does in general depend on $\xmu$.  For example, in the \oaa\ 
calculation, the second order coefficient $\mathcal{B}(y)$
has to be replaced by 
$\mathcal{B}(y)+\beta_0\ln\xmu\mathcal{A}(y)$ where 
$\beta_0=11-\twothirds\nf$ 
is the leading order $\beta$-function coefficient of the 
renormalization group equation and $\nf=5$ is the number of 
active quark flavours.  Similar modifications apply to the 
\nlla\ calculations.   In addition, the 
theoretical cross-sections are usually 
normalized to the Born cross-section $\sigma_0$ while the
experimental distributions are normalized to
the total hadronic cross-section, $\sigtot$, which itself depends 
on $\as$:
\begin{equation}
 \sigtot=\sigma_0\left(1+\frac{\as}{\pi}\right)+\mathcal{O}(\assq)
\end{equation}
and this is taken into account
by means of the replacement $\mathcal{B}(y)\rightarrow\mathcal{B}(y)-2\mathcal{A}(y)$.

The analysis of \as\ presented in this paper incorporates several 
improvements in the theoretical calculations and errors compared 
with previous determinations by \Opal\ based on 
\nlla+\oaa\ \qcd.  The principal changes are as follows:
\begin{itemize}
\item Improvements have been made in the \nlla\ theory predictions
for the jet broadenings~\bt\ and \bw~(ref.~\cite{nllabtbw2} 
superseding ref.~\cite{nllabjet}), 
and for \ytwothree~(ref.~\cite{nllay23}, superseding ref.~\cite{nllad2}, 
which 
was in turn an improvement on ref.~\cite{nlla-r2}). 
\item The \nlla\ calculations for the $C$-parameter~\cite{nllacp}
were not available at the time of the earliest \Opal\ publications.
\item The \nlla\ resummations do not automatically force each event
shape distribution to vanish at the edge of phase space; missing
subleading terms can result in a non-zero prediction outside the
kinematically allowed range of the
observable. A remedy
for this situation involves the substitution
\begin{equation}
L=\ln\left(\frac{1}{y}\right)\;\to\;\widetilde{L}=
\ln \left(\frac{1}{y}-\frac{1}{y_\mathrm{max}}+1\right)\;.
\label{modifiedlog}
\end{equation}
This method~\cite{nllathmh} was known at the time of the original
\Opal\ \Lepone\ analysis~\cite{OPALPR075}, and was investigated as an
alternative to the unmodified \nlla\ prediction. However, this is the
first time it has been
adopted as the standard for \as\ measurements by the \Opal\
Collaboration. 
\item The fixed order coefficients $\mathcal{A}(y)$ and $\mathcal{B}(y)$ 
are now computed using the \Event2 Monte Carlo program~\cite{event2}.
This superseded an earlier program, \Event~\cite{KN}.  
Although both programs are based on the same \oaa\ matrix elements
from ref.~\cite{ERT}, \Event2 includes an improved algorithm to
handle cancellations between real and virtual processes, and therefore
permits a more precise determination of the coefficients. 
\item A more sophisticated technique for assessing the 
theoretical errors associated with missing higher order terms 
in the theory has been adopted, based on the extensive studies 
carried out within the \Lep\ \qcd\ working group~\cite{LEPQCDWG1};
see Sect.~\ref{sec_syserr} for further details.
\end{itemize} 

The theoretical calculations described above
provide predictions of {\em parton-}level distributions, i.e.\ distributions of 
quarks and gluons.  In contrast, the data are corrected to the {\em hadron-}level,
i.e.\ they correspond to the distributions of the stable particles
(including photons and both charged and neutral leptons)
in the event as explained in Sect.~\ref{sec-corrproc}.
In order to confront the
theory with the hadron-level data, it is necessary to 
correct the theory for the effects of soft fragmentation 
and hadronization.  This was done using large samples of events 
(typically $10^7$ events) generated using 
the parton-shower Monte Carlo programs \Pythia\ (used by default),
 \Herwig\ and \Ariadne\ (used for systematic error estimates).  The analytical 
theoretical predictions for the cumulative distribution $R(y)$
were multiplied by the Monte Carlo prediction for the
ratio of $R(y)$ at hadron-level to $R(y)$ at  parton-level\footnote{It should be noted that the
resummed theoretical calculations apply for massless quarks, while the quarks in 
the Monte Carlo models do have masses.  No attempt was made to correct for this.
In previous \Opal\ papers~\cite{OPALPR075,OPALPR158,OPALPR197,OPALPR303}, 
a systematic error was estimated for this effect, and proved
to be smaller than the other hadronization errors.  For consistency with the 
other LEP experiments, we now neglect this.}.

\subsection{QCD predictions for event shape moments}
\label{sec_qcdmon}
The moments of the distribution of an event shape observable $y$ are
defined by 
\begin{equation}\momn{y}{n}=\int_0^{y_{\mathrm{max}}} y^n 
\frac{1}{\sigma} \frac{\dd\sigma}{\dd y} \dd y \;,\end{equation}
where $y_{\mathrm{max}}$ is the kinematically allowed upper limit of the
observable.  The calculations always involve a full integration over
phase space, which implies that comparison with data always probes all
of the available phase space.  This is in contrast to \qcd\ predictions
for distributions; these are commonly only compared with data in
restricted regions, where the theory is able to describe the data
well.  Comparisons of \qcd\ predictions for moments of event shape
distributions with data are thus complementary to tests of the theory
using the differential distributions.

The formula for the \oaa\ \qcd\ prediction of \momn{y}{n}\ is
\begin{equation}
  \momn{y}{n} = {\cal A}_n \asb + {\cal B}_n \asbsq
\label{eq_qcdmom}
\end{equation}
involving 
the \oa\ coefficients ${\cal A}_n$ and \oaa\ coefficients ${\cal
B}_n$.  The values of the coefficients ${\cal A}_n$ and ${\cal B}_n$
can be obtained in the same way as described above by running \Event2.
The renormalization scale dependence and correction to 
$\sigtot$
is implemented in the same way as described above for
distributions.

The \qcd\ predictions were transformed from the parton- to the hadron-level
by multiplying by the ratio of Monte Carlo predictions for the values
of moments at the hadron- and parton-level.  
As for the distributions, \Pythia\ was used as the standard with
\Herwig\ and \Ariadne\ employed for the estimation of systematic errors.  

\section{ Analysis methods }
\label{sec_anal}
\subsection{ Selection of events }
\label{sec_selec}

The selection of events for this analysis consists of 
three main stages: the identification of hadronic event candidates, 
the removal of events with a large amount of initial-state radiation 
(\isr) for events at 130~GeV and above, 
and the removal of four-fermion background events
for events above 160~GeV, i.e.\ above the \ww\ production threshold.

The selection of hadronic events was based on simple cuts on
event multiplicity (to remove leptonic final states) and on
visible energy and longitudinal momentum balance (to remove 
two-photon events).  The cuts used at 91~GeV are documented in
ref.~\cite{OPALPR035}, while those used for higher c.m.\ energies
have reoptimized cut parameters, as described in
ref.~\cite{OPALPR388}.  Those parts of the \Opal\ detector
crucial for the present analysis
(electromagnetic calorimeter, jet chamber and trigger system)
were required to be fully operational.

Standard criteria were used to select good tracks and calorimeter energy 
clusters for subsequent analysis.  Charged particle tracks were 
required to have at least 40 hits in the jet chamber, and at least 
50\% of the maximum possible number of hits given the polar angle
of the track. The momentum transverse to the beam axis was required to
be at least 0.15~GeV.  Furthermore, 
the point of closest approach of the track to the collision 
axis was required to be less than 2~cm from the nominal collision point
in the $x$-$y$ plane and less than 25~cm in the $z$-direction.
Energy clusters in the electromagnetic calorimeter were required 
to have energies exceeding 0.10~GeV (0.25~GeV) in the barrel (endcap)
region of the detector.
The number of good charged particle tracks
was required to be greater than six.  After the above cuts 
the $\tau^+\tau^-$ and two-photon background was negligible.
Furthermore the polar angle of the thrust axis was required to satisfy
$|\cos\theta_{\mathrm{T}}|<0.9$ in order that the events 
be well contained in the detector acceptance.   

At energies significantly above \mz, the process of 
``radiative return'' to the \zzero\ is a common occurrence.
In order to study the properties of hadronic events at a well-defined
energy scale, it is necessary to eliminate events in which a large amount
of energy has been lost to \isr.  The effective centre-of-mass energy 
after \isr, $\sqrt{s'}$, was estimated for each selected event using
the algorithm described in ref.~\cite{OPALPR388}.  At c.m.\ energies
of 130~GeV and above, we demanded 
that $\sqrt{s}-\sqrt{s'}<10$~GeV in order to 
select a sample of predominantly non-radiative events.

At energies above the \ww\ production threshold, four-fermion events, 
especially those involving \qqbar\qqbar\ final states, become a substantial      
background.  These are reduced by using the standard \Opal\ \ww\ selection
procedure, which is based on a relative likelihood method~\cite{OPALPR321}.
The same likelihood technique 
has been applied at each c.m. energy studied, 
with the underlying reference distributions used as 
inputs to the likelihood calculation
recomputed for each energy or 
range of energies.  At  c.m.\ energies of 161~GeV and above,
the \qqbar\qqbar\
likelihood was required to satisfy $\mathcal{L}_{\qqbar\qqbar}<0.25$ and 
the \qqbar\lnu\ likelihood was required 
to satisfy $\mathcal{L}_{\qqbar\lnu}<0.5$.

After applying the above cuts, the numbers of selected non-radiative
\qqbar\ candidate events 
were as given in Table~\ref{datasummary}, and 
were consistent with expectations based on Monte Carlo 
simulated events\footnote{The numbers of events expected on the basis 
of Monte Carlo simulations
are given in all cases except for 91~GeV; to perform an accurate
prediction close to the \zzero\ peak would require a much more 
careful investigation of the beam energy and luminosity 
than is required for the present analysis.}.
After all cuts, the acceptance for non-radiative 
\qqbar\ events (defined for this 
purpose as those having $\sqrt{s}-\sqrt{s'}<1$~GeV) 
ranged from
88.5\% at 91~GeV (where the loss in acceptance is largely geometrical, 
arising from the $|\cos\theta_{\mathrm{T}}|<0.9$ requirement) to 
76.5\% at 207~GeV.  The residual four-fermion background was 
negligible below 161~GeV, and otherwise increased from 2.1\% at 161~GeV
to 6.2\% at 207~GeV.

\subsection{ Correction procedure }
\label{sec-corrproc}
For each accepted event, the value of each of the event shape observables
was computed.  In order to mitigate the effects of double counting of 
energy in tracking and calorimetry, 
a standard algorithm was adopted
which associated charged particle tracks with calorimeter clusters, and 
subtracted the estimated contribution of the charged particles from the 
calorimeter energy.  All selected tracks, and the electromagnetic
calorimeter clusters remaining after this procedure, were used in the 
evaluation of event shapes.  The event shapes were then formed into
histograms at each c.m.\ energy point.    In the cases
where data at more than
one c.m.\ energy were combined (e.g.\ 130 and 136~GeV), the data were simply 
summed, and corresponding Monte Carlo samples were created by combining 
samples generated at each energy weighted to correspond to the
integrated luminosities in data.

The expected number of residual four-fermion background events, $b_i$,
was then subtracted from the number of data events, $N_i$, 
in each bin, $i$, of each distribution.  
The effects of detector acceptance and resolution and of residual \isr\ were then 
accounted for by a simple bin-by-bin correction procedure.
For this procedure to be valid, it is necessary that the Monte Carlo model 
give a good description of the data and that the bin size be sufficiently large that
bin-to-bin migration is reasonably small and symmetric; 
these conditions are 
sufficiently well satisfied for the present
analysis. 
Two event shape distributions were formed for Monte Carlo simulated \qqbar\ events;
the first, at the {\it detector-}level, treated the Monte Carlo identically to the data, 
while the second, at the {\it hadron}-level, was computed using the true momenta of
the stable particles
in the event\footnote{ For this purpose, all particles having proper lifetimes
greater than $3\times10^{-10}$~s were regarded as stable.}, 
and was restricted
to events whose true $s'$ satisfied $\sqrt{s}-\sqrt{s'}<1$~GeV. 
The ratio of the hadron-level to the detector-level for each bin, 
$\alpha_i$, was used as a correction factor for the data, yielding the corrected
bin content $\widetilde{N}_i=\alpha_i(N_i-b_i)$.  
This corrected hadron-level distribution was then normalized to unity:
$P_i=\widetilde{N}_i/N$, where the sum $N=\sum_k\widetilde{N}_k$ 
includes any
underflow and overflow bins. 
Finally, the differential distribution $R'(y)$ was computed by dividing
$P_i$ by the bin width.
The covariance matrix
$V$ for $P_i$ was computed by transforming the diagonal Poisson 
covariance matrix for the uncorrected data~$N_i$:
\begin{equation}
V_{ij} = \sum_k \frac{\partial P_i}{\partial N_k}
                \frac{\partial P_j}{\partial N_k} N_k
       = \frac{1}{N^4} \sum_k \alpha_k^2 N_k
         \left(N\delta_{ik} - \widetilde{N}_i\right)
         \left(N\delta_{jk} - \widetilde{N}_j\right) \;.
\end{equation}
By setting $\alpha_i=1$ and $b_i=0$, i.e.\ $N_i=\widetilde{N}_i$, 
one may obtain the familiar expression for the covariance matrix 
of a multinomial distribution\footnote{  We note therefore 
that the covariance matrix of a multinomial distribution given by
$\widetilde{N}_i$, which has sometimes been used in this context,
is {\em not} identical to the result shown above, 
and hence not strictly correct.}.


The moments were calculated by accumulating sums over all selected
events:
\begin{equation}
N\momn{y}{n} = \sum_{i=1}^{N} y_i^n \;\;,
\end{equation}
where $N$ is the number of events.
These sums were corrected by subtracting the background contribution 
estimated
from simulated events.  Then the correction for experimental and \isr\
effects was performed by multiplying by a detector correction
coefficient obtained by taking the ratio of the
moments at the detector and the
hadron-level using simulated signal events.  The statistical errors of
the moments were calculated at the detector-level including the effects
of Monte Carlo statistics and were subjected to the same detector
correction.

%
%
%
\subsection{ Systematic uncertainties }
\label{sec_syserr}

Contributions to the systematic uncertainties 
affecting the corrected hadron-level 
distributions and moments in data were 
estimated by repeating the analysis with varied cuts or procedures.
In each case, the difference in each bin (or for each moment) with respect 
to the standard analysis was taken as a contribution to the systematic error.
\begin{itemize}
\item The containment cut was tightened to $|\cos\theta_{\mathrm{T}}|<0.7$.
\item The algorithm to compute $s'$ was replaced by a simpler version 
in which at most one initial state photon was accounted for.
\item The event shapes were computed using {\it all} tracks and 
electromagnetic calorimeter clusters.
The effects of double counting are then fully 
taken into account through the detector correction procedure.
\item The bin-by-bin corrections, $\alpha_i$, 
were computed using \Herwig\ instead of
\Pythia\ as the Monte Carlo hadronization model.
\item The cut on the \qqbar\qqbar\ four fermion likelihood
was tightened to $\mathcal{L}_{\qqbar\qqbar}<0.1$ and loosened to
$\mathcal{L}_{\qqbar\qqbar}<0.4$; the larger change resulting
was taken to be a systematic error.   
\item The cut on the \qqbar\lnu\ four fermion likelihood
was tightened to $\mathcal{L}_{\qqbar\lnu}<0.25$ and loosened to
$\mathcal{L}_{\qqbar\lnu}<0.75$; the larger change resulting
was taken to be a systematic error.   
\item The total four fermion background was varied by $\pm5\%$. 
\end{itemize}
The various contributions above, together with the
statistical error on the correction factors arising from limited Monte Carlo
statistics, were summed in quadrature to form the systematic error.
None of these sources of systematic error is dominant, but typically
the larger contributions arise from the use of \Herwig\ and the
use of all tracks and clusters.  In addition, at high \ecm\ and high $y$, 
the variation of the  $\mathcal{L}_{\qqbar\qqbar}$ cut is sometimes 
significant.

In assigning systematic errors to determinations of \as, 
all of the above contributions were taken into account
and are collectively referred to as the {\em experimental} errors.
In addition, two further sources of systematic error were
considered. 
\begin{itemize}
\item As explained above, 
when comparing \qcd\ with the data it is necessary to correct for
the effects of hadronization. The uncertainty associated with this 
{\em hadronization} correction
was assessed by using \Herwig~6.2 and \Ariadne~4.11 instead
of \Pythia.  The larger change in \as\ resulting from these two 
alternatives was taken to define the error.  It should be noted that
these models have already been tuned to similar data to those used here, 
and hence we adopt this, arguably conservative, prescription 
for assessing the error. 
\item The {\em theoretical} error, associated with missing higher order
terms in the theory, has traditionally been assessed by varying the 
renormalization scale factor, $\xmu$, 
described in Sect.~\ref{sec-theocalc}.
The predictions of a complete \qcd\ calculation 
would be independent of \xmu, 
but a finite-order calculation such as that used here retains some 
dependence on \xmu.  In previous \Opal\ analyses a range
$\frac{1}{2}\le\xmu\le2$ has been used, and this is the
procedure we adopt for the analysis of moments.   
Recently, extensive studies 
have been carried out within 
the \Lep\ \qcd\ working group~\cite{LEPQCDWG1}, 
which led to a more elaborate procedure being proposed, which
addresses the question of missing higher orders in further ways.
We adopt this procedure here for the analysis of distributions.  
A generalization of equation~(\ref{modifiedlog}) may be considered
\begin{equation}
\widetilde{L}= \frac{1}{p}
\ln \left(\frac{1}{{(\xl y)}^p}-\frac{1}{{(\xl y_\mathrm{max})}^p}+1\right)\;,
\end{equation}
in which $p$ determines how sharply the kinematic cutoff is applied, and
\xl\ acts as a scale factor on the event shape.
In addition to varying \xmu\ as above, these new parameters are varied in the
ranges $1\le p \le 2$ and $\frac{2}{3} \le \xl \le \frac{3}{2}$ 
(or $\frac{4}{9} \le \xl \le \frac{9}{4}$ for \ytwothree)\footnote{
We note that
\ytwothree\ varies quadratically with the transverse momentum
$k_{\mathrm{T}}$ of a radiated gluon, while the other observables 
vary linearly, so by applying this different scaling 
in the case of \ytwothree\ we ensure that
$\ln(k_{\mathrm{T}})$ is rescaled by the same 
amount for all observables.}.  
An additional
matching scheme ($R$-matching) is also considered.  The maximal uncertainty 
encompassing any of these variations in the theory is taken as the 
systematic error.  We take the average of the upper and lower uncertainty
bands, as defined in Ref.~\cite{LEPQCDWG1}, to define the 
theoretical error.
Ref.~\cite{LEPQCDWG1} should be consulted for 
further details.
\end{itemize}

\section{ Results }
\label{sec_results}
\subsection{ Event shape distributions }
\label{sec_shapes}

\begin{figure}[!htb]
\begin{center}
\resizebox{\textwidth}{!}
{\includegraphics{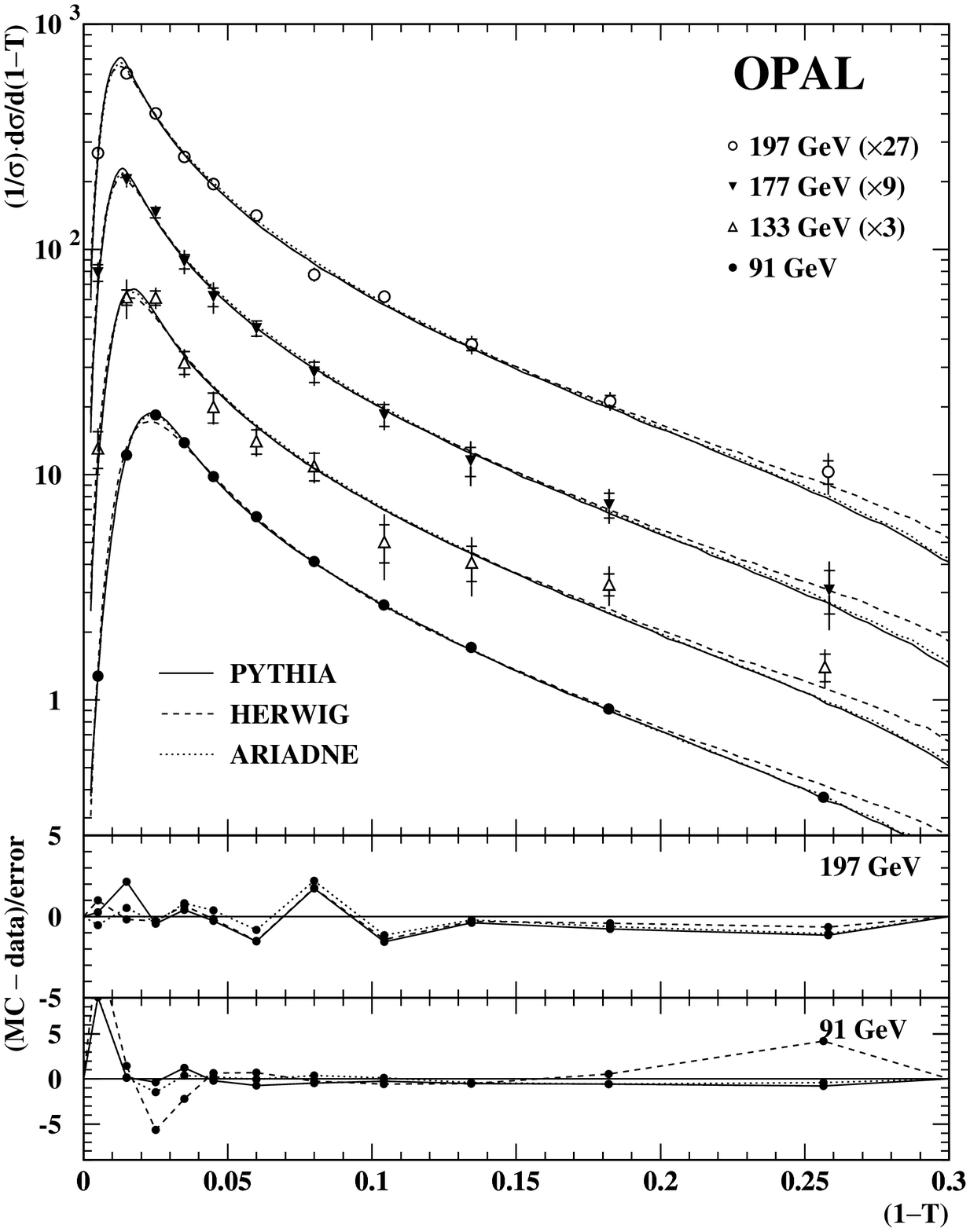}}
\caption[ ]{Distributions of thrust, $(1-T)$, 
at four c.m.\ energy points ---
91~GeV, 133~GeV, 161--183~GeV (labelled 177~GeV) and 189--209~GeV 
(labelled 197~GeV).  The latter three have been multiplied by factors
3, 9 and 27 respectively for the sake of clarity.  The inner error bars
show the statistical errors, while the total errors are indicated by the 
outer error bars.  The predictions of
the \Pythia, \Herwig\ and \Ariadne\ Monte Carlo models as described in 
the text are indicated by curves.  The lower panels of the figure
show the differences between data and Monte Carlo, divided by the 
total errors, at 91 and 197~GeV.}
\label{figdist_t}
\end{center}
\end{figure}

\begin{figure}[!htb]
\begin{center}
\resizebox{\textwidth}{!}
{\includegraphics{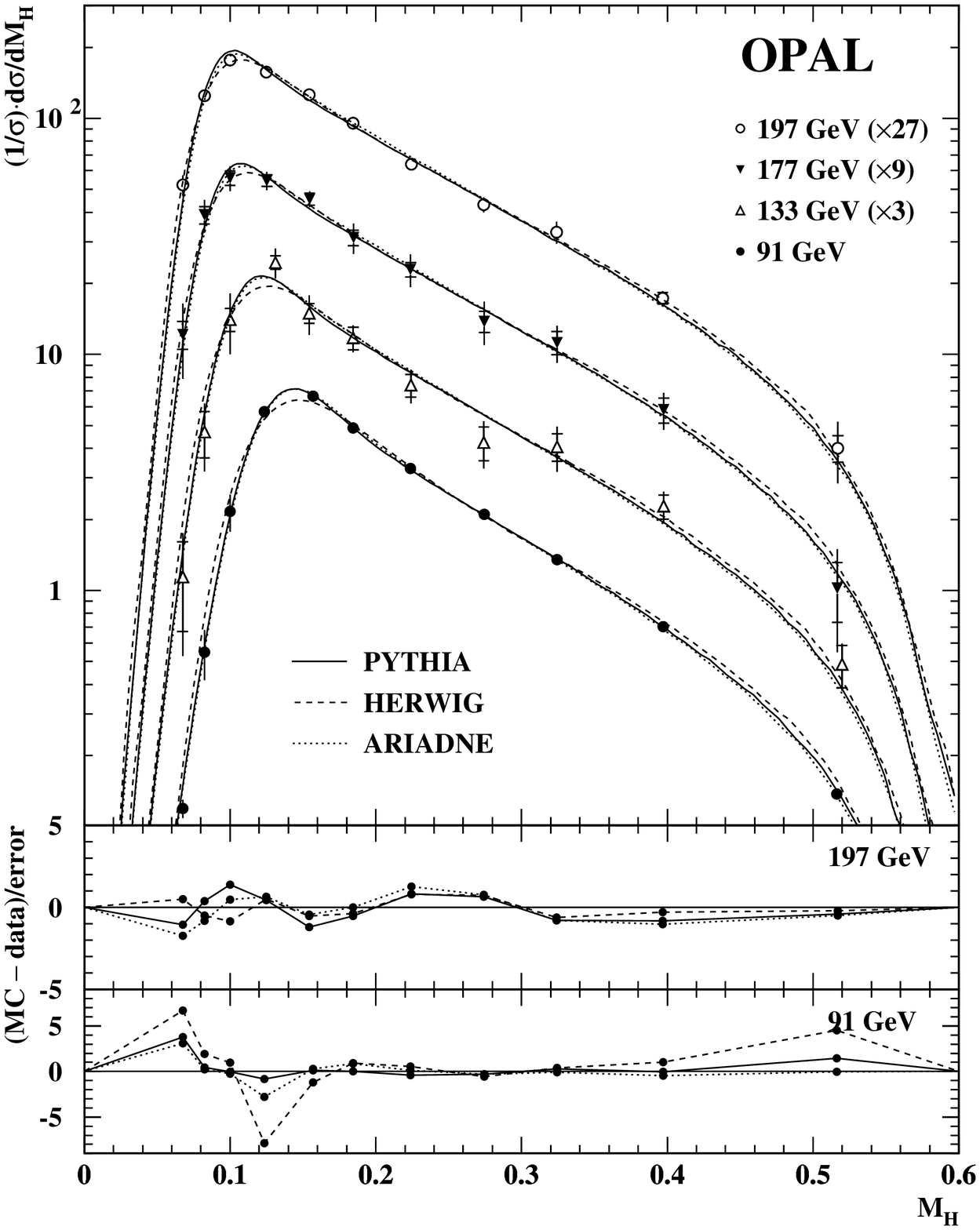}}
\caption[ ]{Distributions of heavy jet mass, \mh, at four c.m.\ energy points ---
91~GeV, 133~GeV, 161--183~GeV (labelled 177~GeV) and 189--209~GeV 
(labelled 197~GeV).  The latter three have been multiplied by factors
3, 9 and 27 respectively for the sake of clarity.  The inner error bars
show the statistical errors, while the total errors are indicated by the 
outer error bars.  The predictions of
the \Pythia, \Herwig\ and \Ariadne\ Monte Carlo models as described in 
the text are indicated by curves.  The lower panels of the figure
show the differences between data and Monte Carlo, divided by the 
total errors, at 91 and 197~GeV.}
\label{figdist_mh}
\end{center}
\end{figure}

\begin{figure}[!htb]
\begin{center}
\resizebox{\textwidth}{!}
{\includegraphics{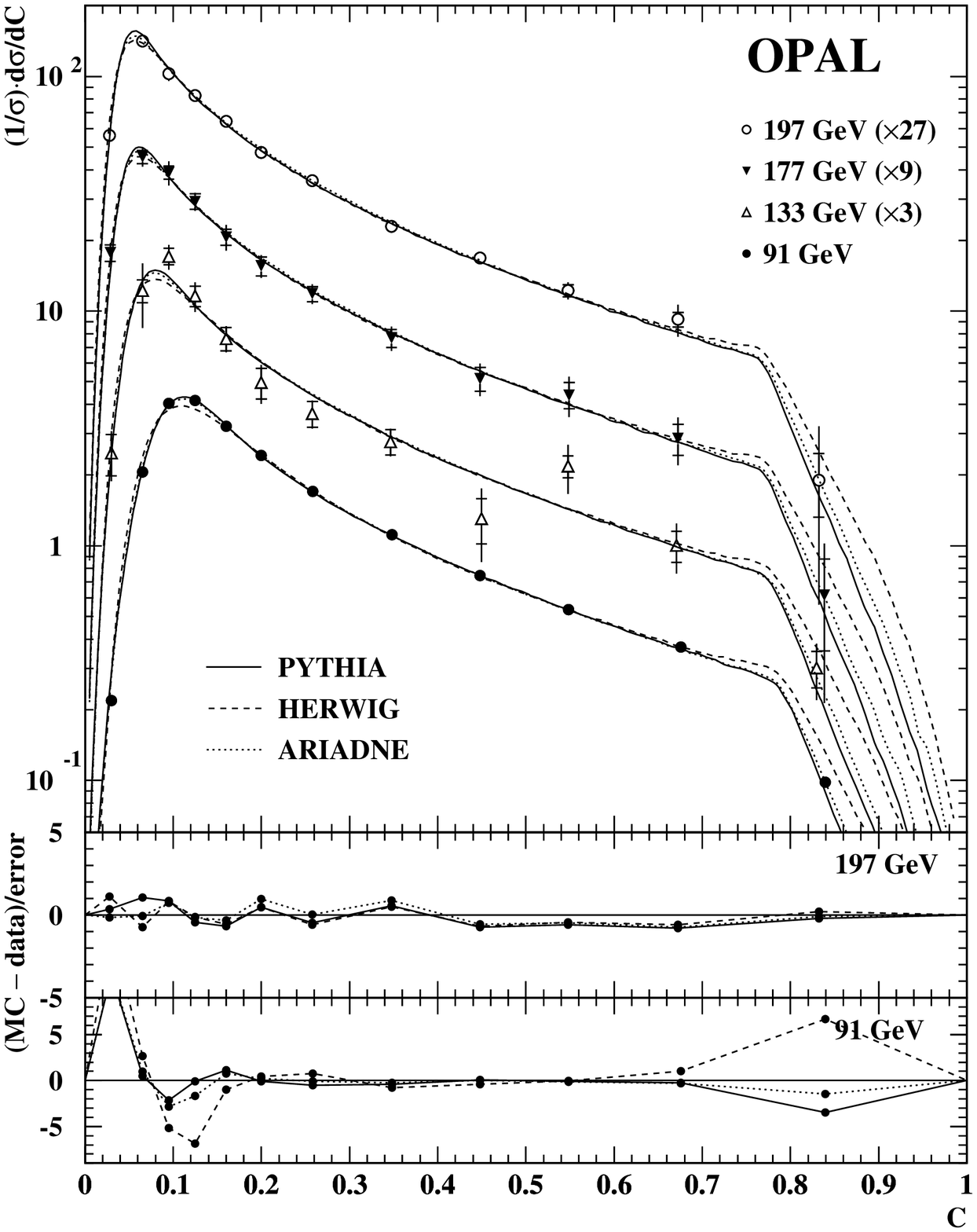}}
\caption[ ]{Distributions of the $C$-parameter 
at four c.m.\ energy points ---
91~GeV, 133~GeV, 161--183~GeV (labelled 177~GeV) and 189--209~GeV 
(labelled 197~GeV).  The latter three have been multiplied by factors
3, 9 and 27 respectively for the sake of clarity.  The inner error bars
show the statistical errors, while the total errors are indicated by the 
outer error bars. The predictions of
the \Pythia, \Herwig\ and \Ariadne\ Monte Carlo models as described in 
the text are indicated by curves.  The lower panels of the figure
show the differences between data and Monte Carlo, divided by the 
total errors, at 91 and 197~GeV.}
\label{figdist_c}
\end{center}
\end{figure}

\begin{figure}[!htb]
\begin{center}
\resizebox{\textwidth}{!}
{\includegraphics{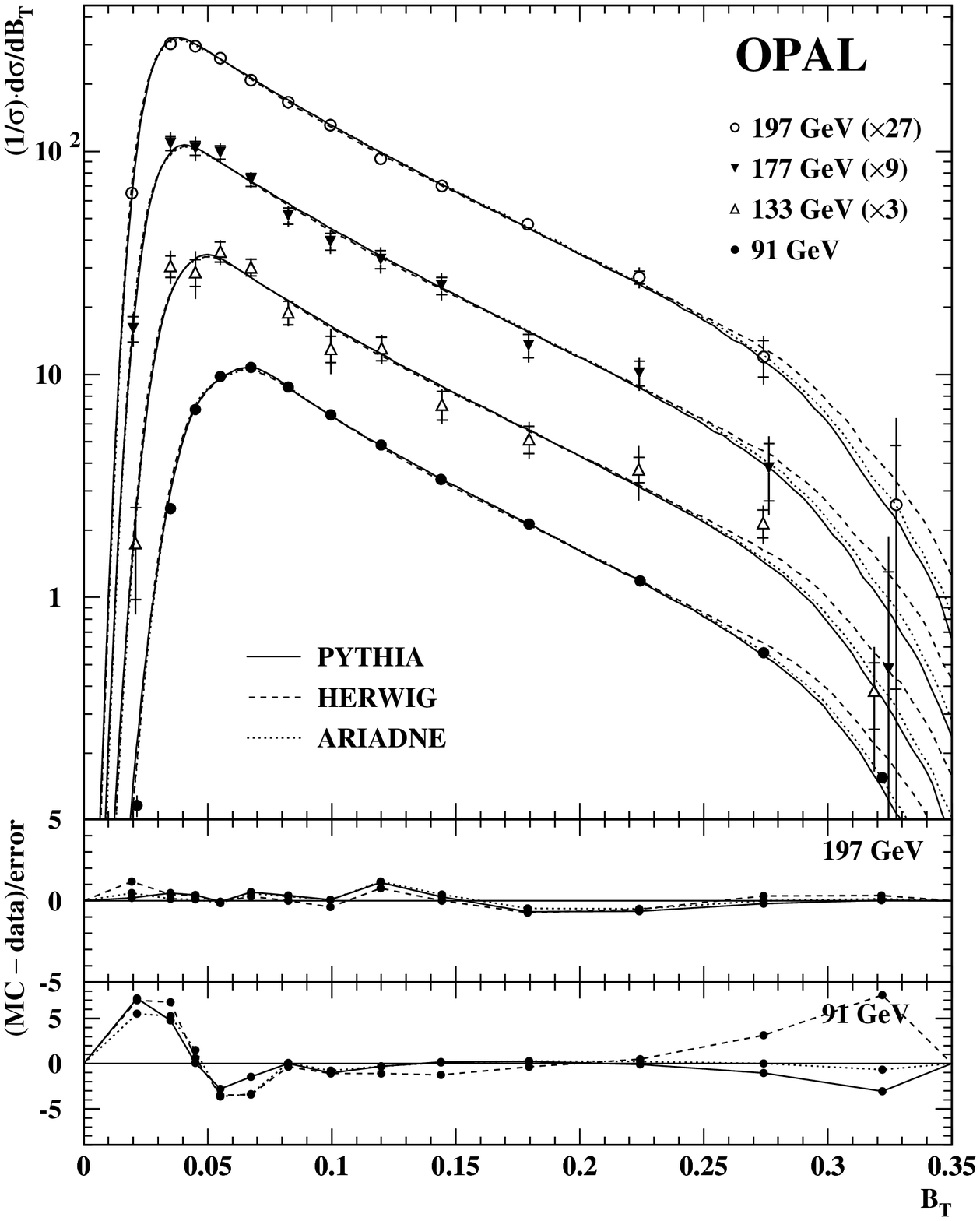}}
\caption[ ]{Distributions of the total jet broadening, \bt, 
at four c.m.\ energy points ---
91~GeV, 133~GeV, 161--183~GeV (labelled 177~GeV) and 189--209~GeV 
(labelled 197~GeV).  The latter three have been multiplied by factors
3, 9 and 27 respectively for the sake of clarity.  The inner error bars
show the statistical errors, while the total errors are indicated by the 
outer error bars.  The predictions of
the \Pythia, \Herwig\ and \Ariadne\ Monte Carlo models as described in 
the text are indicated by curves.  The lower panels of the figure
show the differences between data and Monte Carlo, divided by the 
total errors, at 91 and 197~GeV.}
\label{figdist_bt}
\end{center}
\end{figure}

\begin{figure}[!htb]
\begin{center}
\resizebox{\textwidth}{!}
{\includegraphics{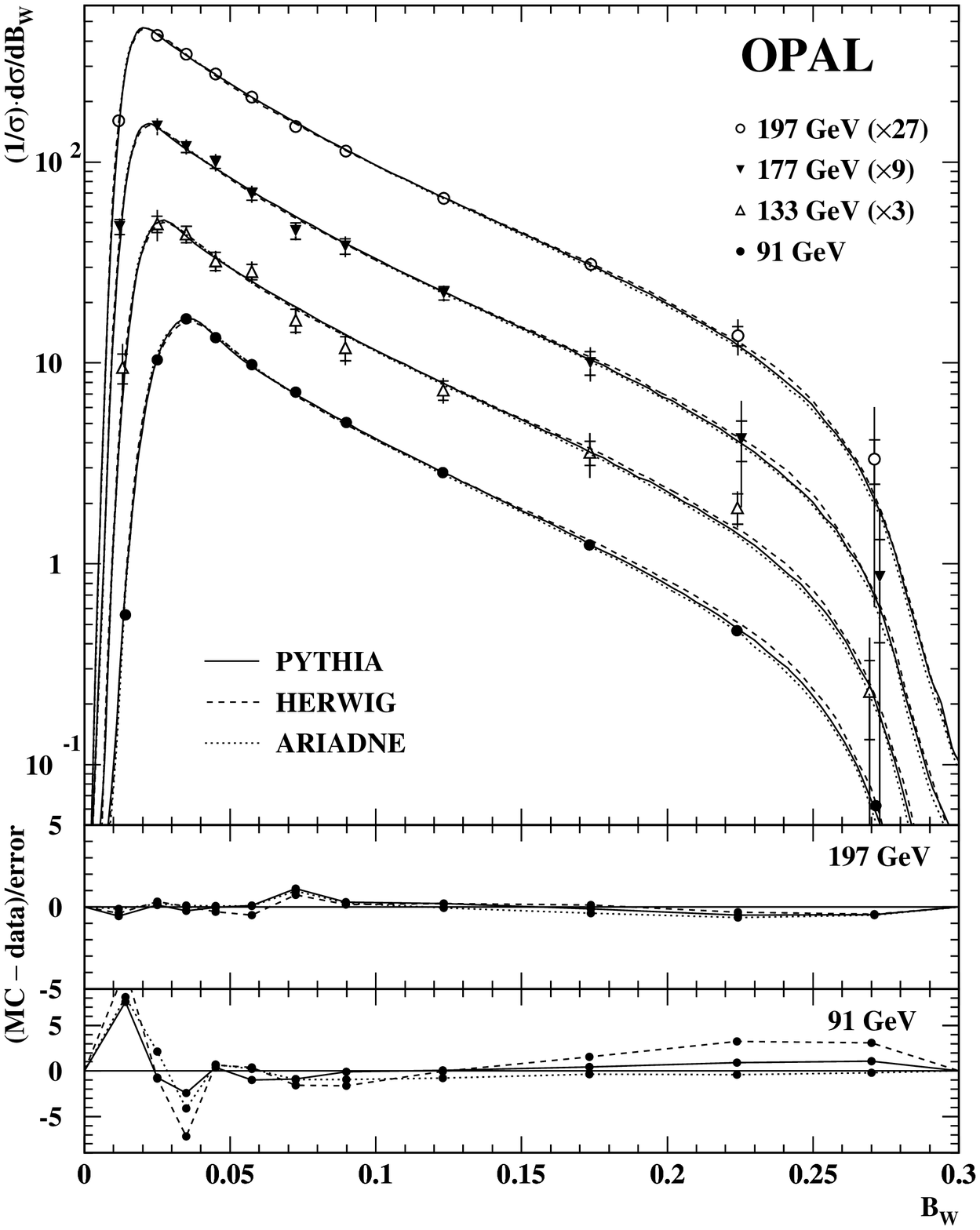}}
\caption[ ]{Distributions of the wide jet broadening, \bw, 
at four c.m.\ energy points ---
91~GeV, 133~GeV, 161--183~GeV (labelled 177~GeV) and 189--209~GeV 
(labelled 197~GeV).  The latter three have been multiplied by factors
3, 9 and 27 respectively for the sake of clarity.  The inner error bars
show the statistical errors, while the total errors are indicated by the 
outer error bars.  The predictions of
the \Pythia, \Herwig\ and \Ariadne\ Monte Carlo models as described in 
the text are indicated by curves.  The lower panels of the figure
show the differences between data and Monte Carlo, divided by the 
total errors, at 91 and 197~GeV.}
\label{figdist_bw}
\end{center}
\end{figure}

\begin{figure}[!htb]
\begin{center}
\resizebox{\textwidth}{!}
{\includegraphics{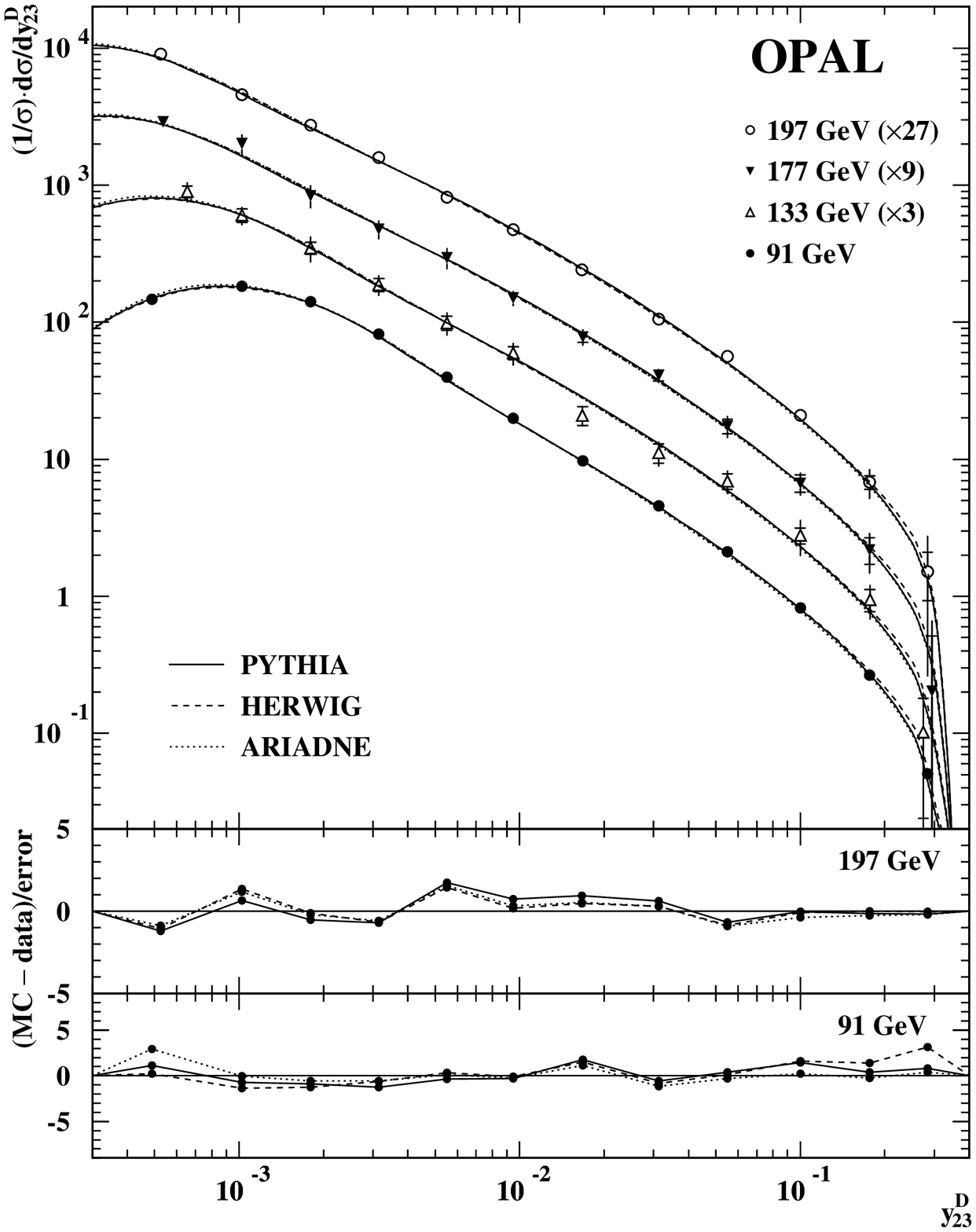}}
\caption[ ]{Distributions of the two- to three-jet transition point,
\ytwothree, at four c.m.\ energy points ---
91~GeV, 133~GeV, 161--183~GeV (labelled 177~GeV) and 189--209~GeV 
(labelled 197~GeV).  The latter three have been multiplied by factors
3, 9 and 27 respectively for the sake of clarity.  The inner error bars
show the statistical errors, while the total errors are indicated by the 
outer error bars.  The predictions of
the \Pythia, \Herwig\ and \Ariadne\ Monte Carlo models as described in 
the text are indicated by curves.  The lower panels of the figure
show the differences between data and Monte Carlo, divided by the 
total errors, at 91 and 197~GeV.}
\label{figdist_y23}
\end{center}
\end{figure}

\begin{figure}[!htb]
\begin{center}
\resizebox{\textwidth}{!}
{\includegraphics{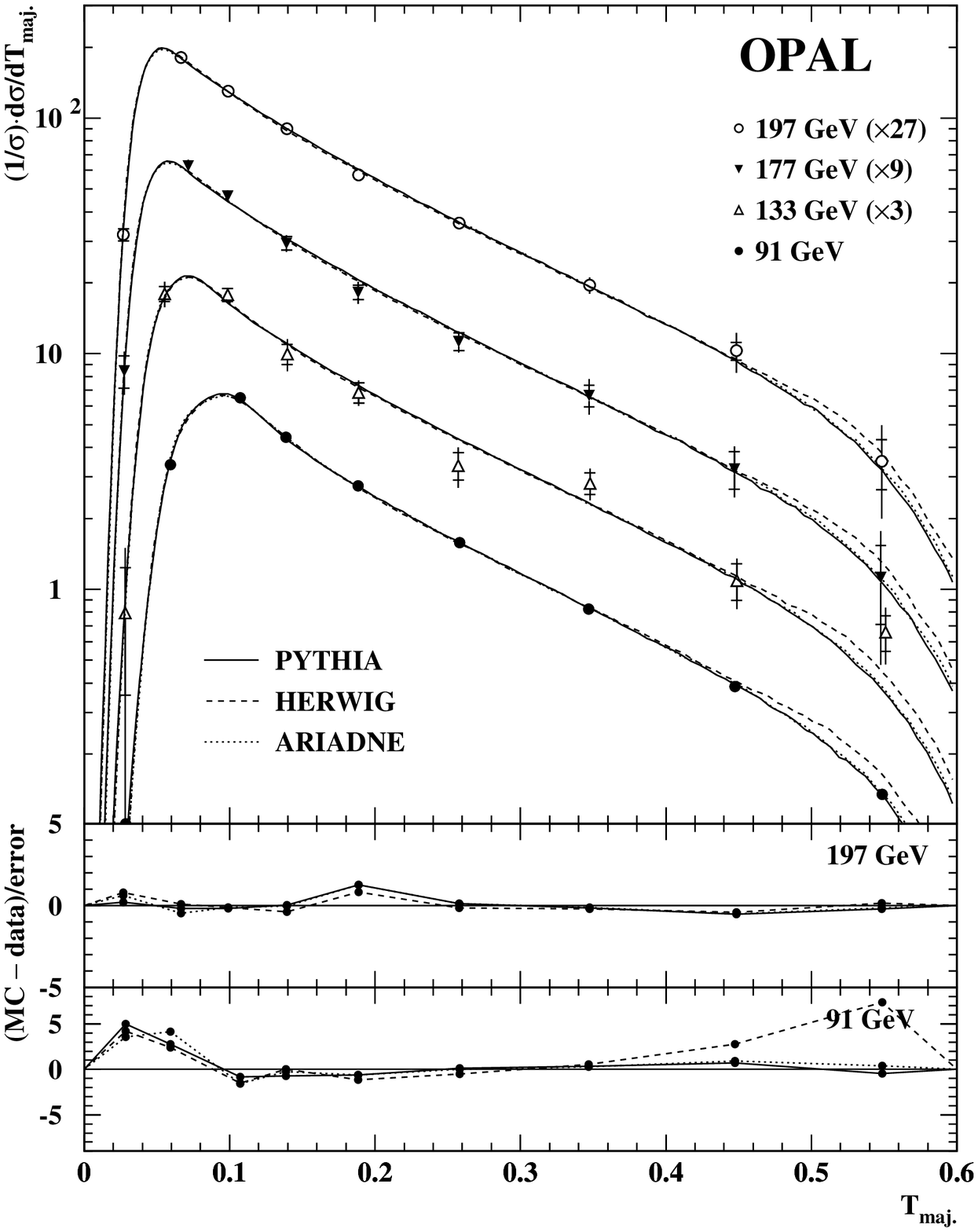}}
\caption[ ]{Distributions of thrust major, \tma,
 at four c.m.\ energy points ---
91~GeV, 133~GeV, 161--183~GeV (labelled 177~GeV) and 189--209~GeV 
(labelled 197~GeV).  The latter three have been multiplied by factors
3, 9 and 27 respectively for the sake of clarity.  The inner error bars
show the statistical errors, while the total errors are indicated by the 
outer error bars.  The predictions of
the \Pythia, \Herwig\ and \Ariadne\ Monte Carlo models as described in 
the text are indicated by curves.  The lower panels of the figure
show the differences between data and Monte Carlo, divided by the 
total errors, at 91 and 197~GeV.}
\label{figdist_tmaj}
\end{center}
\end{figure}

\begin{figure}[!htb]
\begin{center}
\resizebox{\textwidth}{!}
{\includegraphics{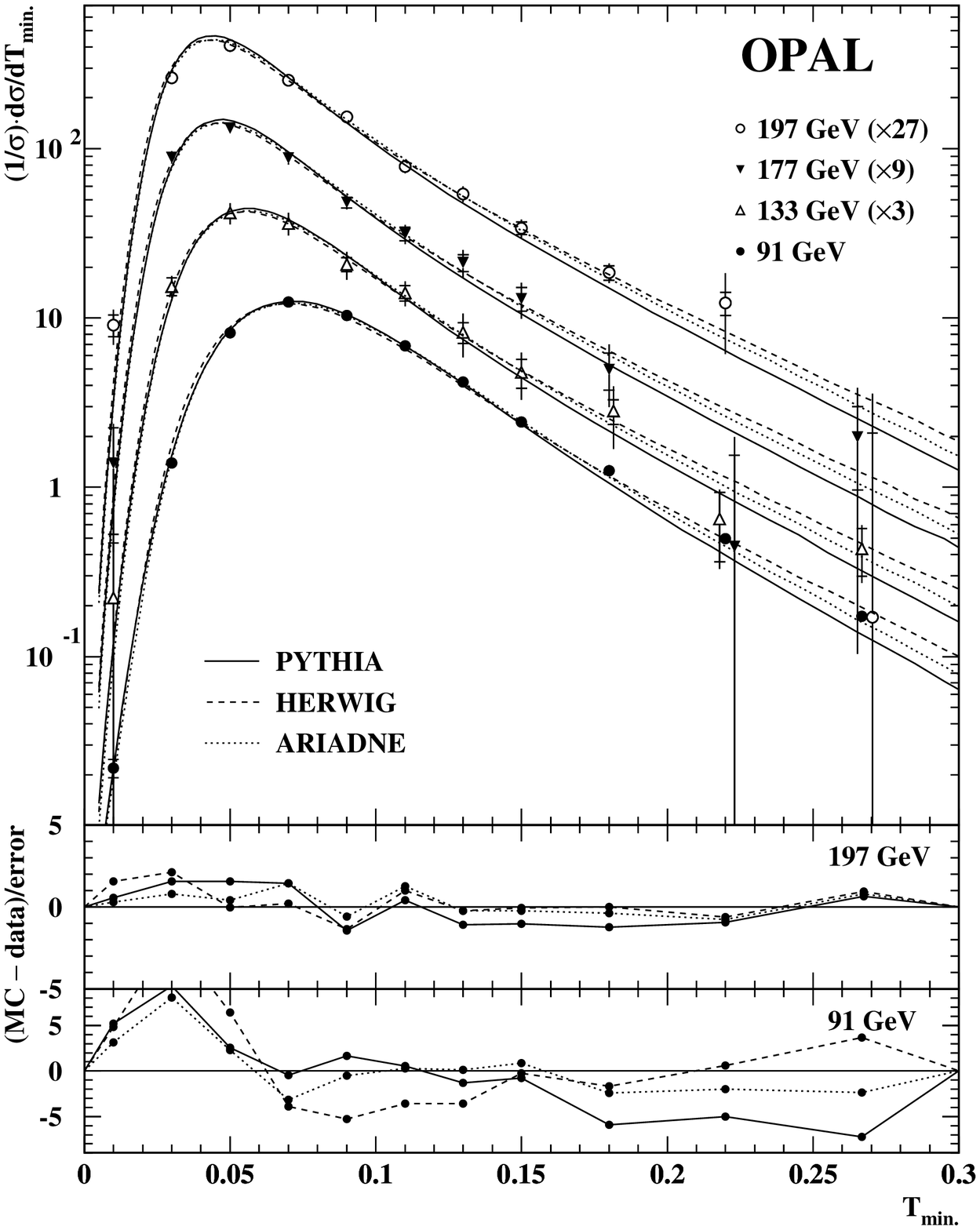}}
\caption[ ]{Distributions of thrust minor, \tmi, 
at four c.m.\ energy points ---
91~GeV, 133~GeV, 161--183~GeV (labelled 177~GeV) and 189--209~GeV 
(labelled 197~GeV).  The latter three have been multiplied by factors
3, 9 and 27 respectively for the sake of clarity.  The inner error bars
show the statistical errors, while the total errors are indicated by the 
outer error bars.  The predictions of
the \Pythia, \Herwig\ and \Ariadne\ Monte Carlo models as described in 
the text are indicated by curves.  The lower panels of the figure
show the differences between data and Monte Carlo, divided by the 
total errors, at 91 and 197~GeV.}
\label{figdist_tmin}
\end{center}
\end{figure}

\begin{figure}[!htb]
\begin{center}
\resizebox{\textwidth}{!}
{\includegraphics{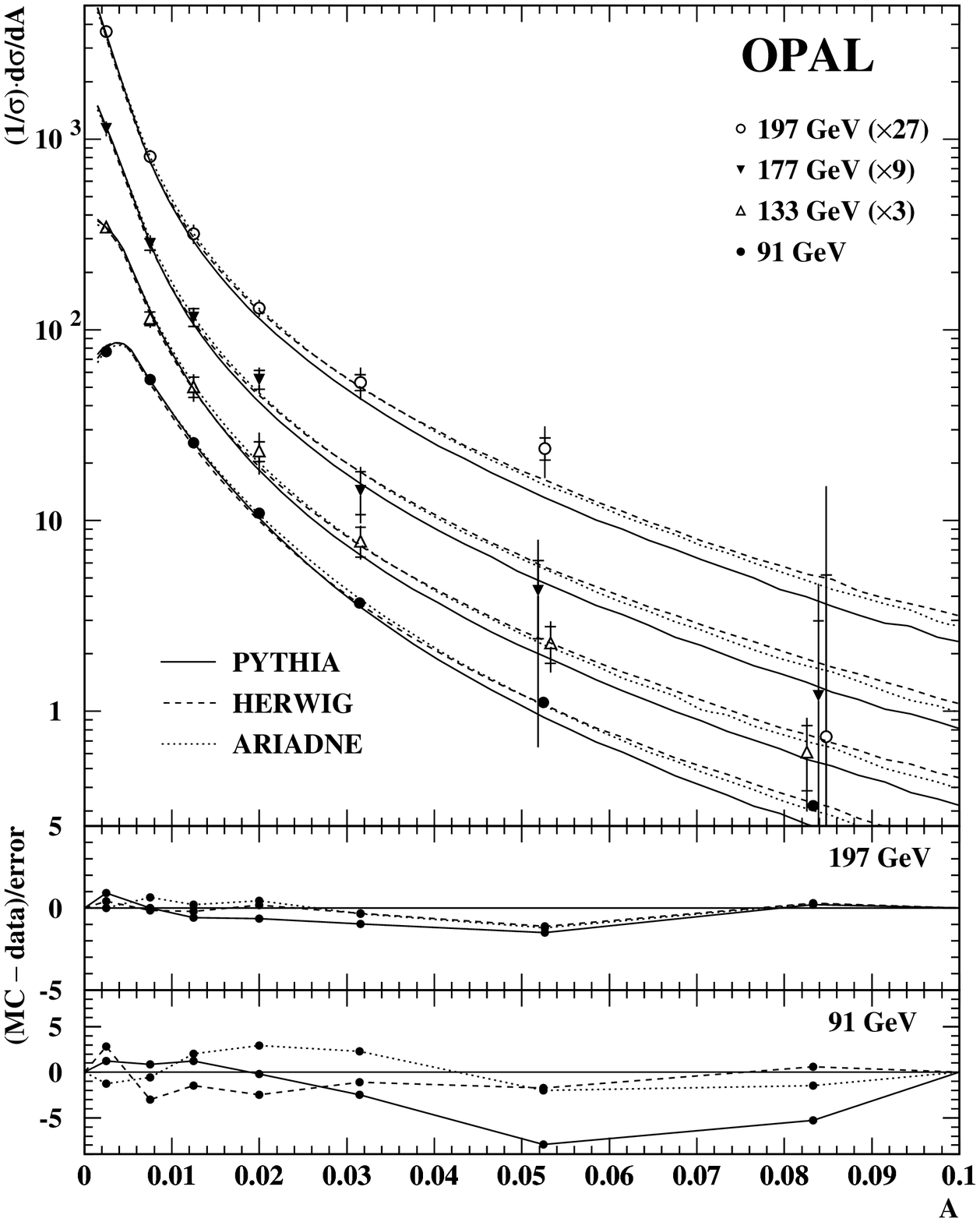}}
\caption[ ]{Distributions of aplanarity, $A$, 
at four c.m.\ energy points ---
91~GeV, 133~GeV, 161--183~GeV (labelled 177~GeV) and 189--209~GeV 
(labelled 197~GeV).  The latter three have been multiplied by factors
3, 9 and 27 respectively for the sake of clarity.  The inner error bars
show the statistical errors, while the total errors are indicated by the 
outer error bars.  The predictions of
the \Pythia, \Herwig\ and \Ariadne\ Monte Carlo models as described in 
the text are indicated by curves.  The lower panels of the figure
show the differences between data and Monte Carlo, divided by the 
total errors, at 91 and 197~GeV.}
\label{figdist_a}
\end{center}
\end{figure}

\begin{figure}[!htb]
\begin{center}
\resizebox{\textwidth}{!}
{\includegraphics{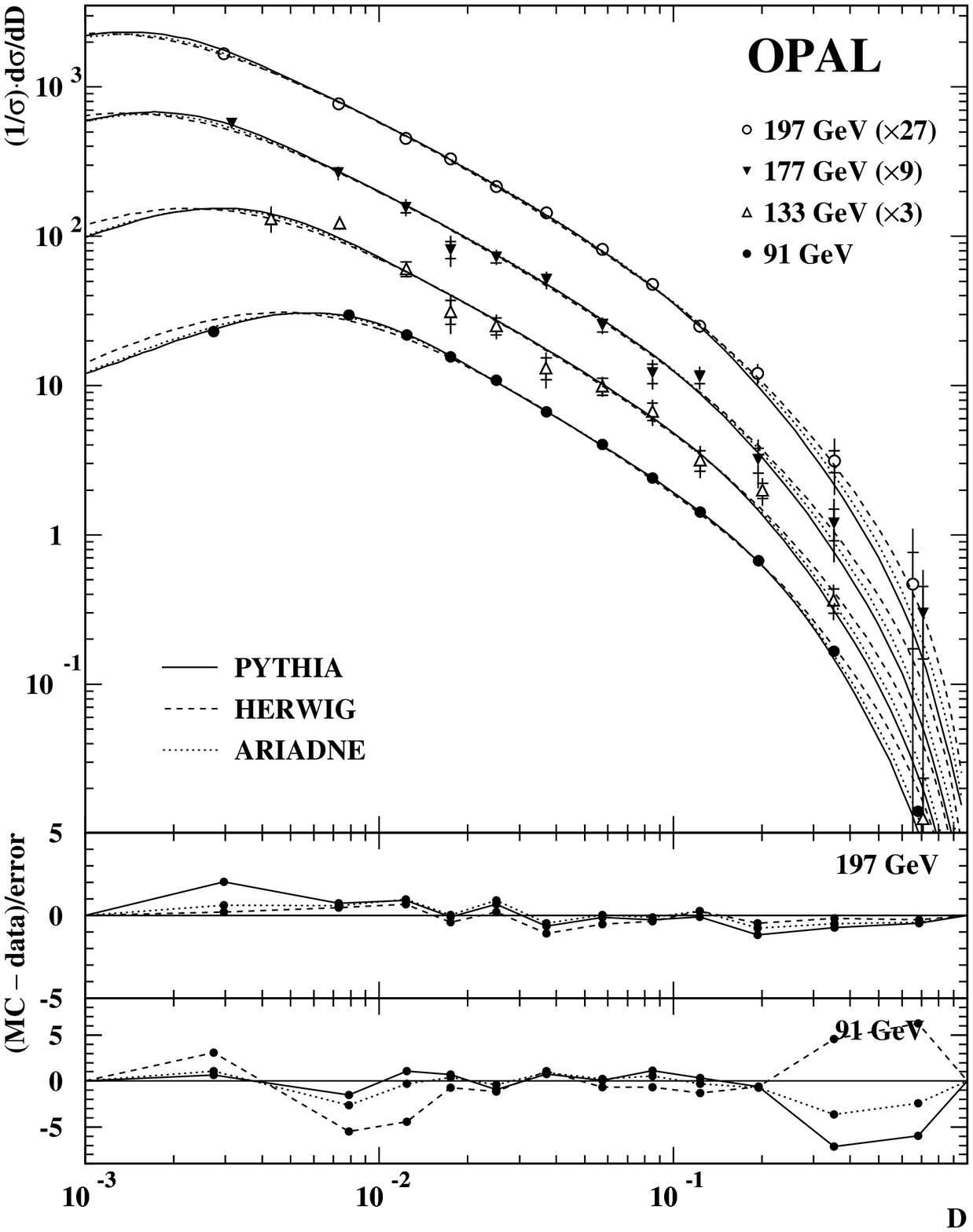}}
\caption[ ]{Distributions of the $D$-parameter at four c.m.\ energy points ---
91~GeV, 133~GeV, 161--183~GeV (labelled 177~GeV) and 189--209~GeV 
(labelled 197~GeV).  The latter three have been multiplied by factors
3, 9 and 27 respectively for the sake of clarity.  The inner error bars
show the statistical errors, while the total errors are indicated by the 
outer error bars.  The predictions of
the \Pythia, \Herwig\ and \Ariadne\ Monte Carlo models as described in 
the text are indicated by curves.  The lower panels of the figure
show the differences between data and Monte Carlo, divided by the 
total errors, at 91 and 197~GeV.}
\label{figdist_d}
\end{center}
\end{figure}

\begin{figure}[!htb]
\begin{center}
\resizebox{\textwidth}{!}
{\includegraphics{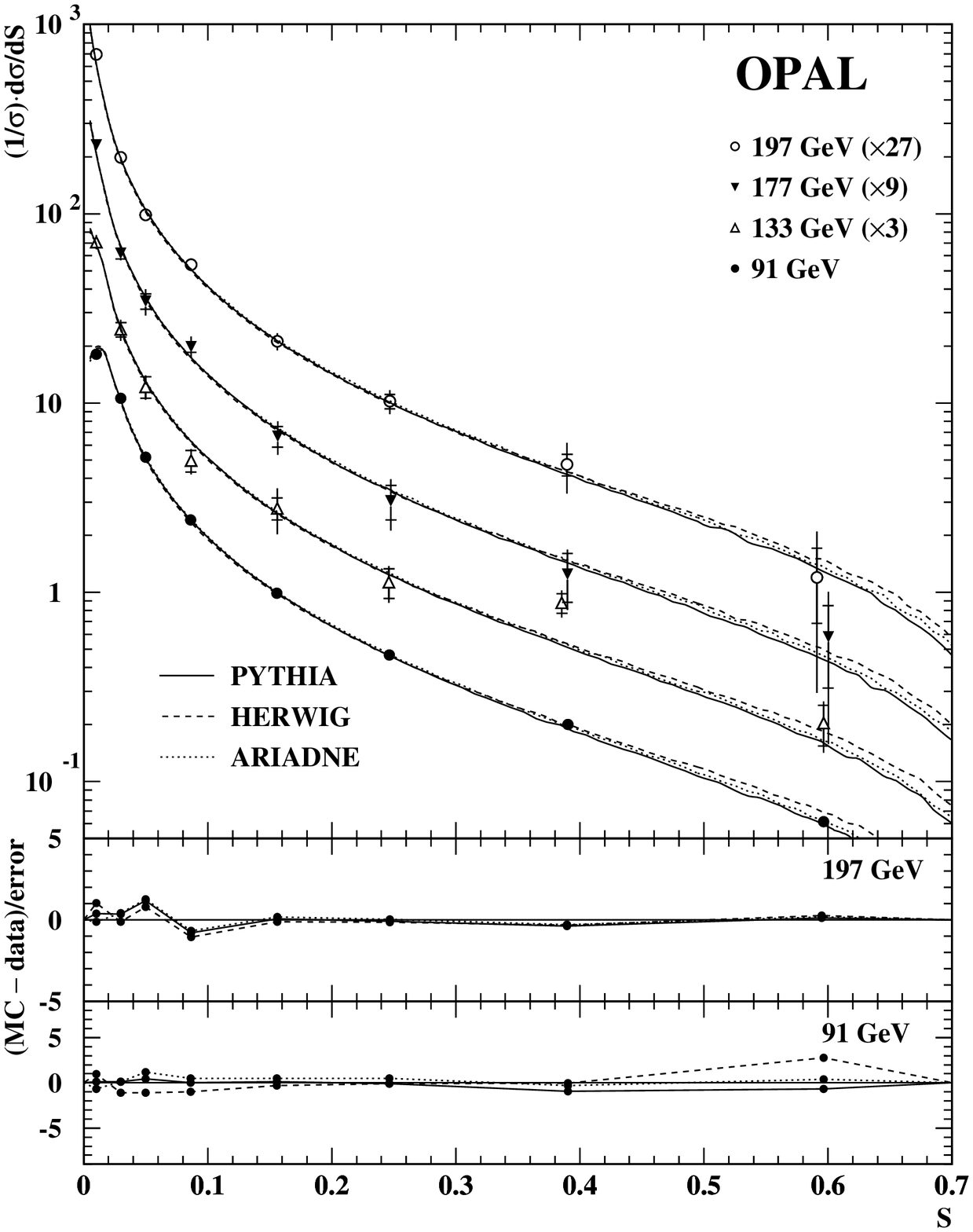}}
\caption[ ]{Distributions of sphericity, $S$, 
at four c.m.\ energy points ---
91~GeV, 133~GeV, 161--183~GeV (labelled 177~GeV) and 189--209~GeV 
(labelled 197~GeV).  The latter three have been multiplied by factors
3, 9 and 27 respectively for the sake of clarity.  The inner error bars
show the statistical errors, while the total errors are indicated by the 
outer error bars.  The predictions of
the \Pythia, \Herwig\ and \Ariadne\ Monte Carlo models as described in 
the text are indicated by curves.  The lower panels of the figure
show the differences between data and Monte Carlo, divided by the 
total errors, at 91 and 197~GeV.}
\label{figdist_s}
\end{center}
\end{figure}

\begin{figure}[!htb]
\begin{center}
\resizebox{\textwidth}{!}
{\includegraphics{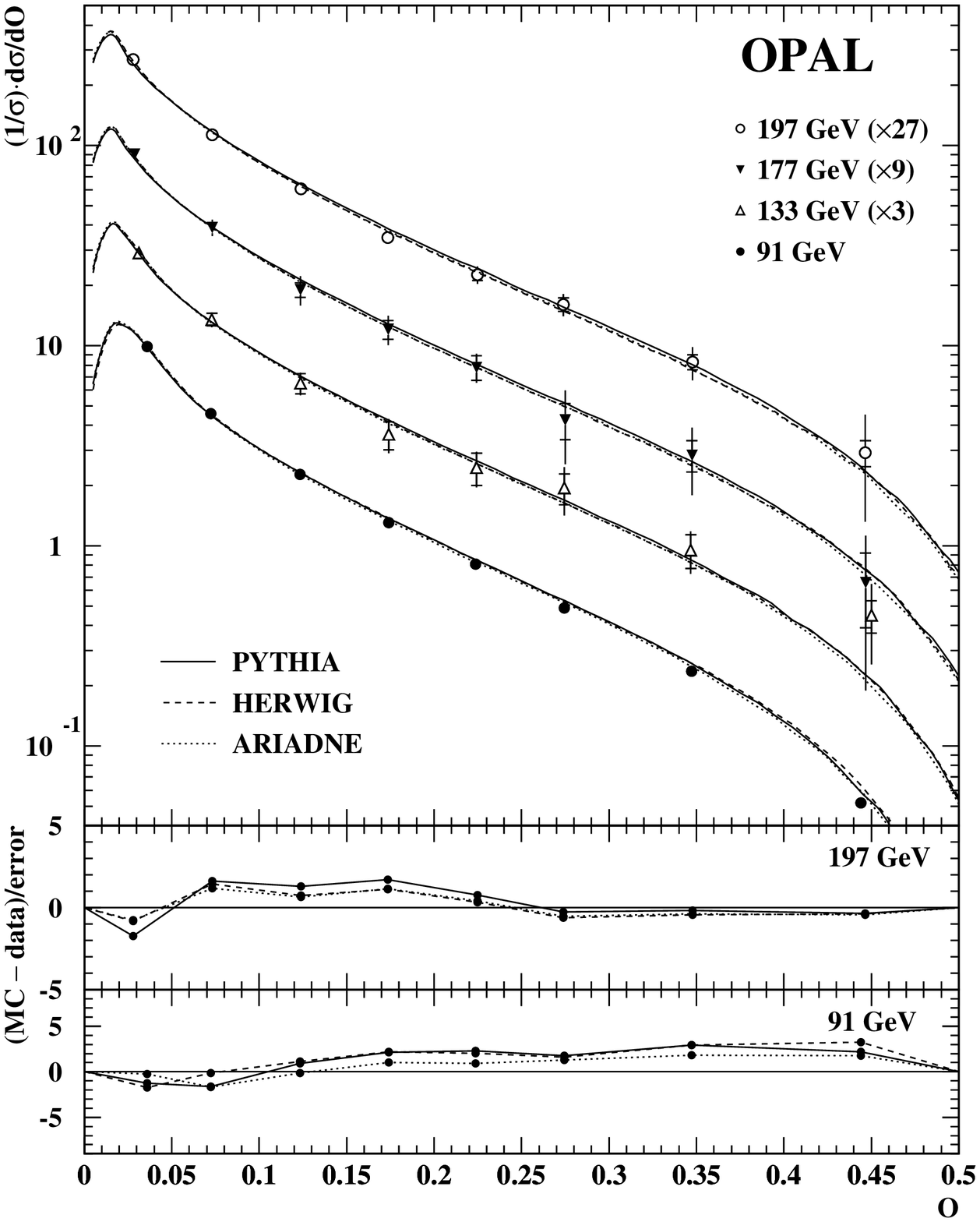}}
\caption[ ]{Distributions of oblateness, $O$, 
at four c.m.\ energy points ---
91~GeV, 133~GeV, 161--183~GeV (labelled 177~GeV) and 189--209~GeV 
(labelled 197~GeV).  The latter three have been multiplied by factors
3, 9 and 27 respectively for the sake of clarity.  The inner error bars
show the statistical errors, while the total errors are indicated by the 
outer error bars.  The predictions of
the \Pythia, \Herwig\ and \Ariadne\ Monte Carlo models as described in 
the text are indicated by curves.  The lower panels of the figure
show the differences between data and Monte Carlo, divided by the 
total errors, at 91 and 197~GeV.}
\label{figdist_o}
\end{center}
\end{figure}

\begin{figure}[!htb]
\begin{center}
\resizebox{0.95\textwidth}{!}
{\includegraphics{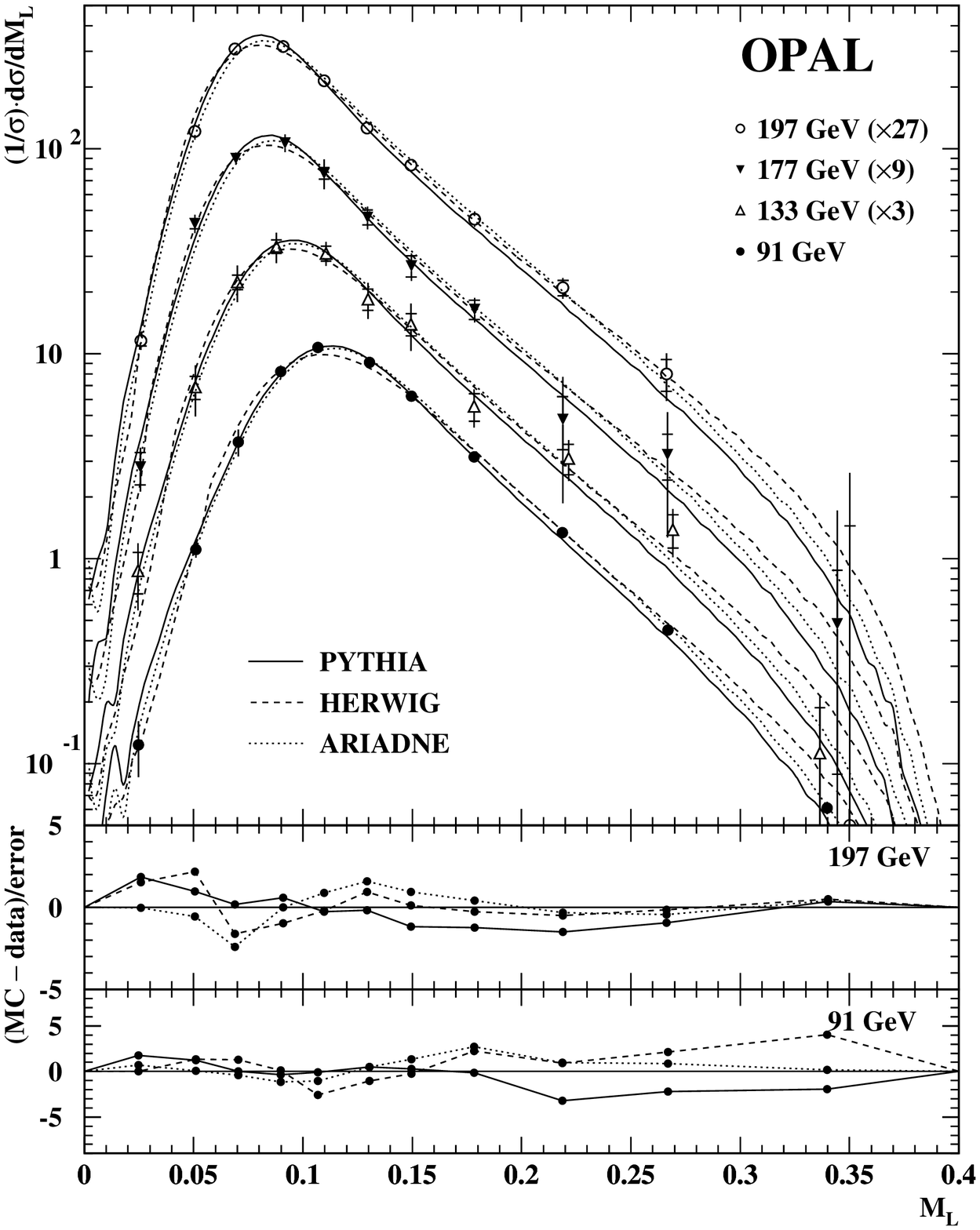}}
\caption[ ]{Distributions of the light jet mass, \ml, 
at four c.m.\ energy points ---
91~GeV, 133~GeV, 161--183~GeV (labelled 177~GeV) and 189--209~GeV 
(labelled 197~GeV).  The latter three have been multiplied by factors
3, 9 and 27 respectively for the sake of clarity.  The inner error bars
show the statistical errors, while the total errors are indicated by the 
outer error bars.  The predictions of
the \Pythia, \Herwig\ and \Ariadne\ Monte Carlo models as described in 
the text are indicated by curves.  The fluctuations seen in the curves at 
low \ml\ are real artefacts of the models, while those at high \ml\ 
are merely caused by statistical fluctuations.
The lower panels of the figure
show the differences between data and Monte Carlo, divided by the 
total errors, at 91 and 197~GeV.}
\label{figdist_ml}
\end{center}
\end{figure}

\begin{figure}[!htb]
\begin{center}
\resizebox{\textwidth}{!}
{\includegraphics{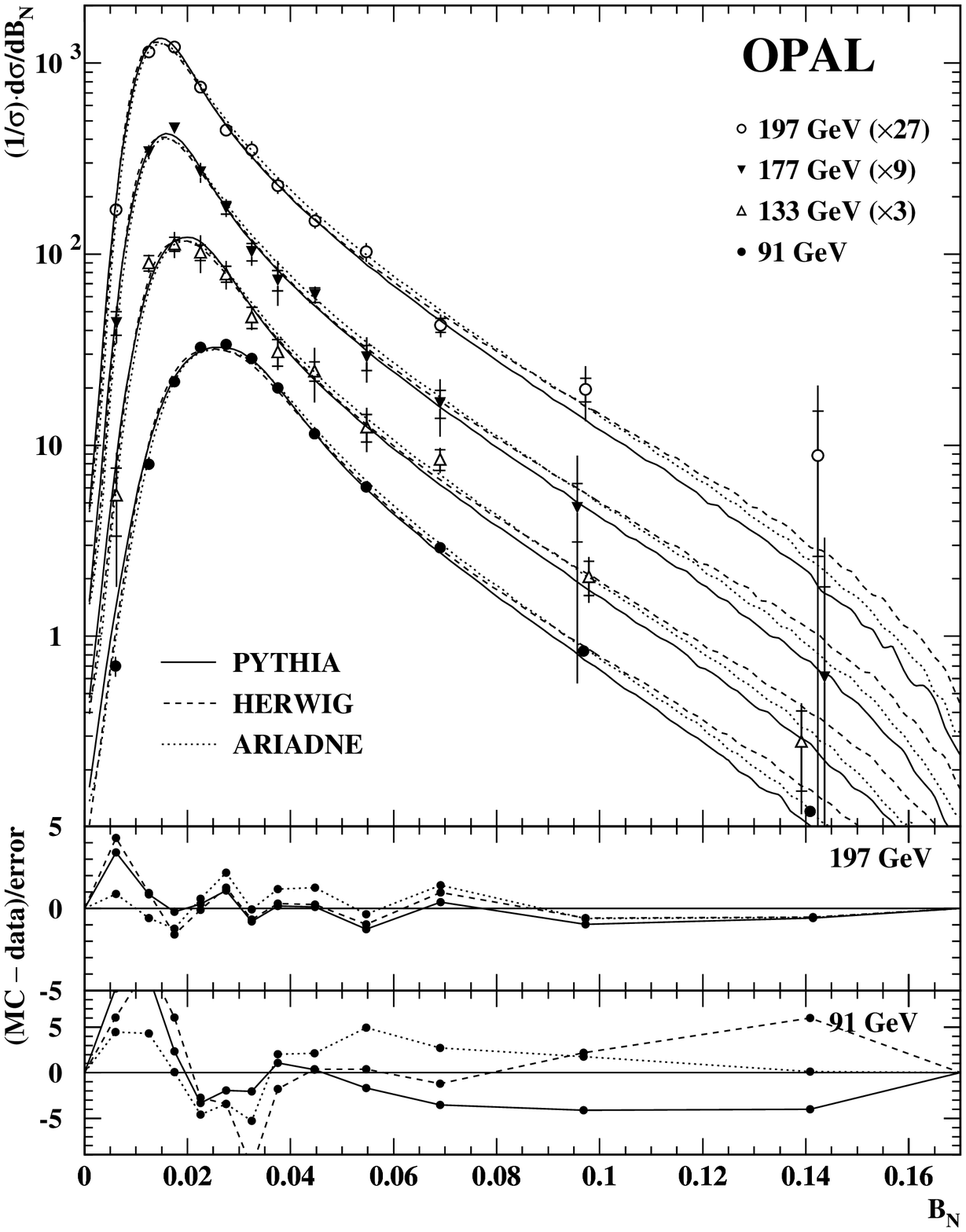}}
\caption[ ]{Distributions of narrow jet broadening, \bn,
at four c.m.\ energy points ---
91~GeV, 133~GeV, 161--183~GeV (labelled 177~GeV) and 189--209~GeV 
(labelled 197~GeV).  The latter three have been multiplied by factors
3, 9 and 27 respectively for the sake of clarity.  The inner error bars
show the statistical errors, while the total errors are indicated by the 
outer error bars.  The predictions of
the \Pythia, \Herwig\ and \Ariadne\ Monte Carlo models as described in 
the text are indicated by curves.  The lower panels of the figure
show the differences between data and Monte Carlo, divided by the 
total errors, at 91 and 197~GeV.}
\label{figdist_bn}
\end{center}
\end{figure}

The measured normalized differential cross-sections, $R'(y)$, for 
each of the 14 event shapes in 
each of four energy ranges are 
given in Tables~\ref{tabdist1}--\ref{tabdist5}.\footnote{Further details 
of the data will be made available in the HEPDATA database, 
{\tt http://durpdg.dur.ac.uk/HEPDATA/}}  
The measurements are also shown in graphical form in 
Figures~\ref{figdist_t}--\ref{figdist_bn}.
In order to clarify the presentation, 
the data from 161 to 183~GeV have been 
combined\footnote{This combination 
also has the advantage of reducing any statistical contributions which
may be present in the systematic error estimates.}, weighted 
by the numbers of events, and likewise the data at 189~GeV and above have been 
combined in order to provide a 
high energy set of data with reasonably high 
statistics.  These two samples of data, which correspond to mean c.m.\ energies
of 177.4 and 197.0~GeV respectively, cover sufficiently small ranges of \ecm\ that 
hard and soft \qcd\ effects should not vary greatly.
The horizontal placement of the data points follows the prescription 
in ref.~\cite{lafferty-wyatt}, except in a few cases where
the 133~GeV and/or 177~GeV data points have been
slightly displaced sideways to avoid overlap.
Superimposed on Figures~\ref{figdist_t}--\ref{figdist_bn} we show the 
distributions predicted by the \Pythia~6.1, \Herwig~6.2 and \Ariadne~4.11 
parton shower models, which in all cases were tuned to other \Opal\
data recorded during the \Lepone\ running at 91~GeV c.m.\ energy.  
The new data appear to be well described by all the models.
   
In order to make a clearer comparison between 
data and models, the lower panels of Figures~\ref{figdist_t}--\ref{figdist_bn}
show the differences between data and each model, divided by the total 
(i.e.\ statistical and experimental) 
error on the data in that bin.  These ratios are shown for 91~GeV
and for the combined high energy data sample at 189~GeV and above.
The sum of squares of these 
differences would, in the absence of correlations, represent a $\chi^2$ 
between  data and the model.  However, since correlations are 
present, such $\chi^2$ values could be 
regarded as giving only a rough indication of the agreement 
between data and the models.
All three models are seen to describe the high energy data well.
Some discrepancies are, however, seen in the more precise 91~GeV data.
In those observables dominated by three-jet production, 
the largest differences are seen when the observable is close
to zero, i.e.\ in the extreme two-jet region. \Herwig\ is generally
apt to give larger $\chi^2$ values than the other two models, 
including some contribution from the extreme three-jet regions also.
The observables dominated by four-jet production (\tmi, \ml, \bn,
$D$ and $A$) are less well modelled, and there is a tendency for 
\Ariadne\ to give the best description of the data.   

%
%
\subsection{ Moments of event shapes }
\label{sec_moments}

\begin{figure}[!htb]
\begin{center}
\resizebox{0.8\textwidth}{!}
{\includegraphics{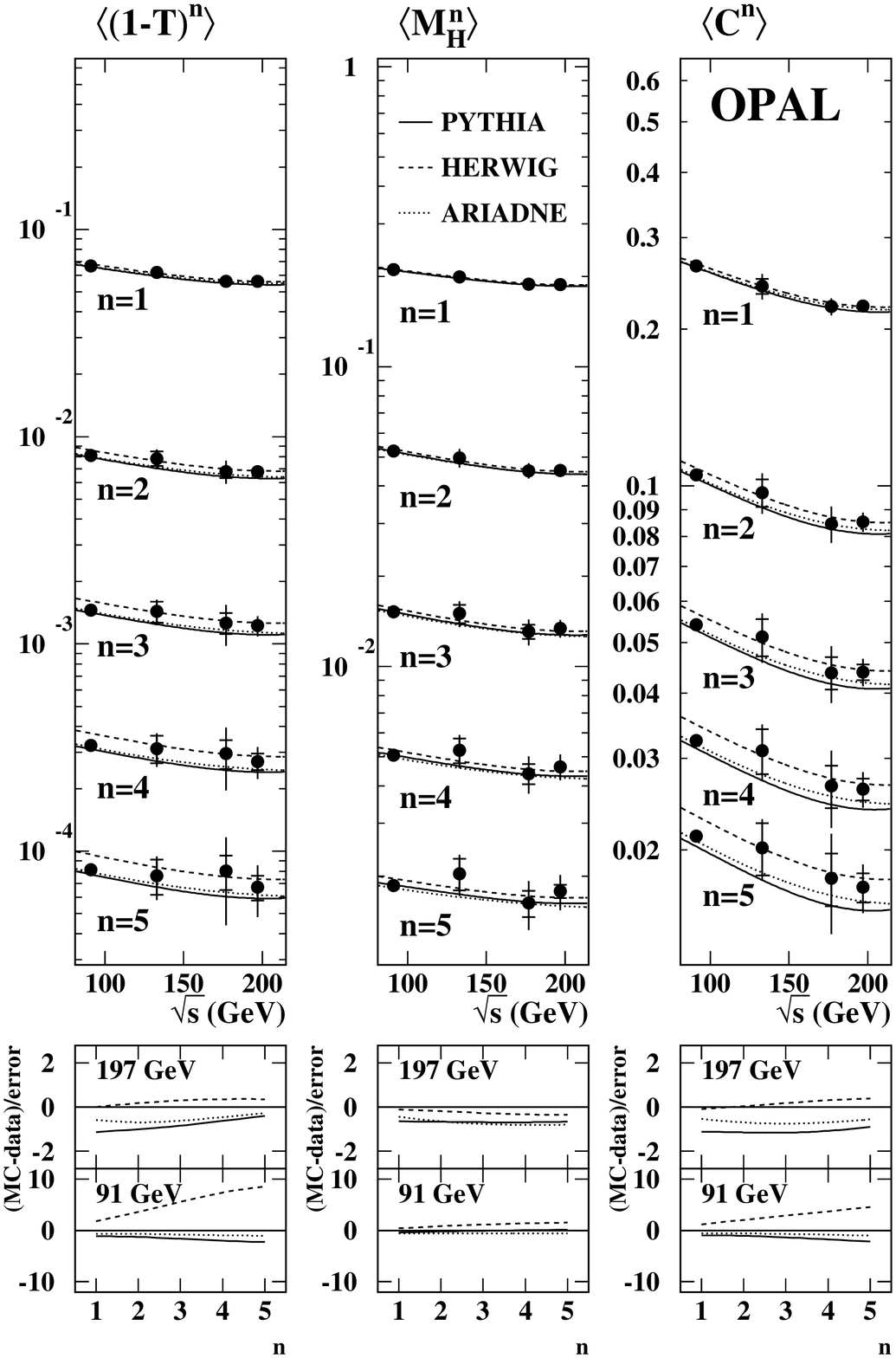}}
\caption[ ]{Moments of thrust, \momn{(1-T)}{n}, 
heavy jet mass, \momn{\mh}{n}, and $C$-parameter, \momn{C}{n}, for
$n=1,\ldots,5$ 
at four c.m.\ energy points ---
91~GeV, 133~GeV, 161--183~GeV (labelled 177~GeV) and 189--209~GeV 
(labelled 197~GeV).   The inner error bars
show the statistical errors, while the total errors are indicated by the 
outer error bars.  The predictions of
the \Pythia, \Herwig\ and \Ariadne\ Monte Carlo models as described in 
the text are indicated by curves.
  The lower panel of the figure
shows the differences between data and Monte Carlo, 
divided by the total errors, as a function of $n$ for the 91~GeV and 
197~GeV data.}
\label{moments_t}
\end{center}
\end{figure}

\begin{figure}[!htb]
\begin{center}
\resizebox{0.8\textwidth}{!}
{\includegraphics{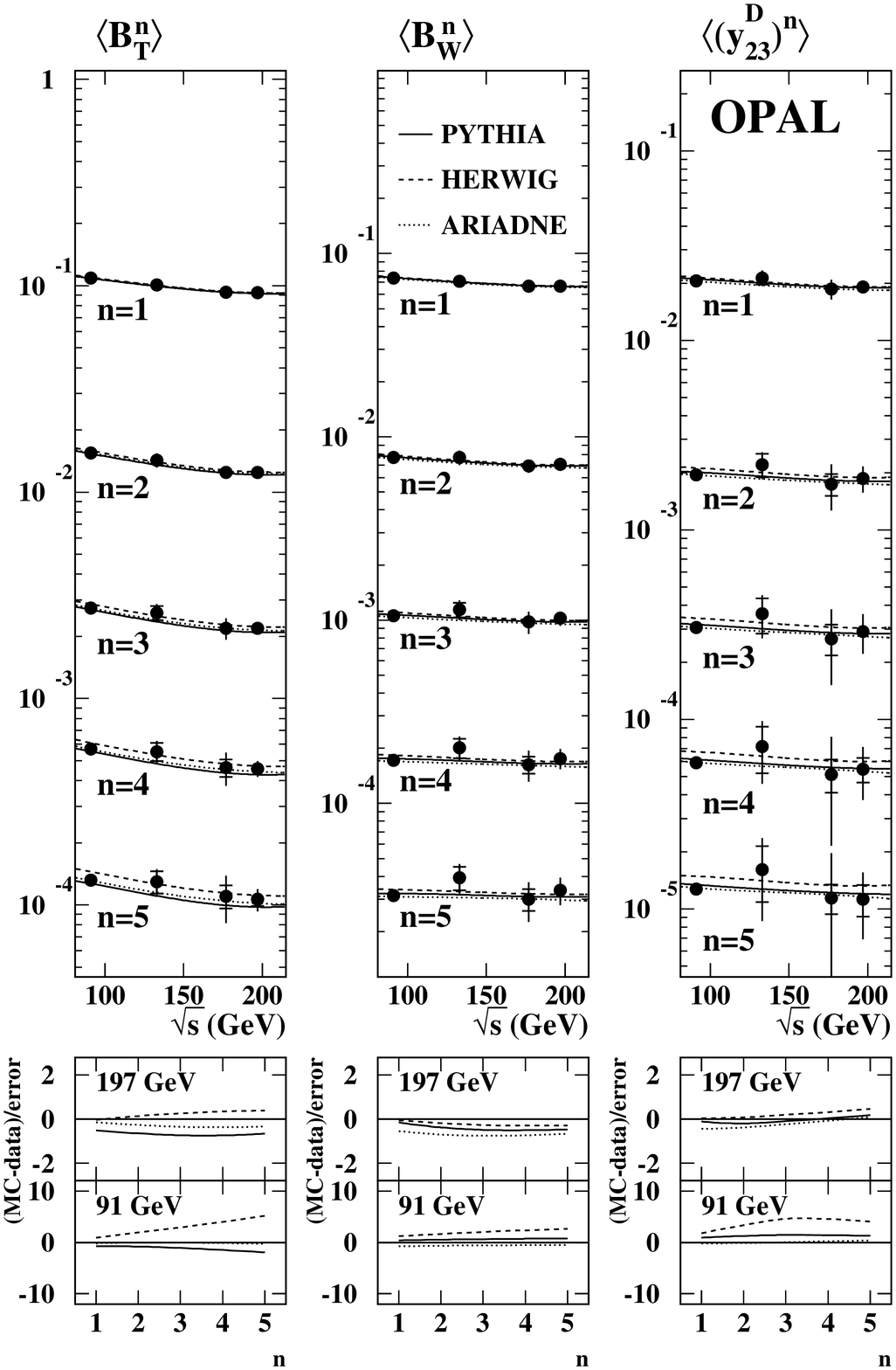}}
\caption[ ]{Moments of total jet broadening, \momn{\bt}{n}, 
wide jet broadening, \momn{\bw}{n}, and Durham jet resolution parameter, \momn{(\ytwothree)}{n}, for
$n=1,\ldots,5$ 
at four c.m.\ energy points ---
91~GeV, 133~GeV, 161--183~GeV (labelled 177~GeV) and 189--209~GeV 
(labelled 197~GeV).   The inner error bars
show the statistical errors, while the total errors are indicated by the 
outer error bars.  The predictions of
the \Pythia, \Herwig\ and \Ariadne\ Monte Carlo models as described in 
the text are indicated by curves.
  The lower panel of the figure
shows the differences between data and Monte Carlo, 
divided by the total errors, as a function of $n$ for the 91~GeV and 
197~GeV data.}
\label{moments_bt}
\end{center}
\end{figure}

\begin{figure}[!htb]
\begin{center}
\resizebox{0.8\textwidth}{!}
{\includegraphics{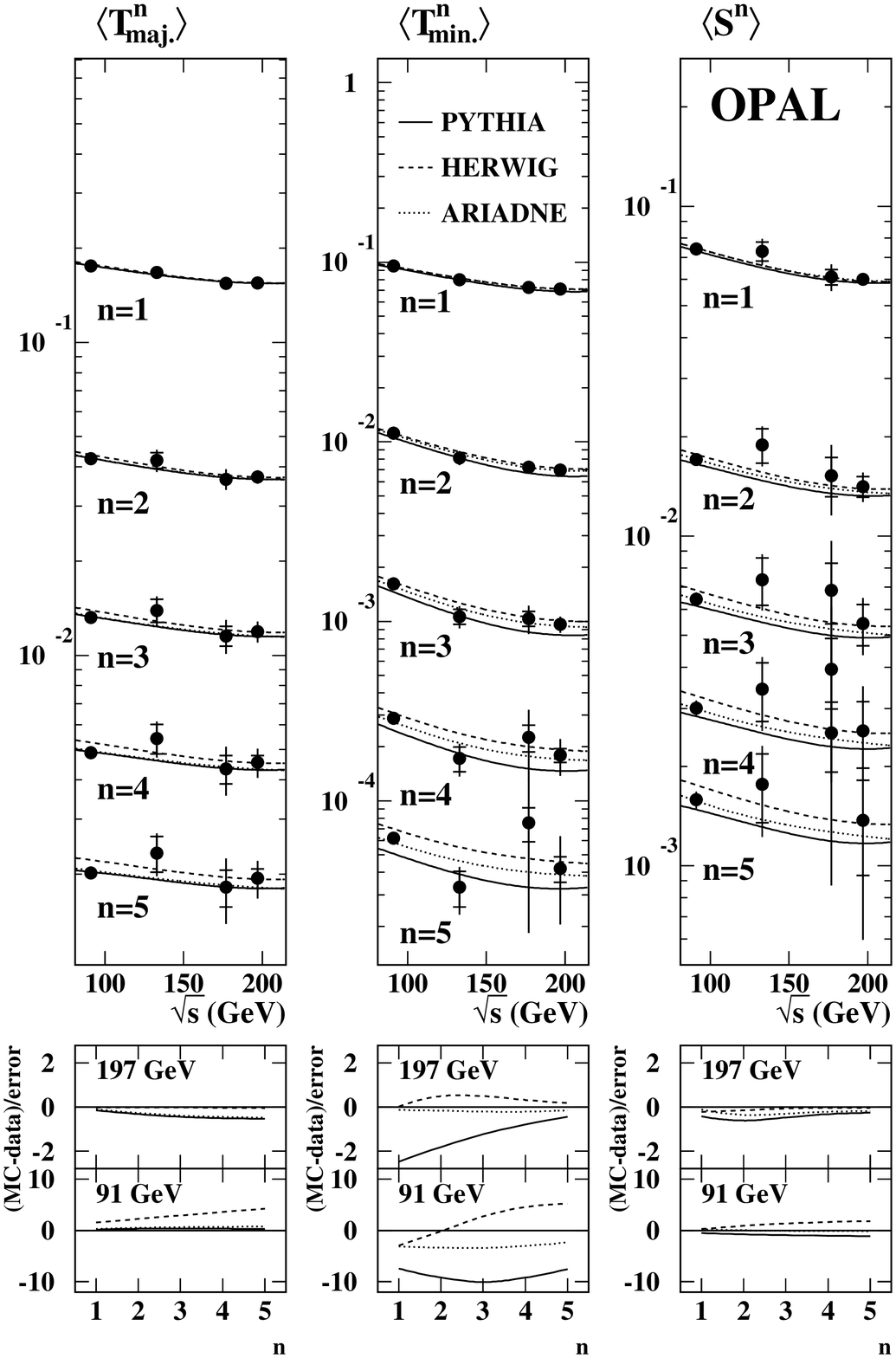}}
\caption[ ]{Moments of thrust major, \momn{\tma}{n}, 
thrust minor, \momn{\tmi}{n}, and sphericity, \momn{$S$}{n}, for
$n=1,\ldots,5$ 
at four c.m.\ energy points ---
91~GeV, 133~GeV, 161--183~GeV (labelled 177~GeV) and 189--209~GeV 
(labelled 197~GeV).   The inner error bars
show the statistical errors, while the total errors are indicated by the 
outer error bars.  The predictions of
the \Pythia, \Herwig\ and \Ariadne\ Monte Carlo models as described in 
the text are indicated by curves.
  The lower panel of the figure
shows the differences between data and Monte Carlo, 
divided by the total errors, as a function of $n$ for the 91~GeV and 
197~GeV data.}
\label{moments_tma}
\end{center}
\end{figure}

\begin{figure}[!htb]
\begin{center}
\resizebox{0.8\textwidth}{!}
{\includegraphics{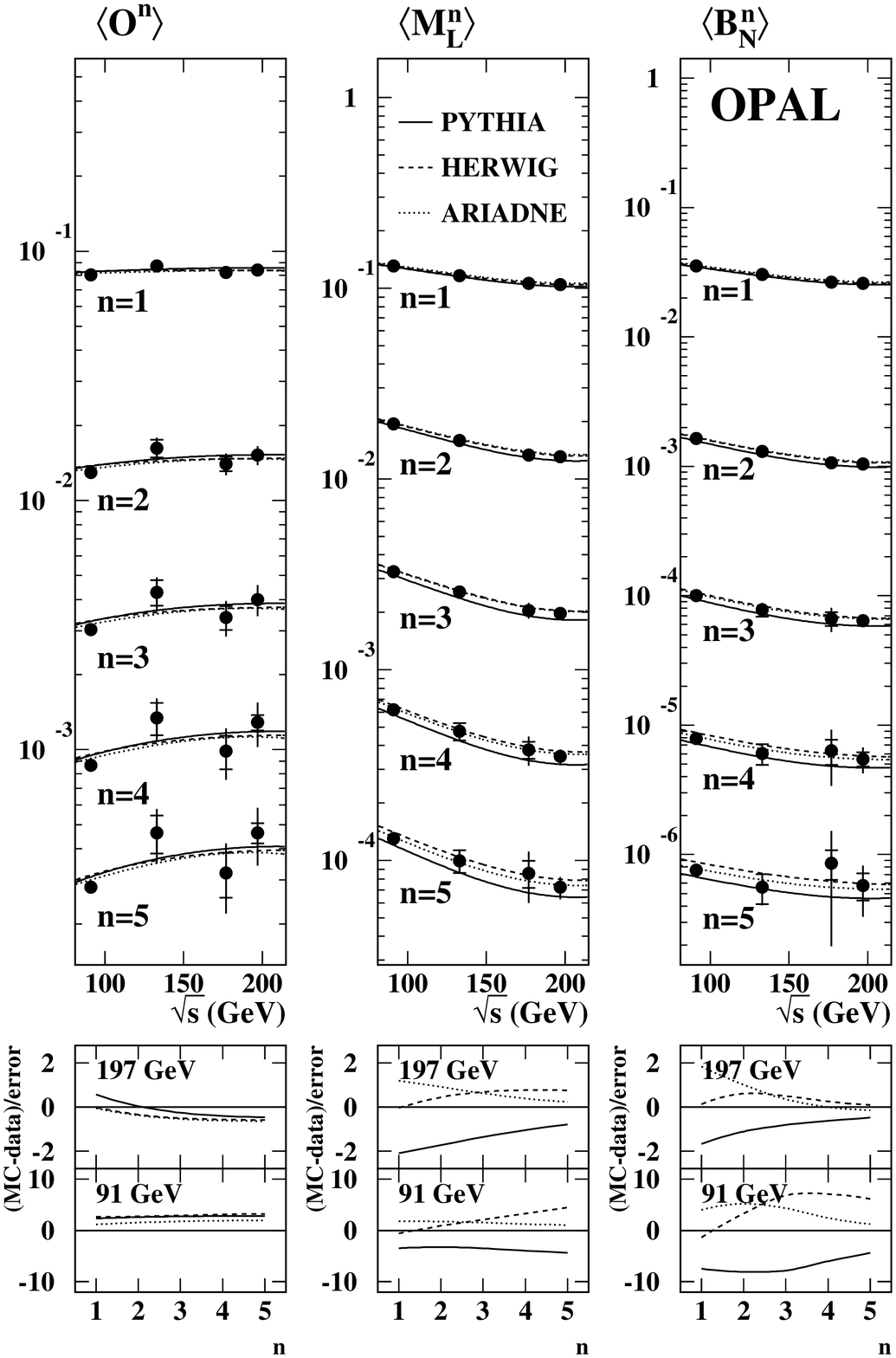}}
\caption[ ]{Moments of oblateness, \momn{$O$}{n}, 
light jet mass, \momn{\ml}{n}, and narrow jet broadening, \momn{\bn}{n}, for
$n=1,\ldots,5$ 
at four c.m.\ energy points ---
91~GeV, 133~GeV, 161--183~GeV (labelled 177~GeV) and 189--209~GeV 
(labelled 197~GeV).   The inner error bars
show the statistical errors, while the total errors are indicated by the 
outer error bars.  The predictions of
the \Pythia, \Herwig\ and \Ariadne\ Monte Carlo models as described in 
the text are indicated by curves.
  The lower panel of the figure
shows the differences between data and Monte Carlo, 
divided by the total errors, as a function of $n$ for the 91~GeV and 
197~GeV data.}
\label{moments_o}
\end{center}
\end{figure}

The measurements of the first five moments of the event shape observables
\thr, \mh, $C$, \bt, \bw, \ytwothree, \tma, \tmi, $S$, $O$, \ml\ and \bn\ 
for the four energy ranges are shown in Tables~\ref{tabmom1} and 
\ref{tabmom2}.\footnote{Further details of the moments
will be made available in the HEPDATA database, 
{\tt http://durpdg.dur.ac.uk/HEPDATA/}}  
The same data are shown
in Figures~\ref{moments_t}--\ref{moments_o}
compared with the same Monte Carlo event generators as
for the distributions.  
The lower parts of the figures show again the
differences between data and model predictions divided by the total
errors.  One observes that
for \Herwig\ the higher moments generally exhibit larger disagreements.  This
observation is consistent with the distributions where \Herwig\ showed 
the most significant
disagreement in the three-jet regions.  The \Pythia\ and \Ariadne\ models
tend to give a better description of the data, the \Ariadne\ model being
somewhat closer to the data than \Pythia.  For most observables \Herwig\
lies above the data at 91~GeV but below at higher energies.  The experimental
precision of the 91~GeV sample is much better than that 
of the other data samples and thus
comparison between data and simulation at 91~GeV is more sensitive.

In order to give an illustration of the sensitivity of the data to \qcd\ effects
like the running of the strong coupling and the changes in hadronization
we compare the first moments of $1-T$ and $C$ measured at 91 and at 197~GeV,
see Table~\ref{tabmom1}.  The two values of \mbox{$\langle 1-T \rangle$} 
are seen to differ by 5.8 standard
deviations, treating the experimental errors as 
uncorrelated
between the measurements.  Using $\langle C \rangle$ we observe an 8.6 standard
deviation difference between the measurements at 91 and 197 GeV.  
This shows that our data are indeed sensitive to perturbative 
\qcd\ effects.

%
%
\subsection{ Determination of \bm{\as} }
\label{sec_alphas}
\subsubsection{Determination of \bm{\as} using event shape distributions}
\label{sec_alphas_dist}
Our measurement of the strong coupling strength \as\ is based on
fits of \qcd\ predictions to the corrected distributions for \thr,
\mh, \cp, \bt, \bw\ and \ytwothree.
The theoretical descriptions of
these six observables are among the most complete, allowing the use of
combined \oaa+\nlla\ \qcd\
calculations~\cite{nllathmh,nllabtbw,nllad2,nllay23,nllabtbw2,nllacp}.
The main improvements to the calculations compared to 
those used in our previous publications were outlined 
in Sect.~\ref{sec-theocalc}.

The value of \as\ was estimated
by comparing theory with data using a minimum-$\chi^2$ method.
In the computation
of $\chi^2$, only the full statistical covariance matrix for the data, 
calculated as explained in Sect.~\ref{sec-corrproc}, was used. 
Separate fits were performed to each of the six observables
at each c.m.\ energy value or range.   
The fit ranges were the same as  
those used in the previous \Opal\ publication~\cite{OPALPR303}, 
and were constrained by the
requirement that the detector and hadronization corrections be reasonably 
small in the fit region, and that both $\chi^2$ and the fitted value of
\as\ be reasonably stable under small variations in the fit range.    

\begin{figure}[p]
 \begin{center}
 \includegraphics[width=0.88\textwidth]{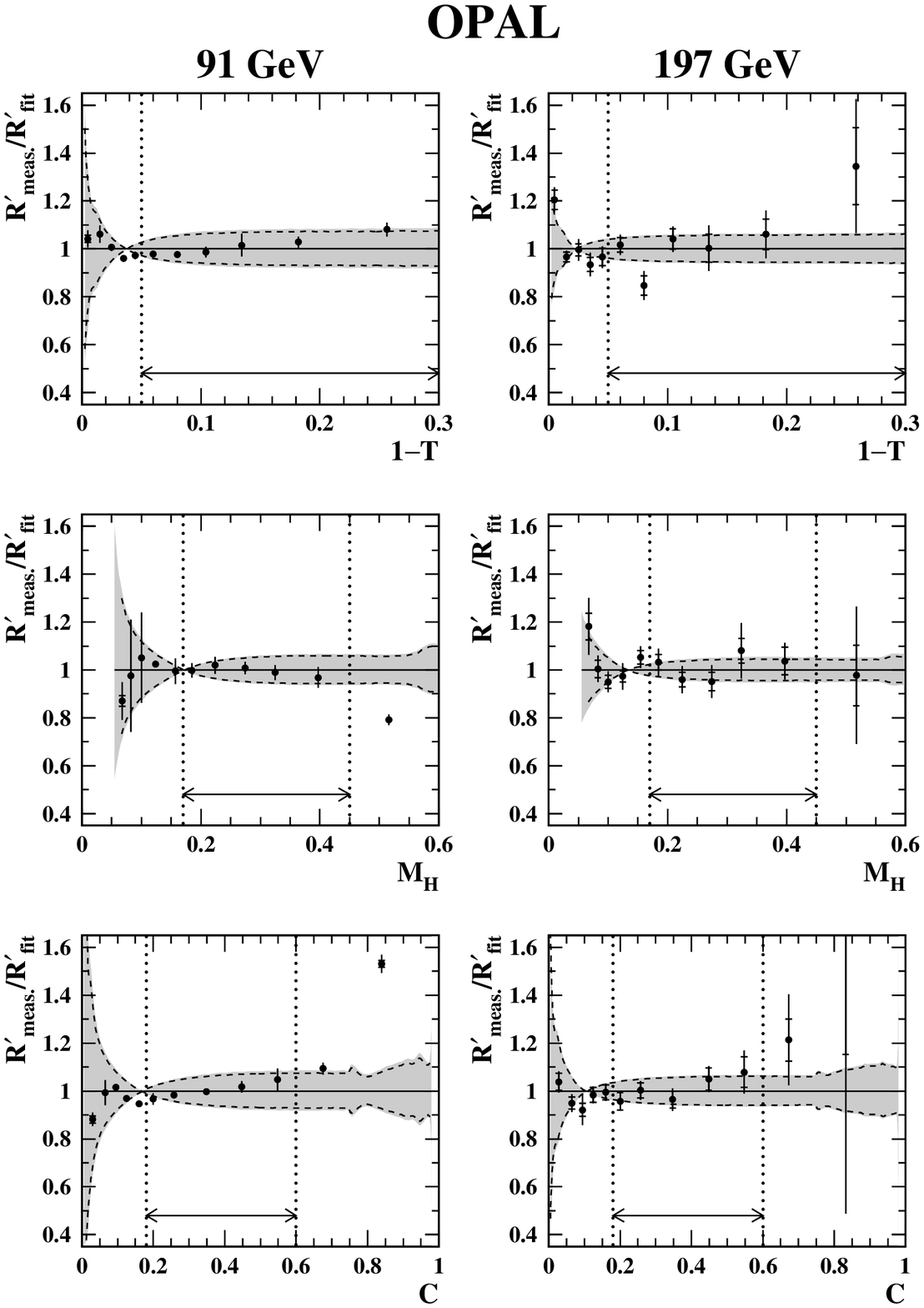}
 \end{center}
 \caption[Ratio of 91 GeV distributions to fitted hadron-level predictions]
 {Fits of \as\ to event shape distributions 
for \thr, \mh\ and $C$ at 91~GeV and
197 GeV (i.e.\ combined 189--209~GeV).
Each data point shows the measured bin contents divided by the integral of
the predicted distribution across the bin; the inner error bars
indicate statistical uncertainties, and the outer bars show the
combined statistical and experimental contributions. The dashed
curves represent fractional variations in the predicted distributions,
corresponding to our perturbative theory uncertainties in~\as. The
slightly wider shaded bands indicate the combined theory and
hadronization uncertainties. The ranges used for fitting each
distribution are shown by horizontal arrows.
 }
\label{fig_fit91}
 \end{figure}

\begin{figure}[p]
 \begin{center}
 \includegraphics[width=0.88\textwidth]{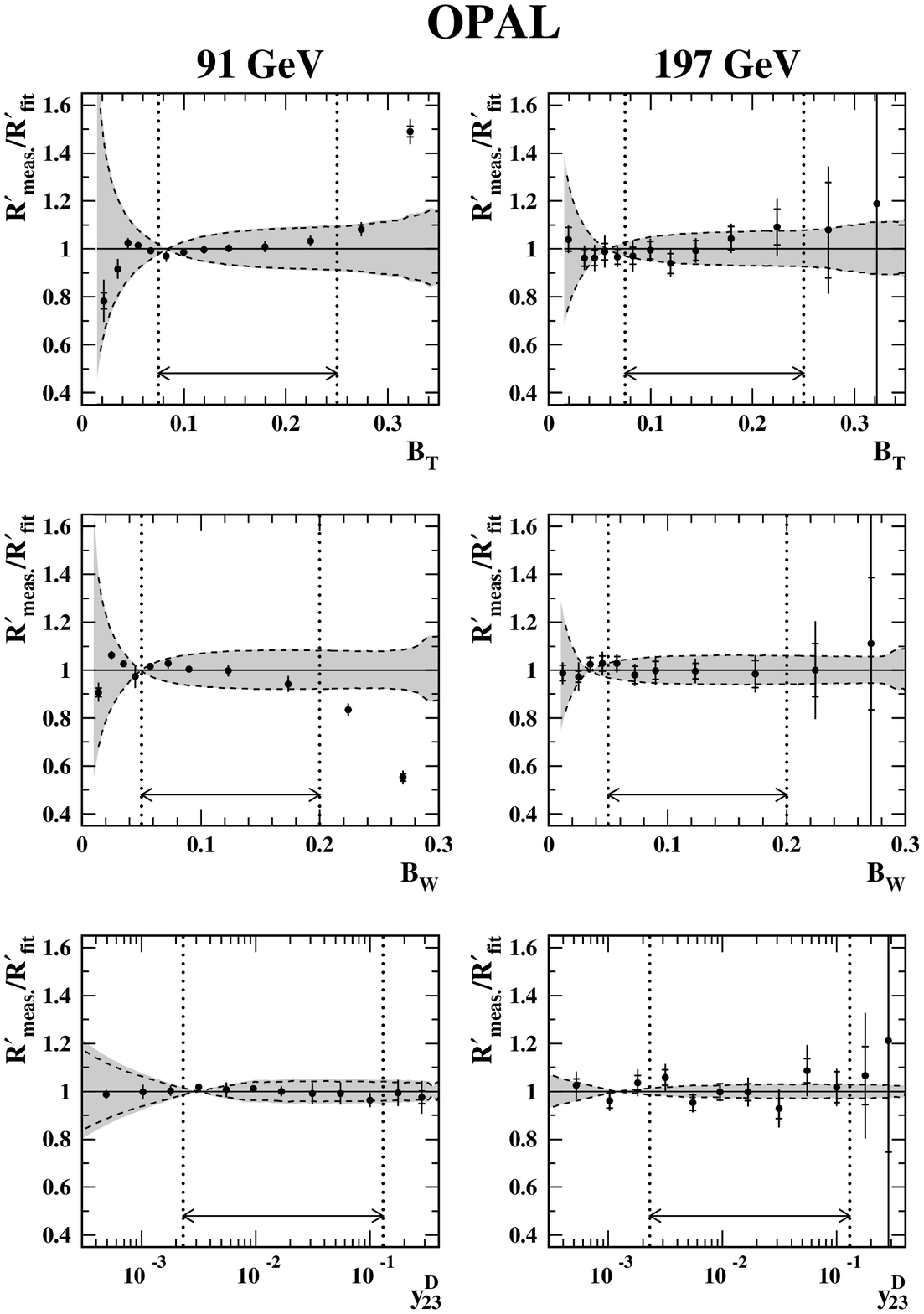}
 \end{center}
 \caption[Ratio of 
 to fitted hadron-level predictions]
 {
Fits of \as\ to event shape distributions 
for \bt, \bw\ and \ytwothree\ at 91~GeV and
197 GeV (i.e.\ combined 189--209~GeV).
Each data point shows the measured bin contents divided by the integral of
the predicted distribution across the bin; the inner error bars
indicate statistical uncertainties, and the outer bars show the
combined statistical and experimental contributions. The dashed
curves represent fractional variations in the predicted distributions,
corresponding to our perturbative theory uncertainties in~\as. The
slightly wider shaded bands indicate the combined theory and
hadronization uncertainties. The ranges used for fitting each
distribution are shown by horizontal arrows.
}
\label{fig_fit197}
 \end{figure}

The fit results are summarized in Table~\ref{tab:alphasfits1} for the
four c.m. energy points presented previously.\footnote{
Note that ref.~\cite{OPALjets} also determines \as\ using
\ytwothree, there called $D_2$.  The small differences between the results
have been investigated in detail, and are not significant.  They may 
be attributed to differences in fit regions, 
the use of statistically different Monte Carlo samples, and
the adoption of slightly different strategies for the assessment of 
theoretical errors.}
In addition, further fit results are presented in  
Table~\ref{tab:alphasfits2} for various other groupings of the data in
c.m.\ energy\footnote{These energy values and ranges 
are those used by the \Lep~\qcd\ working group, namely
161~GeV, 172~GeV, 183~GeV, 189--192~GeV, 196--202~GeV and 
$>202$~GeV.  In the case of the 196--202~GeV point, the value of 
\as\ has been run from the mean c.m. energy of 198.6~GeV
to a nominal value of 200~GeV using the expected \qcd\ behaviour ---
a correction of 0.0001.}.  
The statistical error on the fitted value of \as\ was
estimated from the variation of $\chi^2$ by $\pm1$ about
its minimum.  Systematic uncertainties were assessed using the
techniques described in Sect.~\ref{sec_syserr}.  For each variant of the 
analysis, the corresponding distribution was fitted 
to determine \as, and the difference with respect to 
the value of \as\ from 
the default analysis was taken as a systematic error contribution.
In Figs.~\ref{fig_fit91} and~\ref{fig_fit197} we show the ratio of the 
data to the fitted theory for each of the six event shapes at
91~GeV and 197~GeV respectively.  Because of normalization, the theory predictions 
are seen to ``pivot'' about some value of $y$, which indicates that the data at that
point have no sensitivity to \as.    

\begin{figure}[p]
 \begin{center}
 \includegraphics[width=0.85\textwidth]{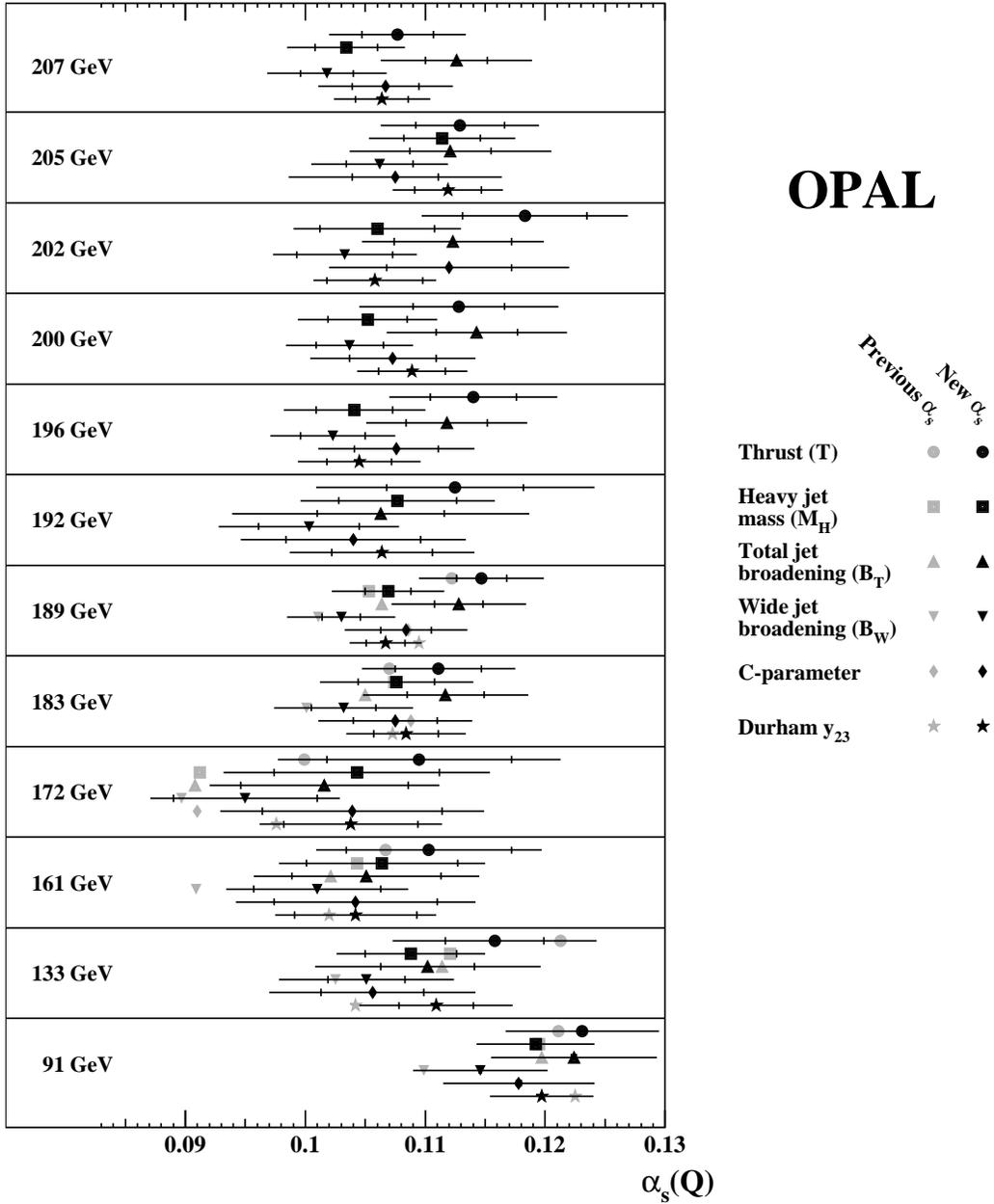}
 \end{center}
 \caption[]
 {Measurements of \as\ using fits to distributions of six event shape
observables. The inner error bars represent statistical
uncertainties and the outer error bars the total uncertainties. 
The grey symbols indicate, without errors,  previously published \Opal\
measurements, which are superseded by our new results.
}
\label{fig_alphas_all}
 \end{figure}

The measurements of \as\ for each observable and c.m.\ energy
are shown in Figure~\ref{fig_alphas_all}.  We note that the 
values of \as\ at 91~GeV are significantly higher than at \Leptwo\
energies, providing evidence for the running of \as.
Systematic differences between the values of \as\ obtained from different 
observables are seen; for example, $(1-T)$ and \bt\ tend to give
higher than average values of \as, whilst \bw\ tends to give the lowest value.
These differences may be ascribed to the differing effects of
uncomputed higher order terms on the various observables; they
are often greater than the statistical errors, but are  
covered by the systematic uncertainties.
There are also significant statistical correlations
between the values of \as\ obtained from the different observables
at a given energy, so that the scatter of the points at
energies where the statistics are low (e.g.\ 161~GeV) tends to 
be smaller than one would expect if the statistical errors  were
uncorrelated. 

It is useful to combine the measurements of
\as\ from different observables and/or c.m.\ energy points,
in order to exploit the data fully.  This problem has been
the subject of extensive study within the \Lep\ \qcd\ 
working group~\cite{LEPQCDWG2},
and we adopt their procedure here.  
In brief, the method is as follows.
The \as\ measurements to be combined are first evolved to 
a common scale, $Q=\mz$, assuming the validity of \qcd.  
Their combination is then performed using
a weighted mean method, such as to minimize the $\chi^2$ between 
the combined value and the measurements.  So, if the
measured values evolved to 
$Q=\mz$ are denoted $\alpha_i$ with covariance matrix $V$, 
the combined value, \asmz, is given by
\begin{equation} \asmz = \sum_i w_i \alpha_i \;\;\;\;\; \mathrm{where}\;\;\;\;\;
w_i = \sum_j (V^{-1})_{ij} / \sum_{j,k} (V^{-1})_{jk} \;.\end{equation}
The combined values may then be evolved back to the 
original scale if required.
The difficulty resides in making a reliable estimate of
$V$ in the presence of dominant and highly correlated systematic
errors; small uncertainties in the estimation of these
correlations can easily cause undesirable features such 
as negative weights.  For this reason, only statistical errors
(estimated using data-sized subsamples of Monte Carlo events to
assess correlations between different observables in a given data sample)
and experimental systematic errors 
(for which the correlations are estimated using the ``minimum overlap''
assumption\footnote{The minimum overlap ansatz involves 
taking the covariance
between any pair of systematic error contributions to be 
equal to the smaller of the two variances.}) 
were allowed to contribute to the off-diagonal elements 
in $V$ when computing the weights, while all error contributions were
included in the diagonal terms.
The hadronization and theoretical uncertainties were 
computed by combining the \as\ values obtained with the 
alternative hadronization models, and from the upper and lower    
theoretical errors, using the same weights.

If the full covariance matrix were used to compute the weights $w_i$, then this
choice of weights would have the effect of minimising the total error.
However, the correlations between the systematic errors
cannot be completely reliably estimated (as manifested by the occurrence of 
negative weights).  The modified procedure adopted here will not in general
minimize the error, and indeed the error on the weighted average may be
greater than one of the inputs; this actually arises in our data
for the case of \ytwothree.  Despite this, we consider the 
combined value to provide the safest estimate of \as, since it cannot be guaranteed 
that the relatively small theoretical error for \ytwothree\ is not fortuitous. 

\begin{figure}[p]
 \begin{center}
 \includegraphics[width=0.95\textwidth]{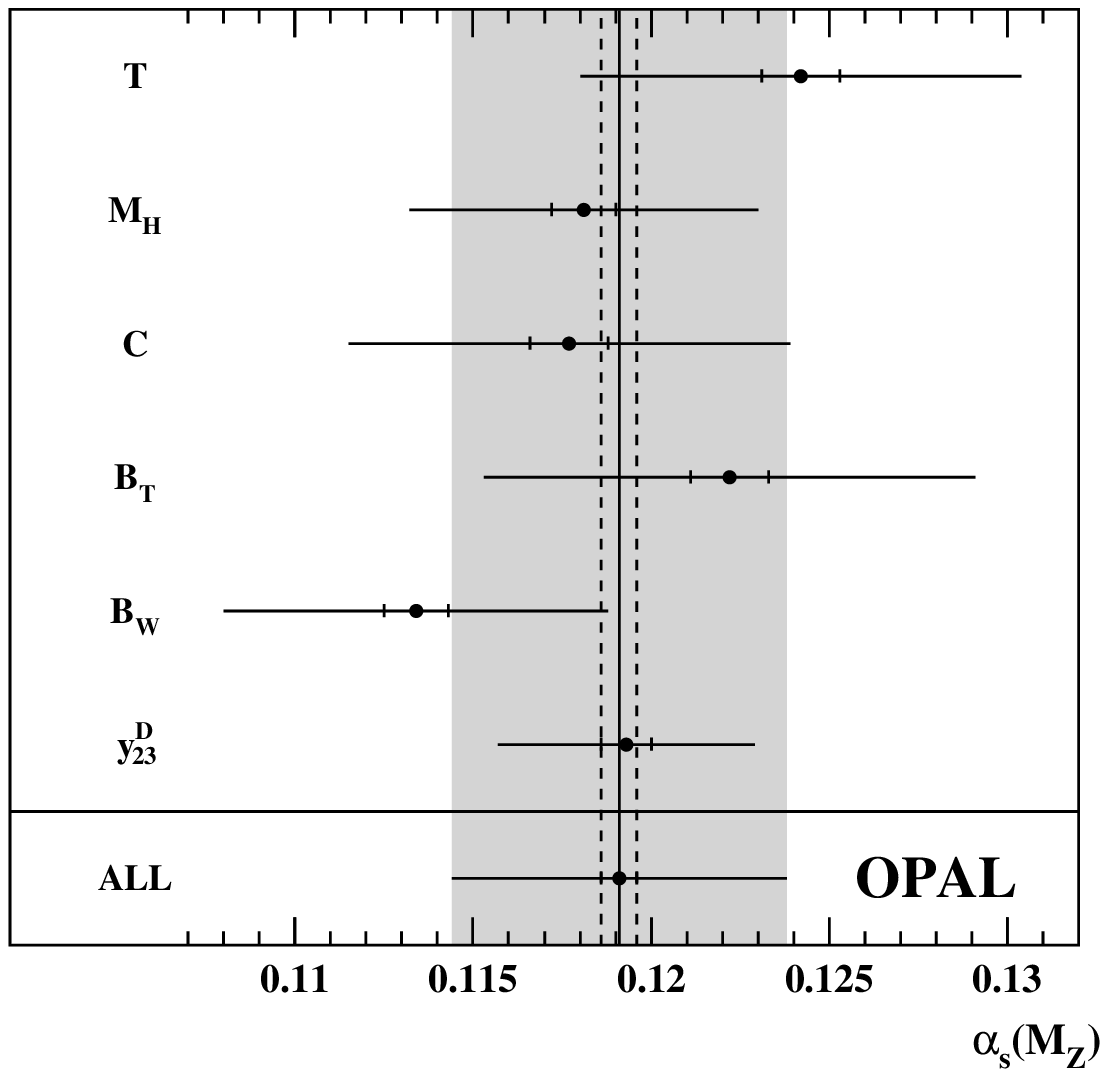}
 \end{center}
 \caption[]
 {\Opal\ combinations of \asmz\ values inferred from distributions of 
individual event shape
observables. The inner error bars are statistical, while the outer
bars represent total uncertainties.
The grey band corresponds to the total uncertainty of
  the combined \asmz\ value, and the dashed lines indicate its
  statistical uncertainty.
}
\label{fig_alphas_by_var}
 \end{figure}

\begin{figure}[p]
 \begin{center}
 \includegraphics[width=0.95\textwidth]{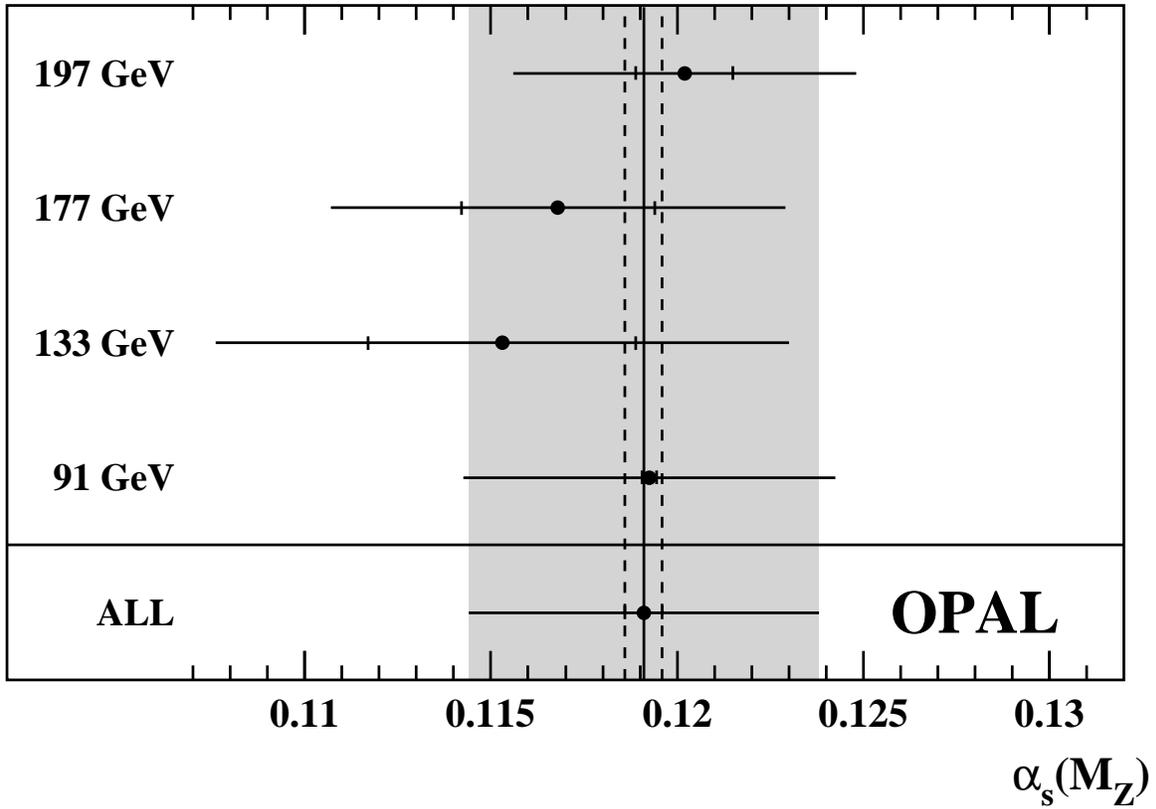}
 \end{center}
 \caption[]
 {\Opal\ \asmz\ combinations based on distributions of
 individual event shape
observables at different energies. 
The inner error bars are statistical, while the outer
bars represent total uncertainties.
The grey band corresponds to the total uncertainty of
the combined \asmz\ value, and the dashed lines indicate its
statistical uncertainty.
}
\label{fig_alphas_by_e}
 \end{figure}

\begin{figure}[p]
 \begin{center}
 \includegraphics[width=0.95\textwidth]{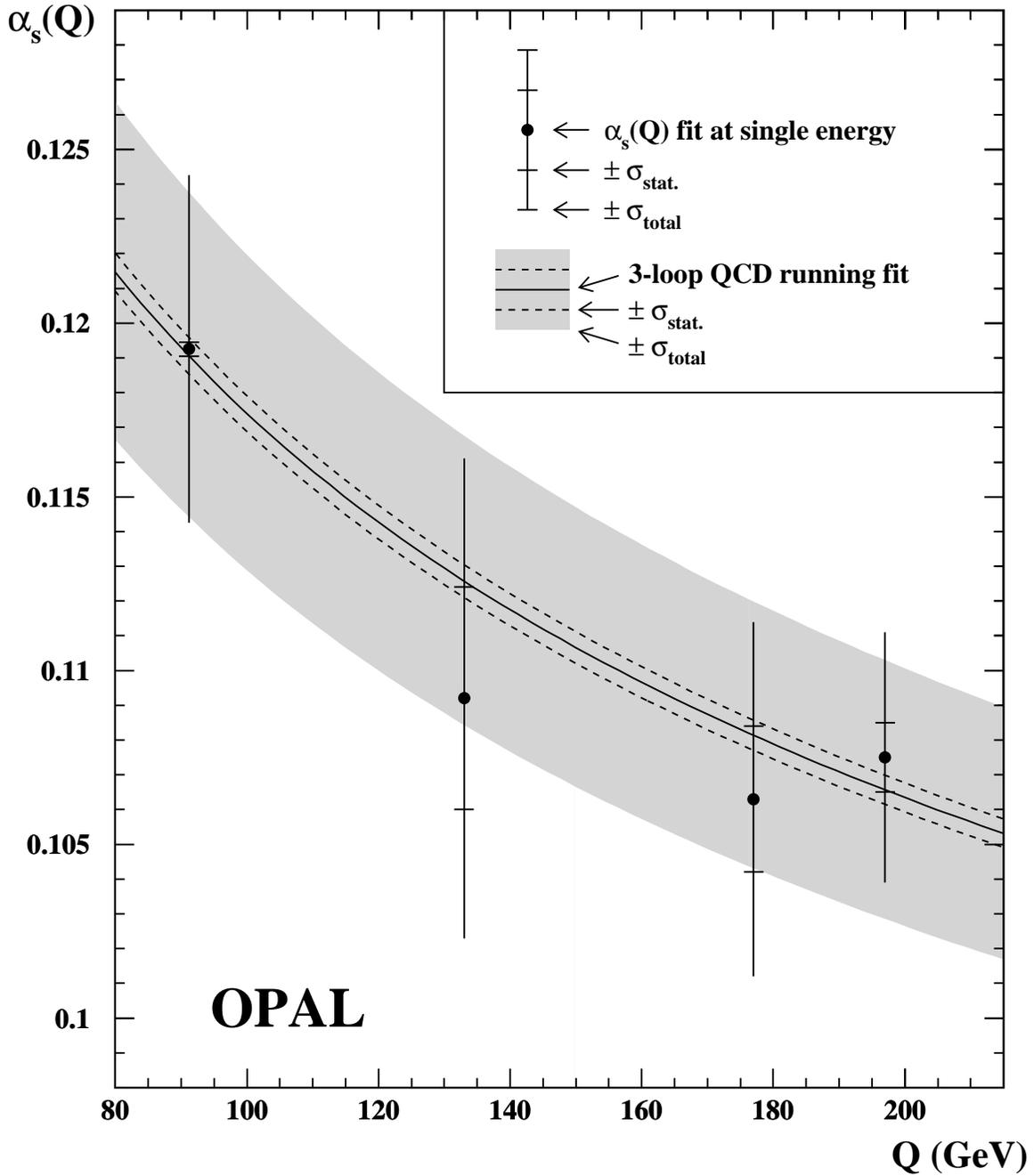}
 \end{center}
 \caption[]
 {A global \qcd\ running fit to the \Opal\ \as\ measurements
based on event shape distributions. Each
  point represents a fit to the six measurements at an individual
  centre-of-mass energy, while the curve represents a global fit to
  all measurements. The form of the curve is determined by the
  $\mathcal{O}(\alpha_{\mathrm{s}}^3)$
  Renormalization Group Equation of \qcd, with \asmz\ as a free
  parameter. The grey band corresponds to the total uncertainty of
  the fitted \asmz\ value, and the dashed curves indicate the
  statistical uncertainty.
}
\label{fig_alphas_run}
 \end{figure}

In Table~\ref{tab:fitsbyobservable_opal} we give the 
values of \as\ for each observable, evolved to a common 
scale \mz, combined over all 
c.m.\ energies.  These results are also summarized in
Figure~\ref{fig_alphas_by_var}.   This shows clearly that the
measurements  from the different observables are far 
from compatible when only statistical errors are
considered, but are consistent with a common mean when 
the systematic errors are included.   
The results of combining the \as\ values for the six observables
are given in the rightmost columns
of Tables~\ref{tab:alphasfits1} and \ref{tab:alphasfits2}.  
In addition, the relative weight, $w_i$,
carried by each observable is given, from which we see that
\ytwothree\ generally carries the greatest weight
(because it has the smallest theoretical uncertainty).
These results are plotted in Figure~\ref{fig_alphas_by_e} where we show the values
at each energy evolved to a common scale, \asmz, and in
Figure~\ref{fig_alphas_run} where we show \as\ as a function of 
the energy scale $Q=\ecm$.
These plots show that the variation of \as\ with \ecm\ is 
consistent with the running predicted by \qcd, and is
incompatible with a constant value of \as.
For example, the two most precise 
values of \as, at 91 and 197~GeV, differ by more than
three standard deviations (applying the minimum overlap ansatz 
to the systematic errors).  

The measurement of \as\ based on the 91~GeV data is
\[
\asmz =  0.1192
\pm0.0002~\mathrm{(stat.)}
\pm0.0008~\mathrm{(expt.)}
\pm0.0015~\mathrm{(hadr.)}
\pm0.0047~\mathrm{(theo.)} 
\]
while the result for \asmz\ combining all 
higher energy data at $Q>\mz$ is
\[
\asmz =  0.1189
\pm0.0011~\mathrm{(stat.)}
\pm0.0015~\mathrm{(expt.)}
\pm0.0008~\mathrm{(hadr.)}
\pm0.0040~\mathrm{(theo.)}\;. 
\]
The consistency between these measurements, which are equal within the 
statistical errors, 
again shows that the data are 
compatible with the running predicted by 
\qcd, since this running was assumed in evolving the
high energy data to $Q=\mz$.  We note that the high energy 
data have significantly smaller theoretical and hadronization
errors, and therefore complement the statistically superior
91~GeV data.
The value of \asmz\ obtained from
all observables and all energies combined 
is\footnote{The $\chi^2/\dof$ for this overall fit is 10.5/23. The value
is significantly smaller than unity because of the necessity to 
neglect correlations in the combination.}:
{\small
\begin{displaymath}
\fbox{$\begin{array}{rcl}
\asmz & = & 0.1191
\pm0.0005~\mathrm{(stat.)}
\pm0.0010~\mathrm{(expt.)}
\pm0.0011~\mathrm{(hadr.)}
\pm0.0044~\mathrm{(theo.)}\\
& = & 0.1191
\pm0.0005~\mathrm{(stat.)}
\pm0.0046~\mathrm{(syst.)} \\
& = & 0.1191
\pm0.0047~\mathrm{(total)}
\end{array}$}
\end{displaymath}
}
\subsubsection{Determination of \bm{\as} using event shape moments}
\label{sec_alphas_mom}

The strong coupling \asmz\ has also been determined from the measured
moments using the \oaa\ \qcd\
predictions explained in Sect.~\ref{sec_qcdmon}.
This is the first such study to be published by the \Opal\ Collaboration.  
The first five
moments of the observables \mbox{\thr}, \mh, \bt, \bw, \cp\ and \ytwothree\ were
studied, i.e.\ the same observables as used for the
determination of \as\ from distributions. 
One might anticipate {\it a priori} that this would yield a less precise 
determination of \as\ than the differential distributions, 
because the theory lacks resummation of large logarithms,
and the moments include regions of phase space where hadronization effects are
large. Nevertheless, the comparison of \as\ determined in this way with 
that obtained from the distributions should provide an illuminating
test of the adequacy of \qcd\ in this area.

The fits proceeded by comparing the data at the four combined energy
points for a given moment of an observable with the theory prediction.
The running of \as\ was implemented in the fit in two-loop precision
using the formula given in ref.~\cite{pdg04}.  A value of 
\chisq\ was calculated
using the statistical errors of the data and minimized to 
extract a value of \asmz.
The fits were repeated for each systematic variation of the analysis.  
The statistical error was found as above by variation of \chisq\ by $\pm1$
 about the minimum.

\begin{figure}[p]
 \begin{center}
 \includegraphics[width=0.95\textwidth]{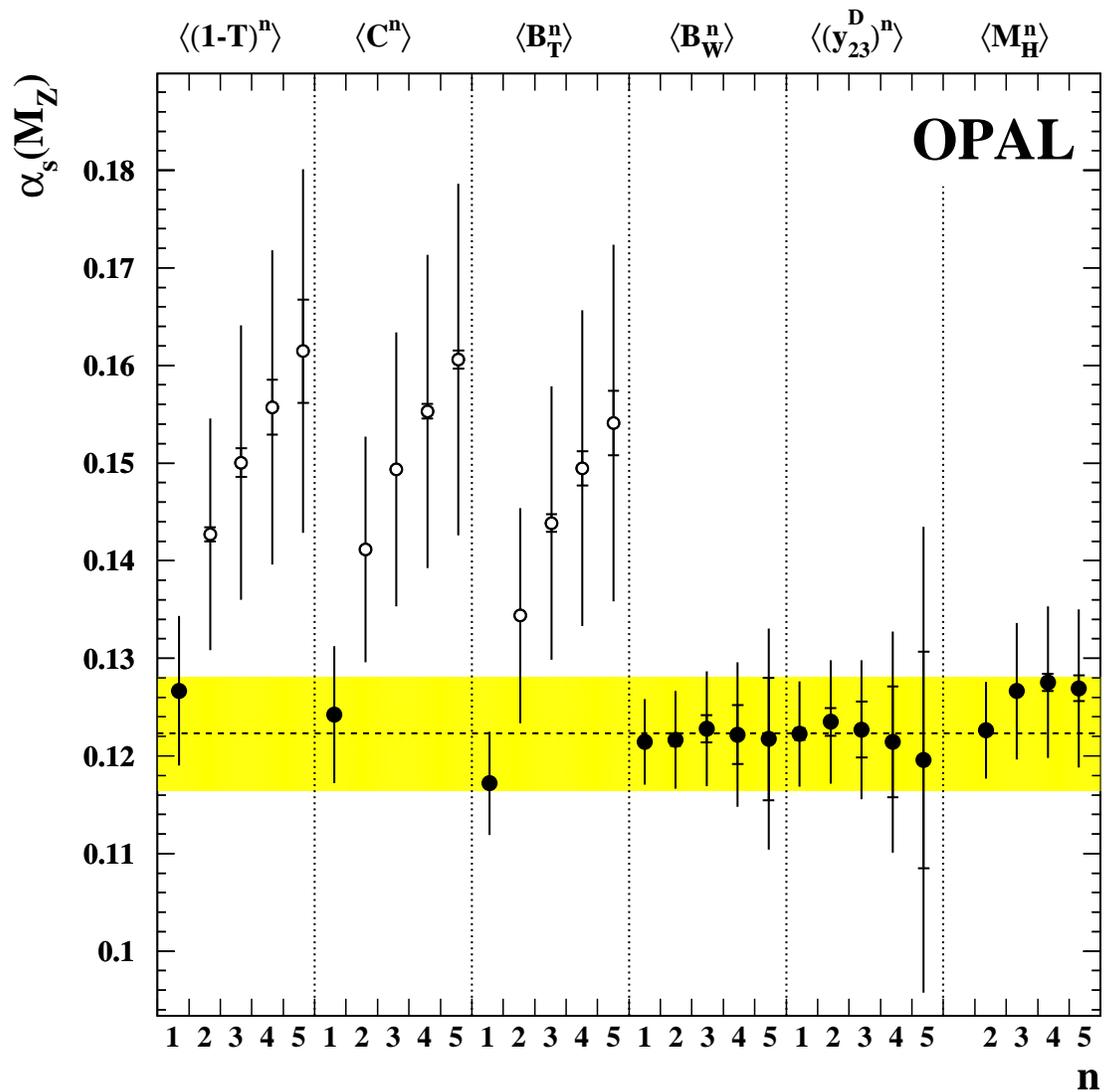}
 \end{center}
 \caption[]
 { Measurements of \as\ using fits to moments of six event shape
observables. The inner error bars represent statistical
uncertainties, and the outer error bars show the total errors.  
The dashed line indicates the weighted average described in the text;
only the measurements indicated by solid symbols were 
used for this purpose.}
\label{fig_momfits}
 \end{figure}

\begin{figure}[p]
 \begin{center}
 \includegraphics[width=0.95\textwidth]{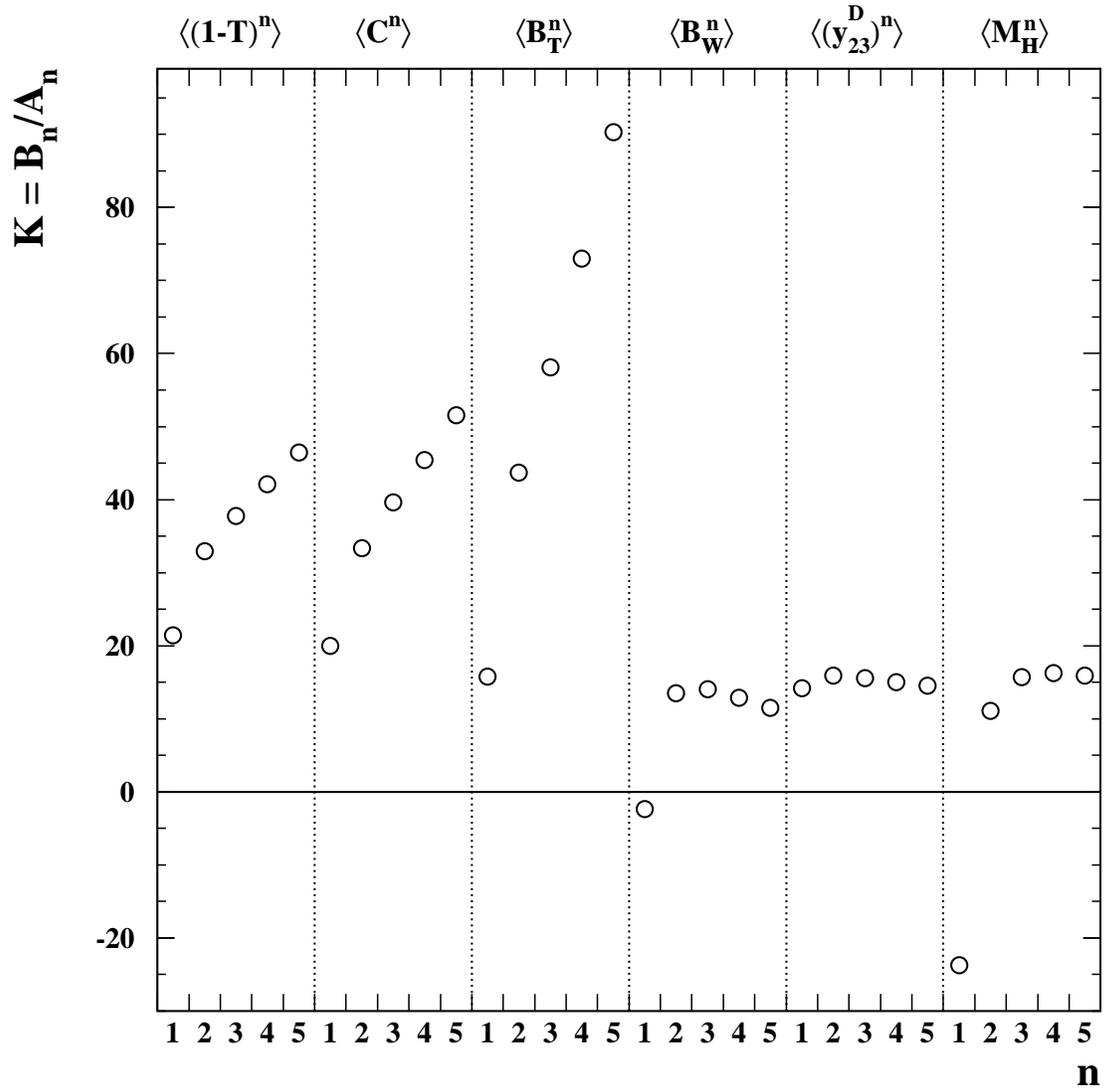}
 \end{center}
 \caption[]
 { The ratio $K=\mathcal{B}_n/\mathcal{A}_n$
of \nlo\ and \lo\ coefficients for the six observables used in our 
determinations of \asmz\ from moments.
 }
\label{fig_boa}
 \end{figure}

The fit results are shown in Table~\ref{tab_momfits} and in
Figure~\ref{fig_momfits}.  The fit to \momone{\mh} was not stable and
therefore no results are shown.  We obtain values of \chisqd\ of
$\mathcal{O}(1)$; the fitted \qcd\ predictions including the running of \as\ are
thus consistent with our data.  However, we find that for
\momn{(\thr)}{n}, \momn{\cp}{n} and \momn{\bt}{n} the fitted values of
\asmz\ increase with the order $n$ of the moment used.  This effect is
not observed for \momn{\bw}{n}, \momn{(\ytwothree)}{n} and \momn{\mh}{n},
$n=2,\ldots,5$.  In order to investigate the origin of this behaviour
we show in Figure~\ref{fig_boa} the ratio $K={\cal B}_n/{\cal A}_n$
of \nlo\ and \lo\ coefficients for the six observables used in our fits.
There is a clear correlation between the increasing values of \asmz\
with moment $n$ and increasing values of $K$ with $n$ for 
\momn{(\thr)}{n}, \momn{\cp}{n} and \momn{\bt}{n}.  The other observables 
\momn{\bw}{n}, \momn{(\ytwothree)}{n} and \momn{\mh}{n}, $n=2,\ldots,5$, 
have fairly
constant values of $K$ and correspondingly stable results for \asmz.  We
also note that \momone{\mh} has a large and negative value of $K$ which is
the cause of the unstable fits\footnote{By reference to 
equation~(\ref{eq_qcdmom}) we can see that there is no real solution for 
\as\ if ${\cal B}_n < -{\cal A}_n^2/4\momn{y}{n}$.  This is 
the case for \momone{\mh}, which accounts for the failure of the fits.
There is one other case where the coefficient ${\cal B}_n$ is negative, 
namely \momone{\bw}, but it is not sufficiently negative to preclude 
a solution for \as.}. 

In order to combine the individual determinations of \asmz\ we used the
same weighted average method as in Sect.~\ref{sec_alphas_dist}.  
We considered only
those results for which the fit was successful and for which
the \nlo\ term in equation~(\ref{eq_qcdmom}) is less than half the \lo\ 
term (i.e.\  $|K\as/2\pi|<0.5$ or $|K|\lesssim 25$), 
namely \momone{\thr},
\momone{\cp}, \momone{\bt}, \momn{\bw}{n} and \momn{(\ytwothree)}{n},
$n=1,\ldots,5$ and \momn{\mh}{n}, $n=2,\ldots,5$; i.e.\ results from 17
observables in total.  The statistical correlations between the 17
results were determined using Monte Carlo simulation at the
hadron-level.  The experimental errors were included in the covariance
matrix using the minimum-overlap assumption while the hadronization
and renormalization scale uncertainties were only added to the diagonal
of the covariance matrix.  The combination was repeated using the same
weights with the \Herwig\ and \Ariadne\ hadronization corrections and for
$\xmu=0.5$ and $\xmu=2.0$ for the calculation of the hadronization
theory systematic errors.  
The weights were generally $\mathcal{O}(10\%)$ for each observable;
the largest weight was 23\% for \momone{\bt} while the only weights below 
3\% were those for \momn{(\ytwothree)}{5} \momn{\mh}{3} and \momn{\mh}{4}. 
The result is
\begin{displaymath}
\begin{array}{rcl}
\asmz & = & 0.1223
\pm0.0005~\mathrm{(stat.)}
\pm0.0014~\mathrm{(expt.)}
\pm0.0016~\mathrm{(hadr.)}
^{+0.0054}
_{-0.0036}
\mathrm{(theo.)}\\
& = & 0.1223
\pm0.0005~\mathrm{(stat.)}
\pm0.0058~\mathrm{(syst.)} \\
& = & 0.1223
\pm0.0059~\mathrm{(total)}
\end{array}
\end{displaymath}
in good agreement with the result from distributions 
presented in Sect.~\ref{sec_alphas_dist}.  

The experimental, hadronization and theory uncertainties are somewhat
larger than for the distributions.  
In an analysis using moments the complete available phase
space is sampled including regions which are more difficult to measure
experimentally and which are less reliably modelled by the hadronization
models.  Also, the \nlo\ \qcd\ prediction is less complete than the
matched \oaa+\nlla\ prediction available for the distributions which we
studied.  
It is nevertheless a remarkable success of \qcd\ together 
with the corresponding 
hadronization models that the \nlo\ theory is able to describe 
successfully
some observables based on the complete available phase space.

\section{ Summary and conclusions }
\label{sec_conc}
In this paper we have presented measurements of the event shapes for
hadronic events produced at \Lep\ at centre-of-mass energies between 91
and 209~GeV.  Both differential distributions and moments have been 
determined.

The predictions
of the \Pythia, \Herwig\ and \Ariadne\ Monte Carlo models are found to be
in general agreement with the measured distributions and moments.  
However, some 
discrepancies are noted in the 91~GeV data where the statistical errors 
are smallest.  The main differences between models and data occur in
the extreme two-jet region, and for observables sensitive to
the production of four or more jets.  In general, \Ariadne\ provides
the best description of the data and \Herwig\ the least good. 

From a fit of \oaa+\nlla\ \qcd\ predictions to the distributions of
six event shape observables,
we have determined the strong coupling parameter $\as$.
The variation of \as\ with energy scale over the range 91 to 209~GeV
is found to be in accordance with the expectations of \qcd.
For example, the measurement of \as\ based on the 91~GeV data is
\[
\asmz =  0.1192
\pm0.0002~\mathrm{(stat.)}
\pm0.0008~\mathrm{(expt.)}
\pm0.0015~\mathrm{(hadr.)}
\pm0.0047~\mathrm{(theo.)} \;.
\]
Assuming the validity of \qcd, the higher energy measurements can all 
be evolved to a common scale $Q=\mz$ and combined, yielding
the following result for \asmz\ combining all 
higher energy data at $Q>\mz$
\[
\asmz =  0.1189
\pm0.0011~\mathrm{(stat.)}
\pm0.0015~\mathrm{(expt.)}
\pm0.0008~\mathrm{(hadr.)}
\pm0.0040~\mathrm{(theo.)}\;. 
\]
Combining all data at all energy scales, the value of \asmz\ is determined to be
\[ \asmz  =  0.1191
\pm0.0005~\mathrm{(stat.)}
\pm0.0010~\mathrm{(expt.)}
\pm0.0011~\mathrm{(hadr.)}
\pm0.0044~\mathrm{(theo.)}
 \;,\]
in good agreement with the world average quoted in ref.~\cite{pdg04}.
The results for \asmz\ derived 
from different event shapes are consistent within
errors.  
\par
Values of \asmz\ have also been derived from the energy dependence
of event shape moments, using \oaa\ \qcd.  Although less complete
than the \oaa+\nlla\ \qcd\ predictions used for the distributions, 
these calculations prove to give a consistent description of many, 
though not all, of the moments.   The combined value obtained from
the moments was
\mbox{$\asmz =  0.1223
\pm0.0005~\mathrm{(stat.)}
\pm0.0014~\mathrm{(expt.)}
\pm0.0016~\mathrm{(hadr.)}
^{+0.0054}
_{-0.0036}
\mathrm{(theo.)}\;,
$}
consistent with that derived from the
distributions. 
However, because the value obtained from
the distributions is based on the more complete theory (and has a smaller
overall error) we consider that to be the most reliable estimate 
from the present data.  

\newpage
\section*{Acknowledgements}
\par

We particularly wish to thank the SL Division for the efficient operation
of the LEP accelerator at all energies
 and for their close cooperation with
our experimental group.  In addition to the support staff at our own
institutions we are pleased to acknowledge the  \\
Department of Energy, USA, \\
National Science Foundation, USA, \\
Particle Physics and Astronomy Research Council, UK, \\
Natural Sciences and Engineering Research Council, Canada, \\
Israel Science Foundation, administered by the Israel
Academy of Science and Humanities, \\
Benoziyo Center for High Energy Physics,\\
Japanese Ministry of Education, Culture, Sports, Science and
Technology (MEXT) and a grant under the MEXT International
Science Research Program,\\
Japanese Society for the Promotion of Science (JSPS),\\
German Israeli Bi-national Science Foundation (GIF), \\
Bundesministerium f\"ur Bildung und Forschung, Germany, \\
National Research Council of Canada, \\
Hungarian Foundation for Scientific Research, OTKA T-038240,
and T-042864,\\
The NWO/NATO Fund for Scientific Research, the Netherlands.\\

%


\clearpage

\begin{table}
\begin{center}
\begin{tabular}{| c | r@{$\;$--$\;$}l | c | c | c |c|}
\hline
Year &
\multicolumn{2}{|c|}{\parbox{2.5cm}{\centering Range of $\sqrt{s}$\\(GeV)}} &
\multicolumn{1}{|c|}{\parbox{2.5cm}{\centering Mean $\sqrt{s}$\\(GeV)}} &
\multicolumn{1}{|c|}{\parbox{2.1cm}{\centering \rule{0pt}{0.4cm}Integrated\\luminosity (pb$^{-1}$)\rule[-0.2cm]{0pt}{0pt}}} &
\multicolumn{1}{|c|}{\parbox{2.5cm}{\centering Number of\\selected events}} & 
\multicolumn{1}{|c|}{\parbox{2.5cm}{\centering Expected number}} 
\\ \hline
1996--2000 &  91.0 &  91.5 &  91.3 &  14.7 & 395695& ---\\
\hline
1995, 1997 & 129.9 & 136.3 & 133.1 &  11.26 & 630 & 698\\
\hline
1996       & 161.2 & 161.6 & 161.3 & 10.06 & 281 & 275\\
1996       & 170.2 & 172.5 & 172.1 & 10.38 & 218 & 232\\
1997       & 180.8 & 184.2 & 182.7 & 57.72 & 1077 & 1084\\
\hline
1998       & 188.3 & 189.1 & 188.6 & 185.2 & 3086 & 3130\\
1999       & 191.4 & 192.1 & 191.6 & 29.53 & 514 & 473\\
1999       & 195.4 & 196.1 & 195.5 & 76.67 & 1137 & 1161\\
1999, 2000 & 199.1 & 200.2 & 199.5 & 79.27 & 1090 & 1131\\
1999, 2000 & 201.3 & 202.1 & 201.6 & 37.75 & 519 & 527\\
2000       & 202.5 & 205.5 & 204.9 & 82.01 & 1130 & 1090\\
2000       & 205.5 & 208.9 & 206.6 & 138.8 & 1717 & 1804\\
\hline
\end{tabular}
\end{center}
\caption{The \Opal\ data samples used for the present analysis.
The horizontal lines separate the data into the four energy ranges 
used for presentation purposes.}
\label{datasummary}
\end{table}

\begin{table}[hb!]
\begin{sideways}
\begin{minipage}[b]{\textheight}
\begin{center}
\scalebox{0.75}{
\begin{tabular}{| r@{ -- }l |
r@{.}l @{ $\pm$ } r@{.}l @{ $\pm$ } r@{.}l @{} l |
r@{.}l @{ $\pm$ } r@{.}l @{ $\pm$ } r@{.}l @{} l |
r@{.}l @{ $\pm$ } r@{.}l @{ $\pm$ } r@{.}l @{} l |
r@{.}l @{ $\pm$ } r@{.}l @{ $\pm$ } r@{.}l @{} l |}
\hline
\multicolumn{2}{|c|}{$1-T$}  
& \multicolumn{7}{|c|}{$R'\,(1-T)$ at 91 GeV}  
& \multicolumn{7}{|c|}{$R'\,(1-T)$ at 133 GeV}  
& \multicolumn{7}{|c|}{$R'\,(1-T)$ at 177 GeV}  
& \multicolumn{7}{|c|}{$R'\,(1-T)$ at 197 GeV}\bigstrut \\ 
\hline \bigstrut[t]
$0.00$ & $0.01$  & 1 & 276 & 0 & 019 & 0 & 040 &  & 4 & 37 & 0 & 82 & 0 & 55 &  & 8 & 76 & 0 & 74 & 0 & 64 &  & 9 & 95 & 0 & 34 & 0 & 27 & \\ 
$0.01$ & $0.02$  & 12 & 25 & 0 & 05 & 0 & 41 &  & 20 & 4 & 1 & 6 & 3 & 8 &  & 22 & 65 & 1 & 11 & 1 & 23 &  & 22 & 44 & 0 & 47 & 0 & 40 & \\ 
$0.02$ & $0.03$  & 18 & 38 & 0 & 06 & 0 & 27 &  & 20 & 4 & 1 & 5 & 1 & 7 &  & 16 & 25 & 0 & 92 & 0 & 76 &  & 14 & 92 & 0 & 37 & 0 & 56 & \\ 
$0.03$ & $0.04$  & 13 & 87 & 0 & 06 & 0 & 14 &  & 10 & 5 & 1 & 2 & 0 & 9 &  & 9 & 84 & 0 & 73 & 0 & 98 &  & 9 & 57 & 0 & 31 & 0 & 41 & \\ 
$0.04$ & $0.05$  & 9 & 83 & 0 & 05 & 0 & 16 &  & 6 & 7 & 1 & 0 & 0 & 4 &  & 6 & 85 & 0 & 65 & 0 & 87 &  & 7 & 23 & 0 & 27 & 0 & 20 & \\ 
$0.05$ & $0.07$  & 6 & 502 & 0 & 026 & 0 & 088 &  & 4 & 70 & 0 & 58 & 0 & 35 &  & 4 & 96 & 0 & 38 & 0 & 20 &  & 5 & 24 & 0 & 16 & 0 & 15 & \\ 
$0.07$ & $0.09$  & 4 & 127 & 0 & 022 & 0 & 041 &  & 3 & 64 & 0 & 51 & 0 & 29 &  & 3 & 18 & 0 & 33 & 0 & 27 &  & 2 & 87 & 0 & 14 & 0 & 15 & \\
\hline
\end{tabular}
}
\end{center}
\caption[]
{Distributions for the thrust $(1-T)$, heavy jet mass (\mh) 
and $C$-parameter  measured by
\Opal\ at centre-of-mass energies $\sqrt{s}=91$, 130--136, 161--183, and
189--209~GeV.
The first uncertainty is statistical, while the second is systematic.}
\label{tabdist1}
\end{minipage}
\end{sideways}
\end{table}

\begin{table}[hb!]
\begin{sideways}
\begin{minipage}[b]{\textheight}
\begin{center}
\scalebox{0.7}{
\begin{tabular}{| r@{ -- }l |
r@{.}l @{ $\pm$ } r@{.}l @{ $\pm$ } r@{.}l @{} l |
r@{.}l @{ $\pm$ } r@{.}l @{ $\pm$ } r@{.}l @{} l |
r@{.}l @{ $\pm$ } r@{.}l @{ $\pm$ } r@{.}l @{} l |
r@{.}l @{ $\pm$ } r@{.}l @{ $\pm$ } r@{.}l @{} l |}
\hline
\multicolumn{2}{|c|}{$B_\mathrm{T}$}  
& \multicolumn{7}{|c|}{$R'\,(B_\mathrm{T})$ at 91 GeV}  
& \multicolumn{7}{|c|}{$R'\,(B_\mathrm{T})$ at 133 GeV}  
& \multicolumn{7}{|c|}{$R'\,(B_\mathrm{T})$ at 177 GeV}  
& \multicolumn{7}{|c|}{$R'\,(B_\mathrm{T})$ at 197 GeV}\bigstrut \\ 
\hline \bigstrut[t]
$0.000$ & $0.030$  & 0 & 117 & 0 & 005 & 0 & 012 &  & 0 & 58 & 0 & 26 & 0 & 16 &  & 1 & 78 & 0 & 23 & 0 & 18 &  & 2 & 40 & 0 & 12 & 0 & 07 & \\ 
$0.030$ & $0.040$  & 2 & 50 & 0 & 03 & 0 & 11 &  & 10 & 2 & 1 & 1 & 1 & 3 &  & 12 & 1 & 0 & 9 & 1 & 2 &  & 11 & 23 & 0 & 41 & 0 & 44 & \\ 
$0.040$ & $0.050$  & 6 & 95 & 0 & 04 & 0 & 13 &  & 9 & 6 & 1 & 3 & 1 & 9 &  & 11 & 5 & 0 & 8 & 1 & 2 &  & 10 & 99 & 0 & 38 & 0 & 49 & \\ 
$0.050$ & $0.060$  & 9 & 806 & 0 & 050 & 0 & 079 &  & 11 & 9 & 1 & 2 & 0 & 9 &  & 11 & 01 & 0 & 75 & 0 & 79 &  & 9 & 67 & 0 & 34 & 0 & 56 & \\ 
$0.060$ & $0.075$  & 10 & 744 & 0 & 040 & 0 & 120 &  & 10 & 06 & 0 & 86 & 0 & 52 &  & 8 & 29 & 0 & 55 & 0 & 43 &  & 7 & 74 & 0 & 25 & 0 & 23 & \\ 
$0.075$ & $0.090$  & 8 & 777 & 0 & 035 & 0 & 211 &  & 6 & 31 & 0 & 76 & 0 & 36 &  & 5 & 71 & 0 & 48 & 0 & 36 &  & 6 & 14 & 0 & 22 & 0 & 35 & \\ 
$0.090$ & $0.110$  & 6 & 597 & 0 & 026 & 0 & 065 &  & 4 & 35 & 0 & 59 & 0 & 82 &  & 4 & 38 & 0 & 37 & 0 & 38 &  & 4 & 85 & 0 & 18 & 0 & 16 & \\
\hline
\end{tabular}
}
\end{center}
\caption[]
{Distributions for the total jet broadening (\bt) wide jet broadening (\bw) and Durham two- 
to three-jet transition parameter (\ytwothree)  measured by
\Opal\ at centre-of-mass energies 
$\sqrt{s}=91$, 130--136, 161--183, and
189--209~GeV.
The first uncertainty is statistical, while the second is systematic.}
\label{tabdist2}
\end{minipage}
\end{sideways}
\end{table}

\begin{table}[hb!]
\begin{sideways}
\begin{minipage}[b]{\textheight}
\begin{center}
\scalebox{0.8}{
\begin{tabular}{| r@{ -- }l |
r@{.}l @{ $\pm$ } r@{.}l @{ $\pm$ } r@{.}l @{} l |
r@{.}l @{ $\pm$ } r@{.}l @{ $\pm$ } r@{.}l @{} l |
r@{.}l @{ $\pm$ } r@{.}l @{ $\pm$ } r@{.}l @{} l |
r@{.}l @{ $\pm$ } r@{.}l @{ $\pm$ } r@{.}l @{} l |}
\hline
\multicolumn{2}{|c|}{$T_\mathrm{maj.}$}  
& \multicolumn{7}{|c|}{$R'\,(T_\mathrm{maj.})$ at 91 GeV}  
& \multicolumn{7}{|c|}{$R'\,(T_\mathrm{maj.})$ at 133 GeV}  
& \multicolumn{7}{|c|}{$R'\,(T_\mathrm{maj.})$ at 177 GeV}  
& \multicolumn{7}{|c|}{$R'\,(T_\mathrm{maj.})$ at 197 GeV}\bigstrut \\ 
\hline \bigstrut[t]
$0.00$ & $0.04$  & 0 & 0363 & 0 & 0028 & 0 & 0073 &  & 0 & 27 & 0 & 15 & 0 & 18 &  & 0 & 94 & 0 & 15 & 0 & 13 &  & 1 & 185 & 0 & 069 & 0 & 025 & \\ 
$0.04$ & $0.08$  & 3 & 373 & 0 & 015 & 0 & 041 &  & 6 & 00 & 0 & 44 & 0 & 60 &  & 6 & 99 & 0 & 29 & 0 & 29 &  & 6 & 70 & 0 & 12 & 0 & 10 & \\ 
$0.08$ & $0.12$  & 6 & 499 & 0 & 018 & 0 & 124 &  & 5 & 93 & 0 & 39 & 0 & 16 &  & 5 & 18 & 0 & 25 & 0 & 15 &  & 4 & 81 & 0 & 10 & 0 & 15 & \\ 
$0.12$ & $0.16$  & 4 & 415 & 0 & 015 & 0 & 065 &  & 3 & 32 & 0 & 33 & 0 & 40 &  & 3 & 27 & 0 & 21 & 0 & 17 &  & 3 & 34 & 0 & 09 & 0 & 10 & \\ 
$0.16$ & $0.22$  & 2 & 753 & 0 & 010 & 0 & 046 &  & 2 & 29 & 0 & 23 & 0 & 18 &  & 2 & 03 & 0 & 15 & 0 & 17 &  & 2 & 124 & 0 & 062 & 0 & 066 & \\ 
$0.22$ & $0.30$  & 1 & 579 & 0 & 007 & 0 & 027 &  & 1 & 12 & 0 & 15 & 0 & 16 &  & 1 & 25 & 0 & 11 & 0 & 08 &  & 1 & 327 & 0 & 044 & 0 & 051 & \\ 
$0.30$ & $0.40$  & 0 & 825 & 0 & 004 & 0 & 023 &  & 0 & 94 & 0 & 10 & 0 & 11 &  & 0 & 74 & 0 & 08 & 0 & 10 &  & 0 & 722 & 0 & 034 & 0 & 047 & \\
\hline
\end{tabular}
}
\end{center}
\caption[]
{Distributions for the thrust major (\tma), thrust minor (\tmi) and 
aplanarity ($A$)  measured by
\Opal\ at centre-of-mass energies 
$\sqrt{s}=91$, 130--136, 161--183, and
189--209~GeV.
The first uncertainty is statistical, while the second is systematic.}
\label{tabdist3}
\end{minipage}
\end{sideways}
\end{table}

\begin{table}[hb!]
\begin{sideways}
\begin{minipage}[b]{\textheight}
\begin{center}
\scalebox{0.8}{
\begin{tabular}{| r@{ -- }l |
r@{.}l @{ $\pm$ } r@{.}l @{ $\pm$ } r@{.}l @{} l |
r@{.}l @{ $\pm$ } r@{.}l @{ $\pm$ } r@{.}l @{} l |
r@{.}l @{ $\pm$ } r@{.}l @{ $\pm$ } r@{.}l @{} l |
r@{.}l @{ $\pm$ } r@{.}l @{ $\pm$ } r@{.}l @{} l |}
\hline
\multicolumn{2}{|c|}{$S$}  
& \multicolumn{7}{|c|}{$R'\,(S)$ at 91 GeV}  
& \multicolumn{7}{|c|}{$R'\,(S)$ at 133 GeV}  
& \multicolumn{7}{|c|}{$R'\,(S)$ at 177 GeV}  
& \multicolumn{7}{|c|}{$R'\,(S)$ at 197 GeV}\bigstrut \\ 
\hline \bigstrut[t]
$0.00$ & $0.02$  & 18 & 06 & 0 & 04 & 0 & 42 &  & 23 & 63 & 0 & 95 & 1 & 81 &  & 25 & 66 & 0 & 70 & 0 & 56 &  & 25 & 67 & 0 & 32 & 0 & 37 & \\ 
$0.02$ & $0.04$  & 10 & 57 & 0 & 03 & 0 & 15 &  & 8 & 15 & 0 & 72 & 0 & 77 &  & 6 & 86 & 0 & 46 & 0 & 29 &  & 7 & 36 & 0 & 20 & 0 & 30 & \\ 
$0.04$ & $0.06$  & 5 & 183 & 0 & 024 & 0 & 074 &  & 4 & 07 & 0 & 53 & 0 & 28 &  & 3 & 83 & 0 & 35 & 0 & 49 &  & 3 & 65 & 0 & 15 & 0 & 18 & \\ 
$0.06$ & $0.12$  & 2 & 409 & 0 & 009 & 0 & 058 &  & 1 & 65 & 0 & 22 & 0 & 13 &  & 2 & 21 & 0 & 15 & 0 & 25 &  & 1 & 99 & 0 & 06 & 0 & 13 & \\ 
$0.12$ & $0.20$  & 0 & 988 & 0 & 005 & 0 & 033 &  & 0 & 93 & 0 & 12 & 0 & 22 &  & 0 & 74 & 0 & 09 & 0 & 12 &  & 0 & 784 & 0 & 041 & 0 & 070 & \\ 
$0.20$ & $0.30$  & 0 & 466 & 0 & 003 & 0 & 014 &  & 0 & 377 & 0 & 068 & 0 & 048 &  & 0 & 338 & 0 & 070 & 0 & 073 &  & 0 & 379 & 0 & 033 & 0 & 044 & \\ 
$0.30$ & $0.50$  & 0 & 2001 & 0 & 0015 & 0 & 0100 &  & 0 & 293 & 0 & 034 & 0 & 033 &  & 0 & 138 & 0 & 040 & 0 & 028 &  & 0 & 175 & 0 & 023 & 0 & 047 & \\
\hline
\end{tabular}
}
\end{center}
\caption[]
{Distributions for the  sphericity ($S$), oblateness ($O$) and light jet mass (\ml) measured by
\Opal\ at centre-of-mass energies 
$\sqrt{s}=91$, 130--136, 161--183, and
189--209~GeV.
The first uncertainty is statistical, while the second is systematic.}
\label{tabdist4}
\end{minipage}
\end{sideways}
\end{table}

\begin{table}[hb!]
\begin{sideways}
\begin{minipage}[b]{\textheight}
\begin{center}
\scalebox{0.8}{
\begin{tabular}{| r@{ -- }l |
r@{.}l @{ $\pm$ } r@{.}l @{ $\pm$ } r@{.}l @{} l |
r@{.}l @{ $\pm$ } r@{.}l @{ $\pm$ } r@{.}l @{} l |
r@{.}l @{ $\pm$ } r@{.}l @{ $\pm$ } r@{.}l @{} l |
r@{.}l @{ $\pm$ } r@{.}l @{ $\pm$ } r@{.}l @{} l |}
\hline
\multicolumn{2}{|c|}{$B_\mathrm{N}$}  
& \multicolumn{7}{|c|}{$R'\,(B_\mathrm{N})$ at 91 GeV}  
& \multicolumn{7}{|c|}{$R'\,(B_\mathrm{N})$ at 133 GeV}  
& \multicolumn{7}{|c|}{$R'\,(B_\mathrm{N})$ at 177 GeV}  
& \multicolumn{7}{|c|}{$R'\,(B_\mathrm{N})$ at 197 GeV}\bigstrut \\ 
\hline \bigstrut[t]
$0.000$ & $0.010$  & 0 & 698 & 0 & 020 & 0 & 081 &  & 1 & 83 & 0 & 71 & 1 & 00 &  & 4 & 86 & 0 & 69 & 0 & 81 &  & 6 & 33 & 0 & 34 & 0 & 29 & \\ 
$0.010$ & $0.015$  & 7 & 96 & 0 & 07 & 0 & 20 &  & 30 & 0 & 2 & 8 & 1 & 6 &  & 38 & 2 & 2 & 2 & 2 & 4 &  & 42 & 4 & 1 & 1 & 1 & 6 & \\ 
$0.015$ & $0.020$  & 21 & 57 & 0 & 11 & 0 & 36 &  & 37 & 7 & 3 & 2 & 4 & 9 &  & 50 & 9 & 2 & 3 & 2 & 3 &  & 45 & 1 & 1 & 1 & 1 & 7 & \\ 
$0.020$ & $0.025$  & 32 & 46 & 0 & 12 & 0 & 47 &  & 34 & 1 & 3 & 2 & 6 & 9 &  & 29 & 8 & 1 & 8 & 3 & 0 &  & 27 & 8 & 0 & 8 & 1 & 6 & \\ 
$0.025$ & $0.030$  & 33 & 65 & 0 & 12 & 0 & 73 &  & 26 & 3 & 2 & 5 & 3 & 7 &  & 19 & 5 & 1 & 4 & 1 & 5 &  & 16 & 6 & 0 & 6 & 0 & 8 & \\ 
$0.030$ & $0.035$  & 28 & 42 & 0 & 11 & 0 & 17 &  & 15 & 6 & 2 & 0 & 1 & 0 &  & 11 & 4 & 1 & 2 & 1 & 3 &  & 13 & 0 & 0 & 5 & 1 & 3 & \\ 
$0.035$ & $0.040$  & 19 & 94 & 0 & 09 & 0 & 31 &  & 10 & 3 & 1 & 7 & 1 & 2 &  & 8 & 1 & 1 & 0 & 1 & 9 &  & 8 & 45 & 0 & 41 & 0 & 66 & \\
\hline
\end{tabular}
}
\end{center}
\caption[]
{Distributions for the narrow jet broadening (\bn) and $D$-parameter  measured by
\Opal\ at centre-of-mass energies 
$\sqrt{s}=91$, 130--136, 161--183, and
189--209~GeV.
The first uncertainty is statistical, while the second is systematic.}
\label{tabdist5}
\end{minipage}
\end{sideways}
\end{table}

\begin{table}[hb!]
\begin{sideways}
\begin{minipage}[b]{\textheight}
\begin{center}
\scalebox{0.75}{
\begin{tabular}{|c| r @{(} r @{.} l @{ $\pm$ } r @{.} l @{ $\pm$ } r @{.} l @{)} l @{} l  | 
                    r @{(} r @{.} l @{ $\pm$ } r @{.} l @{ $\pm$ } r @{.} l @{)} l @{} l  | 
                    r @{(} r @{.} l @{ $\pm$ } r @{.} l @{ $\pm$ } r @{.} l @{)} l @{} l  | 
                    r @{(} r @{.} l @{ $\pm$ } r @{.} l @{ $\pm$ } r @{.} l @{)} l @{} l  |}
 \hline \multicolumn{1}{|c|}{$n$} & \multicolumn{9}{|c|}{$\langle {(1-T)}^n \rangle$ at 91 GeV} & \multicolumn{9}{|c|}{$\langle {(1-T)}^n \rangle$ at 133 GeV} & \multicolumn{9}{|c|}{$\langle {(1-T)}^n \rangle$ at 177 GeV} & \multicolumn{9}{|c|}{$\langle {(1-T)}^n \rangle$ at 197 GeV} \bigstrut \\
 \hline \bigstrut[t]
 $1$  & & 6&671 & 0&017 & 0&066 & $\cdot 10^{-2}$ &  & & 6&20 & 0&26 & 0&32 & $\cdot 10^{-2}$ &  & & 5&60 & 0&17 & 0&22 & $\cdot 10^{-2}$ &  & & 5&60 & 0&08 & 0&15 & $\cdot 10^{-2}$ &  \\
 $2$  & & 8&12 & 0&04 & 0&13 & $\cdot 10^{-3}$ &  & & 7&84 & 0&64 & 0&53 & $\cdot 10^{-3}$ &  & & 6&78 & 0&48 & 0&72 & $\cdot 10^{-3}$ &  & & 6&76 & 0&22 & 0&41 & $\cdot 10^{-3}$ &  \\
 $3$  & & 1&457 & 0&012 & 0&025 & $\cdot 10^{-3}$ &  & & 1&43 & 0&17 & 0&11 & $\cdot 10^{-3}$ &  & & 1&26 & 0&15 & 0&24 & $\cdot 10^{-3}$ &  & & 1&22 & 0&07 & 0&12 & $\cdot 10^{-3}$ &  \\
 $4$  & & 3&236 & 0&037 & 0&053 & $\cdot 10^{-4}$ &  & & 3&13 & 0&48 & 0&33 & $\cdot 10^{-4}$ &  & & 2&96 & 0&47 & 0&87 & $\cdot 10^{-4}$ &  & & 2&71 & 0&25 & 0&41 & $\cdot 10^{-4}$ &  \\
 $5$  & & 8&13 & 0&12 & 0&13 & $\cdot 10^{-5}$ &  & & 7&6 & 1&5 & 1&1 & $\cdot 10^{-5}$ &  & & 8&0 & 1&5 & 3&3 & $\cdot 10^{-5}$ &  & & 6&7 & 0&9 & 1&6 & $\cdot 10^{-5}$ &  \bigstrut[b] \\ \hline
 \hline \multicolumn{1}{|c|}{$n$} & \multicolumn{9}{|c|}{\momn{\mh}{n} at 91 GeV} & \multicolumn{9}{|c|}{\momn{\mh}{n} at 133 GeV} & \multicolumn{9}{|c|}{\momn{\mh}{n} at 177 GeV} & \multicolumn{9}{|c|}{\momn{\mh}{n} at 197 GeV} \bigstrut \\
\hline
\end{tabular}
}
\end{center}
\caption[]
{Moments of the $(1-T)$, \mh, $C$, \bt, \bw\ and \ytwothree\ 
distributions measured by
\Opal\ at 91, 130--136, 161--183, and 189--209~GeV.
The first uncertainty is statistical, while the second is systematic.}
\label{tabmom1}
\end{minipage}
\end{sideways}
\end{table}

\begin{table}[hb!]
\begin{sideways}
\begin{minipage}[b]{\textheight}
\begin{center}
\scalebox{0.75}{
\begin{tabular}{|c| r @{(} r @{.} l @{ $\pm$ } r @{.} l @{ $\pm$ } r @{.} l @{)} l @{} l  | 
                    r @{(} r @{.} l @{ $\pm$ } r @{.} l @{ $\pm$ } r @{.} l @{)} l @{} l  | 
                    r @{(} r @{.} l @{ $\pm$ } r @{.} l @{ $\pm$ } r @{.} l @{)} l @{} l  | 
                    r @{(} r @{.} l @{ $\pm$ } r @{.} l @{ $\pm$ } r @{.} l @{)} l @{} l  |}
 \hline \multicolumn{1}{|c|}{$n$} & \multicolumn{9}{|c|}{\momn{\tma}{n} at 91 GeV} & \multicolumn{9}{|c|}{\momn{\tma}{n} at 133 GeV} & \multicolumn{9}{|c|}{\momn{\tma}{n} at 177 GeV} & \multicolumn{9}{|c|}{\momn{\tma}{n} at 197 GeV} \bigstrut \\
 \hline \bigstrut[t]
 $1$  & & 1&757 & 0&003 & 0&013 & $\cdot 10^{-1}$ &  & & 1&671 & 0&048 & 0&059 & $\cdot 10^{-1}$ &  & & 1&546 & 0&031 & 0&042 & $\cdot 10^{-1}$ &  & & 1&550 & 0&014 & 0&031 & $\cdot 10^{-1}$ &  \\
 $2$  & & 4&244 & 0&015 & 0&059 & $\cdot 10^{-2}$ &  & & 4&20 & 0&24 & 0&24 & $\cdot 10^{-2}$ &  & & 3&65 & 0&16 & 0&22 & $\cdot 10^{-2}$ &  & & 3&71 & 0&08 & 0&17 & $\cdot 10^{-2}$ &  \\
 $3$  & & 1&321 & 0&007 & 0&024 & $\cdot 10^{-2}$ &  & & 1&39 & 0&12 & 0&10 & $\cdot 10^{-2}$ &  & & 1&15 & 0&08 & 0&12 & $\cdot 10^{-2}$ &  & & 1&190 & 0&042 & 0&082 & $\cdot 10^{-2}$ &  \\
 $4$  & & 4&88 & 0&04 & 0&10 & $\cdot 10^{-3}$ &  & & 5&43 & 0&59 & 0&42 & $\cdot 10^{-3}$ &  & & 4&33 & 0&45 & 0&63 & $\cdot 10^{-3}$ &  & & 4&55 & 0&24 & 0&42 & $\cdot 10^{-3}$ &  \\
 $5$  & & 2&019 & 0&019 & 0&044 & $\cdot 10^{-3}$ &  & & 2&33 & 0&30 & 0&20 & $\cdot 10^{-3}$ &  & & 1&81 & 0&25 & 0&35 & $\cdot 10^{-3}$ &  & & 1&94 & 0&14 & 0&23 & $\cdot 10^{-3}$ &  \bigstrut[b] \\ \hline
 \hline \multicolumn{1}{|c|}{$n$} & \multicolumn{9}{|c|}{\momn{\tmi}{n} at 91 GeV} & \multicolumn{9}{|c|}{\momn{\tmi}{n} at 133 GeV} & \multicolumn{9}{|c|}{\momn{\tmi}{n} at 177 GeV} & \multicolumn{9}{|c|}{\momn{\tmi}{n} at 197 GeV} \bigstrut \\
\hline
\end{tabular}
}
\end{center}
\caption[]
{Moments of the \tma, \tmi, $S$, $O$, \ml\ and \bn\ distributions measured by
\Opal\ at 91, 130--136, 161--183,  and 189--209~GeV.
The first uncertainty is statistical, while the second is systematic.}
\label{tabmom2}
\end{minipage}
\end{sideways}
\end{table}

\begin{table}[hb!]
\begin{center}
\small
\resizebox{0.9\textwidth}{!}{
\begin{tabular}{| r | r@{}l r@{}l r@{}l r@{}l r@{}l r@{}l | r@{}l |}
\hline
& \multicolumn{2}{c}{$T$} 
& \multicolumn{2}{c}{\mh} 
& \multicolumn{2}{c}{\bt} 
& \multicolumn{2}{c}{\bw} 
& \multicolumn{2}{c}{$C$} 
& \multicolumn{2}{c|}{\ytwothree}
& \multicolumn{2}{c|}{\parbox{1.6cm}
{\centering \rule{0cm}{0.45cm}Weighted\vspace{-0.15cm}
\\mean\vspace{-0.3cm}\\$\phantom{A}$}}
\bigstrut \\
\hline
 \hline
 {\bf\boldmath $\alpha_{\mathrm{s}}$(91  GeV)}
  & $    0.1231$
  &
  & $    0.1192$
  &
  & $    0.1224$
  &
  & $    0.1146$
  &
  & $    0.1178$
  &
  & $    0.1197$
  &
  & $    0.1192$
  &
 \bigstrut \\ \hline
 Statistical error
  & $\pm 0.0002$
  &
  & $\pm 0.0002$
  &
  & $\pm 0.0002$
  &
  & $\pm 0.0002$
  &
  & $\pm 0.0002$
  &
  & $\pm 0.0002$
  &
  & $\pm 0.0002$
  &
 \bigstrut \\ \hline
 Experimental syst.
  & $\pm 0.0011$
  &
  & $\pm 0.0004$
  &
  & $\pm 0.0007$
  &
  & $\pm 0.0011$
  &
  & $\pm 0.0007$
  &
  & $\pm 0.0010$
  &
  & $\pm 0.0008$
  &
 \bigstrut \\ \hline
 HERWIG hadr.~corr.
  & $   -0.0018$
  &
  & $+   0.0013$
  &
  & $   -0.0023$
  &
  & $   -0.0014$
  & $^*$
  & $   -0.0031$
  & $^*$
  & $   -0.0024$
  & $^*$
  & $   -0.0015$
  & $^*$
 \bigstrut[t] \\ ARIADNE hadr.~corr.
  & $+   0.0031$
  & $^*$
  & $+   0.0021$
  & $^*$
  & $+   0.0024$
  & $^*$
  & $+   0.0007$
  &
  & $+   0.0029$
  &
  & $   -0.0008$
  &
  & $+   0.0013$
  &
 \bigstrut[b] \\
\end{tabular}}
\end{center}
 \caption[ ]{Measurements of \as\ using event shape distributions
in four ranges of c.m.\ energy:
at 91~GeV, 133~GeV, 161--183~GeV (denoted 177~GeV) and
189--209~GeV (denoted 197~GeV).  The hadronization error is taken to 
be the larger of the effects observed using \Herwig\ and \Ariadne;
in each case this is denoted by an asterisk.
The weights and weighted mean are described in the text.
}
 \label{tab:alphasfits1}
 \end{table}

\clearpage

\begin{table}[hb!]
\begin{sideways}
\begin{minipage}[b]{\textheight}
\begin{tabular}{cc}
\small
\resizebox{0.5\textwidth}{!}{
\begin{tabular}{| r | r@{}l r@{}l r@{}l r@{}l r@{}l r@{}l | r@{}l |}
\hline
& \multicolumn{2}{c}{$T$} 
& \multicolumn{2}{c}{\mh} 
& \multicolumn{2}{c}{\bt} 
& \multicolumn{2}{c}{\bw} 
& \multicolumn{2}{c}{$C$} 
& \multicolumn{2}{c|}{\ytwothree}
& \multicolumn{2}{c|}{\parbox{1.6cm}
{\centering \rule{0cm}{0.45cm}Weighted\vspace{-0.15cm}
\\mean\vspace{-0.3cm}\\$\phantom{A}$}}
\bigstrut \\
\hline
 \hline
 {\bf\boldmath $\alpha_{\mathrm{s}}$(161 GeV)}
  & $    0.1103$
  &
  & $    0.1064$
  &
  & $    0.1051$
  &
  & $    0.1010$
  &
  & $    0.1042$
  &
  & $    0.1042$
  &
  & $    0.1046$
  &
 \bigstrut \\ \hline
 Statistical error
  & $\pm 0.0069$
  &
  & $\pm 0.0063$
  &
  & $\pm 0.0062$
  &
  & $\pm 0.0053$
  &
  & $\pm 0.0068$
  &
  & $\pm 0.0051$
  &
  & $\pm 0.0051$
  &
 \bigstrut \\ \hline
 Experimental syst.
  & $\pm 0.0042$
  &
  & $\pm 0.0042$
  &
  & $\pm 0.0045$
  &
  & $\pm 0.0035$
  &
  & $\pm 0.0056$
  &
  & $\pm 0.0033$
  &
  & $\pm 0.0034$
  &
 \bigstrut \\ \hline
 HERWIG hadr.~corr.
  & $   -0.0002$
  &
  & $+   0.0021$
  & $^*$
  & $   -0.0013$
  &
  & $   -0.0002$
  &
  & $   -0.0009$
  &
  & $   -0.0009$
  & $^*$
  & $   -0.0002$
  &
 \bigstrut[t] \\ ARIADNE hadr.~corr.
  & $+   0.0022$
  & $^*$
  & $+   0.0015$
  &
  & $+   0.0018$
  & $^*$
  & $+   0.0007$
  & $^*$
  & $+   0.0022$
  & $^*$
  & $   -0.0005$
  &
  & $+   0.0009$
  & $^*$
 \bigstrut[b] \\ \hline
 Hadronization error
  & $\pm 0.0022$
  &
  & $\pm 0.0021$
  &
  & $\pm 0.0018$
  &
  & $\pm 0.0007$
  &
  & $\pm 0.0022$
  &
  & $\pm 0.0009$
  &
  & $\pm 0.0009$
  &
 \bigstrut[b] \\ \hline
 Theory error
  & $\pm 0.0043$
  &
  & $\pm 0.0034$
  &
  & $\pm 0.0052$
  &
  & $\pm 0.0043$
  &
  & $\pm 0.0044$
  &
  & $\pm 0.0025$
  &
  & $\pm 0.0036$
  &
 \bigstrut \\ \hline
 Weight
  & \multicolumn{2}{c}
 {\centering $     0.12 $}
  & \multicolumn{2}{c}
 {\centering $     0.14 $}
  & \multicolumn{2}{c}
 {\centering $     0.06 $}
  & \multicolumn{2}{c}
 {\centering $     0.21 $}
  & \multicolumn{2}{c}
 {\centering $     0.09 $}
  & \multicolumn{2}{c|}
 {\centering $     0.39 $}
  & \multicolumn{2}{c|}
 {\centering $-         $}
 \bigstrut \\ \hline


%
\hline
 \hline
 {\bf\boldmath $\alpha_{\mathrm{s}}$(172 GeV)}
  & $    0.1095$
  &
  & $    0.1043$
  &
  & $    0.1016$
  &
  & $    0.0950$
  &
  & $    0.1039$
  &
  & $    0.1038$
  &
  & $    0.1011$
  &
 \bigstrut \\ \hline
 Statistical error
  & $\pm 0.0077$
  &
  & $\pm 0.0069$
  &
  & $\pm 0.0070$
  &
  & $\pm 0.0060$
  &
  & $\pm 0.0075$
  &
  & $\pm 0.0056$
  &
  & $\pm 0.0055$
  &
 \bigstrut \\ \hline
 Experimental syst.
  & $\pm 0.0076$
  &
  & $\pm 0.0078$
  &
  & $\pm 0.0037$
  &
  & $\pm 0.0029$
  &
  & $\pm 0.0064$
  &
  & $\pm 0.0046$
  &
  & $\pm 0.0039$
  &
 \bigstrut \\ \hline
 HERWIG hadr.~corr.
  & $   -0.0002$
  &
  & $+   0.0022$
  & $^*$
  & $   -0.0012$
  &
  & $   -0.0002$
  &
  & $   -0.0007$
  &
  & $   -0.0007$
  & $^*$
  & $   -0.0003$
  &
 \bigstrut[t] \\ ARIADNE hadr.~corr.
  & $+   0.0023$
  & $^*$
  & $+   0.0017$
  &
  & $+   0.0018$
  & $^*$
  & $+   0.0008$
  & $^*$
  & $+   0.0022$
  & $^*$
  & $   -0.0003$
  &
  & $+   0.0008$
  & $^*$
 \bigstrut[b] \\ \hline
 Hadronization error
  & $\pm 0.0023$
  &
  & $\pm 0.0022$
  &
  & $\pm 0.0018$
  &
  & $\pm 0.0008$
  &
  & $\pm 0.0022$
  &
  & $\pm 0.0007$
  &
  & $\pm 0.0008$
  &
 \bigstrut[b] \\ \hline
 Theory error
  & $\pm 0.0042$
  &
  & $\pm 0.0033$
  &
  & $\pm 0.0050$
  &
  & $\pm 0.0041$
  &
  & $\pm 0.0042$
  &
  & $\pm 0.0024$
  &
  & $\pm 0.0035$
  &
 \bigstrut \\ \hline
 Weight
  & \multicolumn{2}{c}
 {\centering $     0.06 $}
  & \multicolumn{2}{c}
 {\centering $     0.07 $}
  & \multicolumn{2}{c}
 {\centering $     0.11 $}
  & \multicolumn{2}{c}
 {\centering $     0.32 $}
  & \multicolumn{2}{c}
 {\centering $     0.06 $}
  & \multicolumn{2}{c|}
 {\centering $     0.39 $}
  & \multicolumn{2}{c|}
 {\centering $-         $}
 \bigstrut \\ \hline


%
\hline
 \hline
 {\bf\boldmath $\alpha_{\mathrm{s}}$(183 GeV)}
  & $    0.1111$
  &
  & $    0.1076$
  &
  & $    0.1117$
  &
  & $    0.1032$
  &
  & $    0.1075$
  &
  & $    0.1084$
  &
  & $    0.1079$
  &
 \bigstrut \\ \hline
 Statistical error
  & $\pm 0.0036$
  &
  & $\pm 0.0032$
  &
  & $\pm 0.0032$
  &
  & $\pm 0.0027$
  &
  & $\pm 0.0035$
  &
  & $\pm 0.0027$
  &
  & $\pm 0.0027$
  &
 \bigstrut \\ \hline
 Experimental syst.
  & $\pm 0.0027$
  &
  & $\pm 0.0041$
  &
  & $\pm 0.0033$
  &
  & $\pm 0.0029$
  &
  & $\pm 0.0028$
  &
  & $\pm 0.0034$
  &
  & $\pm 0.0031$
  &
 \bigstrut \\ \hline
 HERWIG hadr.~corr.
  & $   -0.0002$
  &
  & $+   0.0020$
  & $^*$
  & $   -0.0009$
  &
  & $   -0.0002$
  &
  & $   -0.0005$
  &
  & $   -0.0005$
  & $^*$
  & $   -0.0001$
  &
 \bigstrut[t] \\ ARIADNE hadr.~corr.
  & $+   0.0020$
  & $^*$
  & $+   0.0014$
  &
  & $+   0.0014$
  & $^*$
  & $+   0.0006$
  & $^*$
  & $+   0.0020$
  & $^*$
  & $   -0.0003$
  &
  & $+   0.0010$
  & $^*$
 \bigstrut[b] \\ \hline
 Hadronization error
  & $\pm 0.0020$
  &
  & $\pm 0.0020$
  &
  & $\pm 0.0014$
  &
  & $\pm 0.0006$
  &
  & $\pm 0.0020$
  &
  & $\pm 0.0005$
  &
  & $\pm 0.0010$
  &
 \bigstrut[b] \\ \hline
 Theory error
  & $\pm 0.0041$
  &
  & $\pm 0.0033$
  &
  & $\pm 0.0049$
  &
  & $\pm 0.0041$
  &
  & $\pm 0.0042$
  &
  & $\pm 0.0024$
  &
  & $\pm 0.0036$
  &
 \bigstrut \\ \hline
 Weight
  & \multicolumn{2}{c}
 {\centering $     0.15 $}
  & \multicolumn{2}{c}
 {\centering $     0.14 $}
  & \multicolumn{2}{c}
 {\centering $     0.10 $}
  & \multicolumn{2}{c}
 {\centering $     0.20 $}
  & \multicolumn{2}{c}
 {\centering $     0.12 $}
  & \multicolumn{2}{c|}
 {\centering $     0.29 $}
  & \multicolumn{2}{c|}
 {\centering $-         $}
 \bigstrut \\ \hline


%
\end{tabular}}
&
\small
\resizebox{0.5\textwidth}{!}{
\begin{tabular}{| r | r@{}l r@{}l r@{}l r@{}l r@{}l r@{}l | r@{}l |}
\hline
& \multicolumn{2}{c}{$T$} 
& \multicolumn{2}{c}{\mh} 
& \multicolumn{2}{c}{\bt} 
& \multicolumn{2}{c}{\bw} 
& \multicolumn{2}{c}{$C$} 
& \multicolumn{2}{c|}{\ytwothree}
& \multicolumn{2}{c|}{\parbox{1.6cm}
{\centering \rule{0cm}{0.45cm}Weighted\vspace{-0.15cm}
\\mean\vspace{-0.3cm}\\$\phantom{A}$}}
\bigstrut \\
\hline
 \hline
 {\bf\boldmath $\alpha_{\mathrm{s}}$(189 GeV)}
  & $    0.1146$
  &
  & $    0.1070$
  &
  & $    0.1121$
  &
  & $    0.1026$
  &
  & $    0.1079$
  &
  & $    0.1066$
  &
  & $    0.1075$
  &
 \bigstrut \\ \hline
 Statistical error
  & $\pm 0.0020$
  &
  & $\pm 0.0018$
  &
  & $\pm 0.0019$
  &
  & $\pm 0.0015$
  &
  & $\pm 0.0019$
  &
  & $\pm 0.0015$
  &
  & $\pm 0.0016$
  &
 \bigstrut \\ \hline
 Experimental syst.
  & $\pm 0.0019$
  &
  & $\pm 0.0022$
  &
  & $\pm 0.0014$
  &
  & $\pm 0.0011$
  &
  & $\pm 0.0015$
  &
  & $\pm 0.0009$
  &
  & $\pm 0.0012$
  &
 \bigstrut \\ \hline
 HERWIG hadr.~corr.
  & $   -0.0001$
  &
  & $+   0.0022$
  & $^*$
  & $   -0.0009$
  &
  & $   -0.0001$
  &
  & $   -0.0005$
  &
  & $   -0.0004$
  & $^*$
  & $    0.0000$
  &
 \bigstrut[t] \\ ARIADNE hadr.~corr.
  & $+   0.0019$
  & $^*$
  & $+   0.0014$
  &
  & $+   0.0014$
  & $^*$
  & $+   0.0006$
  & $^*$
  & $+   0.0019$
  & $^*$
  & $   -0.0003$
  &
  & $+   0.0007$
  & $^*$
 \bigstrut[b] \\ \hline
 Hadronization error
  & $\pm 0.0019$
  &
  & $\pm 0.0022$
  &
  & $\pm 0.0014$
  &
  & $\pm 0.0006$
  &
  & $\pm 0.0019$
  &
  & $\pm 0.0004$
  &
  & $\pm 0.0007$
  &
 \bigstrut[b] \\ \hline
 Theory error
  & $\pm 0.0041$
  &
  & $\pm 0.0032$
  &
  & $\pm 0.0049$
  &
  & $\pm 0.0040$
  &
  & $\pm 0.0042$
  &
  & $\pm 0.0023$
  &
  & $\pm 0.0033$
  &
 \bigstrut \\ \hline
 Weight
  & \multicolumn{2}{c}
 {\centering $     0.10 $}
  & \multicolumn{2}{c}
 {\centering $     0.13 $}
  & \multicolumn{2}{c}
 {\centering $     0.08 $}
  & \multicolumn{2}{c}
 {\centering $     0.15 $}
  & \multicolumn{2}{c}
 {\centering $     0.11 $}
  & \multicolumn{2}{c|}
 {\centering $     0.42 $}
  & \multicolumn{2}{c|}
 {\centering $-         $}
 \bigstrut \\ \hline


%
\hline
 \hline
 {\bf\boldmath $\alpha_{\mathrm{s}}$(200 GeV)}
  & $    0.1142$
  &
  & $    0.1048$
  &
  & $    0.1128$
  &
  & $    0.1030$
  &
  & $    0.1082$
  &
  & $    0.1064$
  &
  & $    0.1068$
  &
 \bigstrut \\ \hline
 Statistical error
  & $\pm 0.0023$
  &
  & $\pm 0.0021$
  &
  & $\pm 0.0022$
  &
  & $\pm 0.0018$
  &
  & $\pm 0.0023$
  &
  & $\pm 0.0017$
  &
  & $\pm 0.0019$
  &
 \bigstrut \\ \hline
 Experimental syst.
  & $\pm 0.0024$
  &
  & $\pm 0.0021$
  &
  & $\pm 0.0034$
  &
  & $\pm 0.0013$
  &
  & $\pm 0.0028$
  &
  & $\pm 0.0012$
  &
  & $\pm 0.0010$
  &
 \bigstrut \\ \hline
 HERWIG hadr.~corr.
  & $    0.0000$
  &
  & $+   0.0021$
  & $^*$
  & $   -0.0008$
  &
  & $   -0.0001$
  &
  & $   -0.0004$
  &
  & $   -0.0003$
  & $^*$
  & $+   0.0001$
  &
 \bigstrut[t] \\ ARIADNE hadr.~corr.
  & $+   0.0018$
  & $^*$
  & $+   0.0013$
  &
  & $+   0.0013$
  & $^*$
  & $+   0.0006$
  & $^*$
  & $+   0.0018$
  & $^*$
  & $   -0.0003$
  &
  & $+   0.0006$
  & $^*$
 \bigstrut[b] \\ \hline
 Hadronization error
  & $\pm 0.0018$
  &
  & $\pm 0.0021$
  &
  & $\pm 0.0013$
  &
  & $\pm 0.0006$
  &
  & $\pm 0.0018$
  &
  & $\pm 0.0003$
  &
  & $\pm 0.0006$
  &
 \bigstrut[b] \\ \hline
 Theory error
  & $\pm 0.0040$
  &
  & $\pm 0.0032$
  &
  & $\pm 0.0048$
  &
  & $\pm 0.0040$
  &
  & $\pm 0.0041$
  &
  & $\pm 0.0023$
  &
  & $\pm 0.0032$
  &
 \bigstrut \\ \hline
 Weight
  & \multicolumn{2}{c}
 {\centering $     0.09 $}
  & \multicolumn{2}{c}
 {\centering $     0.15 $}
  & \multicolumn{2}{c}
 {\centering $     0.06 $}
  & \multicolumn{2}{c}
 {\centering $     0.16 $}
  & \multicolumn{2}{c}
 {\centering $     0.08 $}
  & \multicolumn{2}{c|}
 {\centering $     0.45 $}
  & \multicolumn{2}{c|}
 {\centering $-         $}
 \bigstrut \\ \hline


%
\hline
 \hline
 {\bf\boldmath $\alpha_{\mathrm{s}}$(206 GeV)}
  & $    0.1098$
  &
  & $    0.1067$
  &
  & $    0.1125$
  &
  & $    0.1036$
  &
  & $    0.1070$
  &
  & $    0.1086$
  &
  & $    0.1078$
  &
 \bigstrut \\ \hline
 Statistical error
  & $\pm 0.0023$
  &
  & $\pm 0.0020$
  &
  & $\pm 0.0021$
  &
  & $\pm 0.0017$
  &
  & $\pm 0.0022$
  &
  & $\pm 0.0017$
  &
  & $\pm 0.0017$
  &
 \bigstrut \\ \hline
 Experimental syst.
  & $\pm 0.0008$
  &
  & $\pm 0.0019$
  &
  & $\pm 0.0017$
  &
  & $\pm 0.0009$
  &
  & $\pm 0.0019$
  &
  & $\pm 0.0015$
  &
  & $\pm 0.0013$
  &
 \bigstrut \\ \hline
 HERWIG hadr.~corr.
  & $    0.0000$
  &
  & $+   0.0021$
  & $^*$
  & $   -0.0007$
  &
  & $   -0.0001$
  &
  & $   -0.0003$
  &
  & $   -0.0002$
  &
  & $+   0.0001$
  &
 \bigstrut[t] \\ ARIADNE hadr.~corr.
  & $+   0.0019$
  & $^*$
  & $+   0.0013$
  &
  & $+   0.0012$
  & $^*$
  & $+   0.0006$
  & $^*$
  & $+   0.0018$
  & $^*$
  & $   -0.0002$
  & $^*$
  & $+   0.0007$
  & $^*$
 \bigstrut[b] \\ \hline
 Hadronization error
  & $\pm 0.0019$
  &
  & $\pm 0.0021$
  &
  & $\pm 0.0012$
  &
  & $\pm 0.0006$
  &
  & $\pm 0.0018$
  &
  & $\pm 0.0002$
  &
  & $\pm 0.0007$
  & 
 \bigstrut[b] \\ \hline
 Theory error
  & $\pm 0.0040$
  &
  & $\pm 0.0031$
  &
  & $\pm 0.0048$
  &
  & $\pm 0.0039$
  &
  & $\pm 0.0041$
  &
  & $\pm 0.0023$
  &
  & $\pm 0.0033$
  &
 \bigstrut \\ \hline
 Weight
  & \multicolumn{2}{c}
 {\centering $     0.11 $}
  & \multicolumn{2}{c}
 {\centering $     0.13 $}
  & \multicolumn{2}{c}
 {\centering $     0.09 $}
  & \multicolumn{2}{c}
 {\centering $     0.18 $}
  & \multicolumn{2}{c}
 {\centering $     0.09 $}
  & \multicolumn{2}{c|}
 {\centering $     0.41 $}
  & \multicolumn{2}{c|}
 {\centering $-         $}
 \bigstrut \\ \hline


%
\end{tabular}}

\end{tabular}
 \caption[ ]{Measurements of \as\ using event shape distributions
at various values
or ranges of c.m.\ energy.  The values labelled 189~GeV
correspond to the data samples at 189 and 192~GeV, 
the values labelled 200~GeV combine the data at 196, 200 and 202~GeV,
and those labelled 206~GeV include all data above 202~GeV.
The hadronization error is taken to 
be the larger of the effects observed using \Herwig\ and \Ariadne;
in each case this is denoted by an asterisk.
The weights and weighted mean are described in the text.
}
 \label{tab:alphasfits2}
\end{minipage}
\end{sideways}
 \end{table}

\begin{table}[p]
\begin{center}
{\small
\begin{tabular}{|r||r r r r r r|| r|}
\hline
 & \multicolumn{1}{|c}{\boldmath $T$}
 & \multicolumn{1}{c}{\boldmath $M_\mathrm{H}$}
 & \multicolumn{1}{c}{\boldmath $C$}
 & \multicolumn{1}{c}{\boldmath $B_\mathrm{T}$}
 & \multicolumn{1}{c}{\boldmath $B_\mathrm{W}$}
 & \multicolumn{1}{c||}{\boldmath $\ytwothree$}
 & \\
 & & & & & & & \vspace{-0.55cm} \\
 & \multicolumn{1}{|c}{\textbf{only}}
 & \multicolumn{1}{c}{\textbf{only}}
 & \multicolumn{1}{c}{\textbf{only}}
 & \multicolumn{1}{c}{\textbf{only}}
 & \multicolumn{1}{c}{\textbf{only}}
 & \multicolumn{1}{c||}{\textbf{only}}
 & \multicolumn{1}{c|}{\textbf{All}} 

\bigstrut[b]
\\
\hline
\bigstrut[t]

\bigstrut[b] $\asmz$
 & $0.1242$
 & $0.1181$
 & $0.1177$
 & $0.1222$
 & $0.1134$
 & $0.1193$
 & $0.1191$
 \\ 
\hline \bigstrut Stat. error
 & $\pm 0.0011$
 & $\pm 0.0009$
 & $\pm 0.0011$
 & $\pm 0.0011$
 & $\pm 0.0009$
 & $\pm 0.0007$
 & $\pm 0.0005$
 \\ 
\hline \vspace{-0.15cm}{\scriptsize Expt. error}
 & {\scriptsize $\pm 0.0018$}
 & {\scriptsize $\pm 0.0014$}
 & {\scriptsize $\pm 0.0015$}
 & {\scriptsize $\pm 0.0017$}
 & {\scriptsize $\pm 0.0015$}
 & {\scriptsize $\pm 0.0014$}
 & {\scriptsize $\pm 0.0010$}
 \\ 
\vspace{-0.15cm}{\scriptsize Hadr. error}
 & {\scriptsize $\pm 0.0027$}
 & {\scriptsize $\pm 0.0020$}
 & {\scriptsize $\pm 0.0026$}
 & {\scriptsize $\pm 0.0020$}
 & {\scriptsize $\pm 0.0007$}
 & {\scriptsize $\pm 0.0011$}
 & {\scriptsize $\pm 0.0011$}
 \\ 
{\scriptsize Theory }
 & {\scriptsize $\pm0.0053$}
 & {\scriptsize $\pm0.0042$}
 & {\scriptsize $\pm0.0053$}
 & {\scriptsize $\pm0.0062$}
 & {\scriptsize $\pm0.0051$}
 & {\scriptsize $\pm0.0031$}
 & {\scriptsize $\pm0.0044$}
 \\ 
\bigstrut[b] Syst. error
 & $\pm 0.0061$
 & $\pm 0.0048$
 & $\pm 0.0061$
 & $\pm 0.0068$
 & $\pm 0.0053$
 & $\pm 0.0035$
 & $\pm 0.0046$
 \\ 
\hline \bigstrut Total error
 & $\pm 0.0062$
 & $\pm 0.0049$
 & $\pm 0.0062$
 & $\pm 0.0069$
 & $\pm 0.0054$
 & $\pm 0.0036$
 & $\pm 0.0047$
 \\ 
%
\hline
\end{tabular}
}
\end{center}
\caption{Combined \asmz\ fit results based on distributions of 
different observables, averaged
 over all c.m.\ energies, using \Opal\ data. }
\label{tab:fitsbyobservable_opal}
\end{table}

\clearpage
\begin{table}[hb!]
\begin{center}
\small
\resizebox{0.7\textwidth}{!}{
\begin{tabular}{| r | r@{}l r@{}l r@{}l r@{}l r@{}l r@{}l | r@{}l |}
\hline
& \multicolumn{2}{c}{$\langle (1-T)^1 \rangle$} 
& \multicolumn{2}{c}{$\langle C^1 \rangle$} 
& \multicolumn{2}{c}{\momn{\bt}{1}} 
& \multicolumn{2}{c}{\momn{\bw}{1}} 
& \multicolumn{2}{c}{$\langle \ytwothree{}^1 \rangle$} 
& \multicolumn{2}{c|}{}
\bigstrut \\
\hline
 \hline
 {\bf\boldmath \asmz)}
  & $0.1267$  & &  $0.1242$  & &  $0.1172$  & &  $0.1214$  & &  $0.1223$  & &           &
 \bigstrut \\ \hline
 Statistical error
  & $\pm 0.0003$  & &  $\pm 0.0002$  & &  $\pm 0.0002$  & &  $\pm 0.0003$  & &  $\pm 0.0006$  & &           &
 \bigstrut \\ \hline
 Experimental syst.
  & $\pm 0.0010$  & &  $\pm 0.0008$  & &  $\pm 0.0006$  & &  $\pm 0.0013$  & &  $\pm 0.0027$  & &           &
 \bigstrut \\ \hline
 HERWIG hadr.~corr.
  & $-0.0017$  & &  $-0.0022$  & &  $-0.0023$  & &  $-0.0022$  &$^*$ &  $+0.0005$  &$^*$ &   &
 \bigstrut[t] \\ ARIADNE hadr.~corr.
  & $+0.0038$  &$^*$ &  $+0.0035$  &$^*$ &  $+0.0029$  &$^*$ &  $+0.0010$  & &  $+0.0001$  & &  &
 \bigstrut[b] \\ \hline
 Hadronization error
  &  $\pm 0.0038$  & &  $\pm 0.0035$  & &  $\pm 0.0029$  & &  $\pm 0.0022$  & &  $\pm 0.0005$  & &    &
 \bigstrut \\ \hline
 $x_\mu$ variation
  & $^{+0.0072}_{-0.0058}$ & & $^{+0.0066}_{-0.0053}$ & & $^{+0.0050}_{-0.0038}$ & & $^{+0.0050}_{-0.0004}$ & & $^{+0.0052}_{-0.0039}$ &  &    &
 \bigstrut \\ \hline
 $\chi^2 / dof.$
 &  $4.2/3$  & &  $6.2/3$  & &  $11.6/3$  & &  $2.2/3$  & &  $1.0/3$  & &    &
 \bigstrut \\ \hline
\hline
\hline
& \multicolumn{2}{c}{$\langle (1-T)^2 \rangle$} 
& \multicolumn{2}{c}{$\langle C^2 \rangle$} 
& \multicolumn{2}{c}{\momn{\bt}{2}} 
& \multicolumn{2}{c}{\momn{\bw}{2}} 
& \multicolumn{2}{c}{$\langle \ytwothree{}^2 \rangle$} 
& \multicolumn{2}{c|}{\momn{\mh}{2}}
\bigstrut \\
\hline
 \hline
 {\bf\boldmath \asmz}
  &  $0.1427$  & &  $0.1412$  & &  $0.1344$  & &  $0.1216$  & &  $0.1235$  & &  $0.1226$  &
 \bigstrut \\ \hline
 Statistical error
  &  $\pm 0.0007$  & &  $\pm 0.0004$  & &  $\pm 0.0004$  & &  $\pm 0.0006$  & &  $\pm 0.0014$  & &  $\pm 0.0003$  &
 \bigstrut \\ \hline
 Experimental syst.
  &  $\pm 0.0017$  & &  $\pm 0.0015$  & &  $\pm 0.0012$  & &  $\pm 0.0020$  & &  $\pm 0.0031$  & &  $\pm 0.0018$  &
 \bigstrut \\ \hline
 HERWIG hadr.~corr.
  &  $+0.0006$  & &  $-0.0003$  & &  $-0.0011$  & &  $-0.0012$  &$^*$ &  $+0.0015$  &$^*$ &  $+0.0025$  &$^*$
 \bigstrut[t] \\ ARIADNE hadr.~corr.
  &  $+0.0041$  &$^*$ &  $+0.0040$  &$^*$ &  $+0.0029$  &$^*$ &  $+0.0002$  & &  $+0.0005$  & &  $+0.0020$  &
 \bigstrut[b] \\ \hline
 Hadronization error
  &  $\pm 0.0041$  & &  $\pm 0.0040$  & &  $\pm 0.0029$  & &  $\pm 0.0012$  & &  $\pm 0.0015$  & &  $\pm 0.0025$  &
 \bigstrut \\ \hline
 $x_\mu$ variation
  & $^{+0.0120}_{-0.0098}$ & & $^{+0.0118}_{-0.0096}$ & & $^{+0.0115}_{-0.0095}$ & & $^{+0.0050}_{-0.0036}$ & & $^{+0.0057}_{-0.0044}$ & & $^{+0.0045}_{-0.0030}$ & 
 \bigstrut \\ \hline
 $\chi^2 / dof.$
   &  $1.8/3$  & &  $2.5/3$  & &  $3.1/3$  & &  $1.3/3$  & &  $0.7/3$  & &  $0.8/3$  &
 \bigstrut \\ \hline
\hline
\hline
& \multicolumn{2}{c}{$\langle (1-T)^3 \rangle$} 
& \multicolumn{2}{c}{$\langle C^3 \rangle$} 
& \multicolumn{2}{c}{\momn{\bt}{3}} 
& \multicolumn{2}{c}{\momn{\bw}{3}} 
& \multicolumn{2}{c}{$\langle \ytwothree{}^3 \rangle$} 
& \multicolumn{2}{c|}{\momn{\mh}{3}}
\bigstrut \\
\hline
 \hline
 {\bf\boldmath \asmz}
  &  $0.1501$  & &  $0.1494$  & &  $0.1439$  & &  $0.1228$  & &  $0.1227$  & &  $0.1266$  &
 \bigstrut \\ \hline
 Statistical error
  &  $\pm 0.0015$  & &  $\pm 0.0006$  & &  $\pm 0.0009$  & &  $\pm 0.0014$  & &  $\pm 0.0029$  & &  $\pm 0.0006$  &
 \bigstrut \\ \hline
 Experimental syst.
  &  $\pm 0.0020$  & &  $\pm 0.0018$  & &  $\pm 0.0017$  & &  $\pm 0.0032$  & &  $\pm 0.0040$  & &  $\pm 0.0026$  &
 \bigstrut \\ \hline
 HERWIG hadr.~corr.
  &  $+0.0018$  & &  $+0.0009$  & &  $-0.0002$  & &  $-0.0010$  &$^*$ &  $+0.0016$  &$^*$ &  $+0.0035$  &$^*$
 \bigstrut[t] \\ ARIADNE hadr.~corr.
  &  $+0.0044$  &$^*$ &  $+0.0043$  &$^*$ &  $+0.0028$  &$^*$ &  $-0.0002$  & &  $+0.0006$  & &  $+0.0015$  &
 \bigstrut[b] \\ \hline
 Hadronization error
  &  $\pm 0.0044$  & &  $\pm 0.0043$  & &  $\pm 0.0028$  & &  $\pm 0.0010$  & &  $\pm 0.0016$  & &  $\pm 0.0035$  &
 \bigstrut \\ \hline
 $x_\mu$ variation
 & $^{+0.0144}_{-0.0117}$ & & $^{+0.0145}_{-0.0118}$ & & $^{+0.0149}_{-0.0121}$ & & $^{+0.0052}_{-0.0039}$ & & $^{+0.0056}_{-0.0042}$ & & $^{+0.0061}_{-0.0047}$ &
 \bigstrut \\ \hline
 $\chi^2 / dof.$
 &  $1.0/3$  & &  $1.7/3$  & &  $1.5/3$  & &  $1.2/3$  & &  $0.4/3$  & &  $1.0/3$  &
 \bigstrut \\ \hline
\hline
\hline
& \multicolumn{2}{c}{$\langle (1-T)^4 \rangle$} 
& \multicolumn{2}{c}{$\langle C^4 \rangle$} 
& \multicolumn{2}{c}{\momn{\bt}{4}} 
& \multicolumn{2}{c}{\momn{\bw}{4}} 
& \multicolumn{2}{c}{$\langle \ytwothree{}^4 \rangle$} 
& \multicolumn{2}{c|}{\momn{\mh}{4}}
\bigstrut \\
\hline
 \hline
 {\bf\boldmath \asmz}
  &  $0.1557$  & &  $0.1553$  & &  $0.1495$  & &  $0.1222$  & &  $0.1214$  & &  $0.1276$  &
 \bigstrut \\ \hline
 Statistical error
  &  $\pm 0.0028$  & &  $\pm 0.0007$  & &  $\pm 0.0017$  & &  $\pm 0.0030$  & &  $\pm 0.0057$  & &  $\pm 0.0009$  &
 \bigstrut \\ \hline
 Experimental syst.
  &  $\pm 0.0021$  & &  $\pm 0.0019$  & &  $\pm 0.0020$  & &  $\pm 0.0051$  & &  $\pm 0.0085$  & &  $\pm 0.0034$  &
 \bigstrut \\ \hline
 HERWIG hadr.~corr.
  &  $+0.0026$  & &  $+0.0016$  & &  $+0.0004$  & &  $-0.0011$  &$^*$ &  $+0.0015$  &$^*$ &  $+0.0040$  &$^*$
 \bigstrut[t] \\ ARIADNE hadr.~corr.
  &  $+0.0050$  &$^*$ &  $+0.0048$  &$^*$ &  $+0.0027$  &$^*$ &  $-0.0005$  & &  $+0.0006$  & &  $+0.0010$  &
 \bigstrut[b] \\ \hline
 Hadronization error
  &  $\pm 0.0050$  & &  $\pm 0.0048$  & &  $\pm 0.0027$  & &  $\pm 0.0011$  & &  $\pm 0.0015$  & &  $\pm 0.0040$  &
 \bigstrut[b] \\ \hline
 $x_\mu$ variation
 & $^{+0.0164}_{-0.0132}$ & & $^{+0.0168}_{-0.0134}$ & & $^{+0.0173}_{-0.0139}$ & & $^{+0.0049}_{-0.0035}$ & & $^{+0.0053}_{-0.0040}$ & & $^{+0.0063}_{-0.0049}$ &
 \bigstrut \\ \hline
 $\chi^2 / dof.$
  &  $0.6/3$  & &  $1.3/3$  & &  $0.9/3$  & &  $1.0/3$  & &  $0.3/3$  & &  $1.2/3$  &
 \bigstrut \\ \hline
\hline
\hline
& \multicolumn{2}{c}{$\langle (1-T)^5 \rangle$} 
& \multicolumn{2}{c}{$\langle C^5 \rangle$} 
& \multicolumn{2}{c}{\momn{\bt}{5}} 
& \multicolumn{2}{c}{\momn{\bw}{5}} 
& \multicolumn{2}{c}{$\langle \ytwothree{}^5 \rangle$} 
& \multicolumn{2}{c|}{\momn{\mh}{5}}
\bigstrut \\
\hline
 \hline
 {\bf\boldmath \asmz}
  &  $0.1615$  & &  $0.1606$  & &  $0.1541$  & &  $0.1217$  & &  $0.1196$  & &  $0.1269$  &
 \bigstrut \\ \hline
 Statistical error
  &  $\pm 0.0053$  & &  $\pm 0.0009$  & &  $\pm 0.0033$  & &  $\pm 0.0063$  & &  $\pm 0.0111$  & &  $\pm 0.0013$  &
 \bigstrut \\ \hline
 Experimental syst.
  &  $\pm 0.0027$  & &  $\pm 0.0019$  & &  $\pm 0.0026$  & &  $\pm 0.0086$  & &  $\pm 0.0207$  & &  $\pm 0.0040$  &
 \bigstrut \\ \hline
 HERWIG hadr.~corr.
  &  $+0.0032$  & &  $+0.0022$  & &  $+0.0007$  & &  $-0.0011$  &$^*$ &  $+0.0013$  &$^*$ &  $+0.0042$  &$^*$
 \bigstrut[t] \\ ARIADNE hadr.~corr.
  &  $+0.0058$  &$^*$ &  $+0.0052$  &$^*$ &  $+0.0026$  &$^*$ &  $-0.0008$  & &  $+0.0006$  & &  $+0.0007$  &
 \bigstrut[b] \\ \hline
 Hadronization error
  &  $\pm 0.0058$  & &  $\pm 0.0052$  & &  $\pm 0.0026$  & &  $\pm 0.0011$  & &  $\pm 0.0013$  & &  $\pm 0.0042$  &
 \bigstrut[b] \\ \hline
 $x_\mu$ variation
 & $^{+0.0185}_{-0.0147}$ & & $^{+0.0189}_{-0.0150}$ & & $^{+0.0195}_{-0.0154}$ & & $^{+0.0044}_{-0.0030}$ & & $^{+0.0049}_{-0.0037}$ & & $^{+0.0062}_{-0.0047}$ & 
 \bigstrut \\ \hline
 $\chi^2 / dof.$
  &  $0.4/3$  & &  $1.0/3$  & &  $0.5/3$  & &  $0.7/3$  & &  $0.2/3$  & &  $1.3/3$  &
 \bigstrut \\ \hline
\end{tabular}}
\end{center}
 \caption[ ]{Measurements of \asmz\  from event shape moments
over the full range of c.m.\ energy,
 91--209~GeV.  The hadronization error is taken to 
be the larger of the effects observed using \Herwig\ and \Ariadne;
in each case this is denoted by an asterisk.}
 \label{tab_momfits} 
 \end{table}

\clearpage

\clearpage

\clearpage

\clearpage

\clearpage

 \clearpage

  \clearpage

\clearpage

 \clearpage

 \clearpage

 \clearpage

 \clearpage

 \clearpage

\end{document}